\documentclass[11pt]{article}
\usepackage{jheppub}

\usepackage[space]{grffile}
\usepackage{pdfpages}

\usepackage{slashed}
\usepackage{graphicx,subfigure}
\usepackage{amsmath,amssymb}
\usepackage{gensymb} 
\usepackage{epsf,color}
\usepackage[dvipsnames]{xcolor}
\usepackage{hyperref}
\usepackage{cancel}
\usepackage{framed}
\usepackage{float}
\usepackage{multirow}
\usepackage{physics}
\usepackage{orcidlink,xspace}

\usepackage{diagbox}
\usepackage{makecell}
\usepackage{mathrsfs}
\usepackage{comment}

\definecolor{nicered}{rgb}{0.5,0.,0.}
\definecolor{nicegreen}{rgb}{0.,0.5,0.}
\definecolor{niceblue}{rgb}{0.,0.,0.5}
\hypersetup{colorlinks,citecolor=nicegreen,linkcolor=nicered,urlcolor=niceblue}
\numberwithin{equation}{section}
\newcommand{\beq}{\begin{equation}}
\newcommand{\eeq}{\end{equation}}
\newcommand{\bea}{\begin{eqnarray}}
\newcommand{\eea}{\end{eqnarray}}
\newcommand{\bear}{\begin{eqnarray}}
\newcommand{\eear}{\end{eqnarray}}

\newcommand{\GeV}{{\rm GeV}}

\newcommand{\TeV}{{\rm TeV}}

\newcommand{\ba}{\begin{array}}
\newcommand{\ea}{\end{array}}

\usepackage{enumitem}
\setlist{nolistsep}

\makeatletter
\gdef\@fpheader{}
\makeatother
\begin{document}
\preprint{COMETA-2025-02, IRMP-CP3-25-09, MSUHEP-25-002, CPTNP-2025-011}

\title{Light Axion-Like Particles at Future Lepton Colliders}
\newcommand{\lae}{\begin{array}{c}\,\sim\vspace{-21pt}\\<
\end{array}}
\newcommand{\gae}{\begin{array}{c}\,\sim\vspace{-21pt}\\>
\end{array}}

\arxivnumber{2505.10023}

\author[a,\orcidlink{0000-0002-0921-5329}]{Shou-shan Bao,}
\affiliation[a]{Institute of Frontier and Interdisciplinary Science,
MOE Key Laboratory of Particle Physics and Particle Irradiation, Shandong University (Qingdao Campus), Qingdao, Shandong, 266237, China}
\author[b,\orcidlink{0000-0002-9419-6598}]{Yang Ma,}
\affiliation[b]{Center for Cosmology, Particle Physics and Phenomenology, Universit\'e catholique de Louvain, B-1348 Louvain-la-Neuve, Belgium}
\author[c,d,\orcidlink{0000-0002-1835-7660}]{Yongcheng Wu,}
\affiliation[c]{Department of Physics, Institute of Theoretical Physics and Institute of Physics Frontiers and Interdisciplinary Sciences, Nanjing Normal University, Nanjing, 210023, China}
\affiliation[d]{Nanjing Key Laboratory of Particle Physics and Astrophysics, Nanjing, 210023, China}
\author[e,\orcidlink{0000-0003-4261-3393}]{Keping Xie,}
\affiliation[e]{Department of Physics and Astronomy, Michigan State University, East Lansing, MI 48824, USA}
\author[a,\orcidlink{0000-0003-2720-5404}]{Hong Zhang}

\emailAdd{ssbao@sdu.edu.cn}
\emailAdd{yang.ma@uclouvain.be}
\emailAdd{ycwu@njnu.edu.cn}
\emailAdd{xiekepi1@msu.edu}
\emailAdd{hong.zhang@sdu.edu.cn}
\date{\today}

\abstract{
Axion-like particles (ALPs) are well-motivated extensions of the Standard Model (SM) that appear in many new physics scenarios, with masses spanning a broad range. In this work, we systematically study the production and detection prospects of light ALPs at future lepton colliders, including electron-positron and multi-TeV muon colliders. At lepton colliders, light ALPs can be produced in association with a photon or a $Z$ boson. For very light ALPs ($m_a < 1$ MeV), the ALPs are typically long-lived and escape detection, leading to a mono-$V$ ($V = \gamma, Z$) signature. In the long-lived limit, we find that the mono-photon channel at the Tera-$Z$ stage of future electron-positron colliders provides the strongest constraints on ALP couplings to SM gauge bosons, $g_{aVV}$, thanks to the high luminosity, low background, and resonant enhancement from on-shell $Z$ bosons. At higher energies, the mono-photon cross section becomes nearly energy-independent, and the sensitivity is governed by luminosity and background. At multi-TeV muon colliders, the mono-$Z$ channel can yield complementary constraints. For heavier ALPs ($m_a > 100$ MeV) that decay promptly, mono-$V$ signatures are no longer valid. In this case, ALPs can be probed via non-resonant vector boson scattering (VBS) processes, where the ALP is exchanged off-shell, leading to kinematic deviations from SM expectations. We analyze constraints from both light-by-light scattering and electroweak VBS, the latter only accessible at TeV-scale colliders. While generally weaker, these constraints are robust and model-independent. Our combined analysis shows that mono-$V$ and non-resonant VBS channels provide powerful and complementary probes of ALP-gauge boson interactions.
}

\maketitle

\section{Introduction}
\label{sec:intro}

The axion was originally proposed to solve the strong CP problem within the Standard Model (SM) of particle physics \cite{Peccei:1977hh, Peccei:1977ur, Weinberg:1977ma, Wilczek:1977pj, Kim:1979if, Shifman:1979if, Zhitnitsky:1980tq, Dine:1981rt}. 
Over time, the concept has evolved, leading to the broader notion of axion-like particles (ALPs), which can emerge in various scenarios beyond the Standard Model (BSM). These ALPs are pseudo-Nambu-Goldstone bosons, arising from the spontaneous breaking of global $U(1)$ symmetries and behaving as singlet pseudoscalars under the SM gauge groups.
Their mass spectrum is notably expansive, ranging from $10^{-21}$ eV to the TeV scale, motivated by a diverse array of theoretical considerations and experimental data \cite{Chang:2000ii, Boehm:2014hva, Berlin:2014tja, Krasznahorkay:2015iga, Marciano:2016yhf, Feng:2016ysn, Ellwanger:2016wfe, Han:2022kvj}.
This broad range highlights the versatility of ALPs in addressing various theoretical challenges and accommodating different observational phenomena.

In general, there are five coupling constants to describe the interactions between the ALPs and the SM gauge bosons, \emph{i.e.}, $g_{a\gamma\gamma}$, $g_{a\gamma Z}$, $g_{a Z Z}$, $g_{a W W}$, and $g_{a G G}$, where $G$ denotes the gluon field.
Various experiments are in process or proposed to measure these couplings \cite{PrimEx:2010fvg,Proceedings:2012ulb,Irastorza:2018dyq, Graham:2015ouw,Semertzidis:2021rxs,Adams:2022pbo}.
Low energy electron-positron collider experiments, electron beam dump experiments, Electron-Ion Collider (EIC), and flavor physics experiments could provide good constraints on $g_{a\gamma\gamma}$ and the ALP mass $m_a$ \cite{Dolan:2017osp,Jiang:2018jqp,Belle-II:2020jti,Bauer:2021mvw,Liu:2023bby,Balkin:2023gya}. 
However, due to the suppression by the weak boson ($W/Z$) masses, the three couplings $g_{a\gamma Z}$, $g_{a Z Z}$, and $g_{a W W}$ could not be extracted reliably in the low-energy experiments. Thus, high-energy colliders are essential to explore these interactions and extract the constraints on these couplings~\cite{Mimasu:2014nea,Jaeckel:2015jla,Brivio:2017ije,Bauer:2017ris,Bauer:2018uxu,dEnterria:2021ljz}.

ALP signals at the Large Hadron Collider (LHC) have been extensively studied to probe the ALP-gluon coupling $g_{aGG}$~\cite{Bauer:2018uxu,Lee:2018pag,Frugiuele:2018coc,Lanfranchi:2020crw,Haghighat:2020nuh,dEnterria:2021ljz,Ghebretinsaea:2022djg}.  
The ALP-photon coupling $g_{a\gamma\gamma}$ can also be investigated at the LHC, \emph{e.g.}, via photon-jet production~\cite{Wang:2021uyb,Ren:2021prq} or light-by-light scattering in Pb-Pb collisions~\cite{CMS:2018erd,ATLAS:2020hii}.  
The ALP-weak-boson couplings $g_{a\gamma Z}$, $g_{aZZ}$, and $g_{aWW}$ can be constrained through mono-$Z/W$ production~\cite{Brivio:2017ije} and the non-observation of $Z$ decays into a photon and invisible particles ($Z \to \gamma+$inv.)~\cite{ATLAS:2020uiq}.  
Compared to the LHC, high-energy lepton colliders have the advantage of probing ALP interactions with electroweak (EW) gauge bosons in a cleaner environment, free from the hadronic background noise.
The ALPs have been previously searched at LEP~\cite{OPAL:2000puu,L3:1994shn,OPAL:2002vhf}, Babar~\cite{BaBar:2010eww}, and Belle II~\cite{Belle-II:2020jti}, providing foundational insights into their interactions with SM particles.
Future electron-positron colliders, including circular type, {\emph e.g.},~the Future Circular Collider (FCC-ee)~\cite{FCC:2018evy,Bernardi:2022hny} 
and the Circular Electron-Positron Collider (CEPC) \cite{CEPCStudyGroup:2018ghi,CEPCStudyGroup:2018rmc,An:2018dwb,CEPCAcceleratorStudyGroup:2019myu,CEPCPhysicsStudyGroup:2022uwl,CEPCStudyGroup:2023quu}, and the linear ones, {\emph e.g.}, the International Linear Collider (ILC)~\cite{ILC:2013jhg,ILCInternationalDevelopmentTeam:2022izu} and Compact Linear Collider (CLIC)~\cite{Aicheler:2012bya,Linssen:2012hp,Lebrun:2012hj,CLIC:2016zwp,CLICdp:2018cto,Brunner:2022usy}, are expected to offer new opportunities to study ALP couplings to SM particles with even greater precision~\cite{Bose:2022obr}.
Additionally, the proposed multi-TeV muon colliders~\cite{MuonCollider:2022xlm,Aime:2022flm,Black:2022cth,Accettura:2023ked,Delahaye:2019omf,Bartosik:2020xwr,Schulte:2021hgo,Long:2020wfp,MuonCollider:2022nsa,MuonCollider:2022ded,MuonCollider:2022glg,InternationalMuonCollider:2025sys} represent a novel opportunity at the energy frontier, merging the benefits of both electron-positron and hadron colliders~\cite{Costantini:2020stv,Han:2020uid,Han:2021kes,BuarqueFranzosi:2021wrv,Ruiz:2021tdt}. These colliders are particularly well-suited for probing EW processes and exploring BSM physics~\cite{Maltoni:2022bqs,Belloni:2022due}. 

At high-energy colliders, the ALPs can be produced directly in association with photons, jets, and EW gauge bosons \cite{Bauer:2018uxu, Lee:2018pag, Frugiuele:2018coc, Lanfranchi:2020crw, Haghighat:2020nuh, dEnterria:2021ljz, Ghebretinsaea:2022djg, Mimasu:2014nea, Jaeckel:2015jla,Brivio:2017ije, Bauer:2017ris, Wang:2021uyb, Ren:2021prq, Knapen:2016moh, Hook:2019qoh, Ebadi:2019gij, Yue:2019gbh,Wang:2022ock,Blasi:2023hvb,Bhattacharya:2025hme}, or from the decay of $Z$ boson \cite{Kim:1989xj,Djouadi:1990ms,Rupak:1995kg,Jaeckel:2015jla,Liu:2017zdh,Calibbi:2022izs,Yue:2022ash} and Higgs boson \cite{Dobrescu:2000jt,Dobrescu:2000yn,Chang:2006bw,Draper:2012xt,CMS:2012qms,Curtin:2013fra,ATLAS:2015rsn,CMS:2015twz,CMS:2017dmg,Bauer:2017nlg,Bauer:2017ris,An:2018myv,Li:2021ygc, CMS:2021pcy,Cepeda:2021rql,Davoudiasl:2021haa,Cepeda:2025diq}.
Generally, a heavy ALP is short-lived and can decay into SM particles within the detector, leaving a resonant signal. In this case, the ALP mass $m_a$ can be accurately determined by reconstructing the invariant mass spectrum.  
Recent studies highlight the potential of lepton colliders to search for resonant ALPs~\cite{Inan:2020aal, Inan:2020kif, Zhang:2021sio, Yue:2021iiu, Steinberg:2021wbs, Liu:2021lan, Han:2022mzp, Bao:2022onq, Inan:2022rcr, Yue:2023mew,RebelloTeles:2023uig, Buttazzo:2018qqp, Yue:2023mjm, Li:2024zhw, Polesello:2025gwj} and to probe ALP couplings to leptons and neutrinos~\cite{Haghighat:2021djz,Cheung:2021mol, Lu:2023ryd, Calibbi:2024rcm, Batell:2024cdl, Yue:2024xrc}.  
On the other hand, light ALPs can have long lifetimes, behaving as invisible recoils at colliders, which can also be used to study their interactions with SM particles~\cite{Mimasu:2014nea, Brivio:2017ije, Bauer:2017ris, Bauer:2018uxu, dEnterria:2021ljz,Cheung:2022umw, Wang:2024zky,Wang:2025xyh}.  
Moreover, processes involving off-shell ALP mediators play an important role, providing insights into ALP interactions that are independent of the ALP mass and decay width~\cite{Gavela:2019cmq, Florez:2021zoo, Carra:2021ycg, CMS:2021xor, Bonilla:2022pxu,Biswas:2023ksj}. 

The purpose of this work is to study the constraints that future lepton colliders can impose on the couplings between SM electroweak gauge bosons and ALPs. We consider both the proposed $e^+ e^-$ colliders~\cite{CEPCStudyGroup:2018rmc,CEPCStudyGroup:2018ghi,An:2018dwb,CEPCAcceleratorStudyGroup:2019myu,CEPCPhysicsStudyGroup:2022uwl,CEPCStudyGroup:2023quu,FCC:2018evy,Bernardi:2022hny} and the multi-TeV muon colliders~\cite{MuonCollider:2022xlm,Aime:2022flm,Black:2022cth,Accettura:2023ked,Delahaye:2019omf,Bartosik:2020xwr,Schulte:2021hgo,Long:2020wfp,MuonCollider:2022nsa,MuonCollider:2022ded,MuonCollider:2022glg,InternationalMuonCollider:2025sys}, focusing on ALPs with masses much smaller than the center-of-mass energy of the collider, i.e., $m_a \ll \sqrt{s}$.
In the case that ALP is produced in association with an SM gauge boson ($V=\gamma,Z)$, the long-lived ALP can fly out of the detector and leave a missing energy signature, \emph{e.g.}, in mono-boson (mono-$V$) production. 
By assuming that the ALP only interacts with the EW vector bosons, i.e., $\gamma$, $Z$, and $W^\pm$, we take the assumption ${\rm BR}(a\to \gamma\gamma) =1$ for the analysis.
The analysis is valid when the ALP is long-lived, for which we also need to make an assumption on the ALP lifetime.
Meanwhile, ALPs can participate in the vector boson scattering (VBS), including both light-by-light and EW VBS~\cite{BuarqueFranzosi:2021wrv}, in an off-shell and non-resonant way. The analysis for the non-resonant VBS channel does not rely on any assumption on the ALP's decay width or branch fraction.
Both the mono-$V$ and the off-shell scatterings are insensitive to the ALP mass $m_a$, and probe the ALP couplings uniquely. 
As will be shown later, the Tera-$Z$ phase of future electron-positron colliders can play a strong role in constraining the $g_{aVV}$ couplings through the mono-photon measurement, thanks to its high luminosity and clean environment.
In addition to the traditional mono-photon channel~\cite{OPAL:2000puu}, the mono-$Z$ channel and VBS processes at future high-energy lepton colliders could also provide complementary constraints on the ALP couplings to weak gauge bosons. 

The $a{\text -}V{\text -}V$ coupling has been studied at future lepton colliders, including FCC-ee~\cite{RebelloTeles:2023uig,Yue:2021iiu}, the ILC~\cite{Yue:2023mjm}, and multi-TeV muon colliders~\cite{Casarsa:2021rud,Han:2022mzp,Bao:2022onq,Inan:2022rcr,Li:2024zhw}. Most of these analyses focus on ALP production followed by its decay into a pair of vector bosons within the detector, which is in contrast to the focus of our study. Furthermore, previous works typically assume specific benchmark scenarios for the two operators $(C_W,\,C_B)$, and present constraints in terms of the effective coupling $g_{a\gamma\gamma}$ or the decay constant $f_a$. In our analysis, we treat both Wilson coefficients as independent parameters, so that the constraints on the $g_{aVV}$ couplings are more model-independent.

For long-lived ALPs, we consider not only the conventional mono-photon channel but also the mono-$Z$ process, and identify the region in ALP mass $m_a$ where the long-lived assumption remains valid. In the regime of heavier ALPs, where this assumption breaks down, we show that \textit{non-resonant vector boson scattering (VBS)} can provide robust constraints that are independent of the ALP mass and width.

The organization of this article is as follows. 
We will introduce a general parameterization of ALP models using an effective Lagrangian and review the current constraints on ALP couplings to SM gauge bosons from existing data in Sec.~\ref{sec:eft2exist}.
In Sec.~\ref{sec:monoV}, we analyze the long-lived ALP production in association with a gauge boson ($V=\gamma,Z)$ at future lepton colliders with the ALP flying out of the detector and leaving a mono-$V$ signature. In Sec.~\ref{sec:VBS}, we study the non-resonant VBS processes, including light-by-light and EW boson scatterings. Finally, the constraints from the mono-$V$ and non-resonant VBS channels are combined and compared with existing ones in Sec.~\ref{sec:comb2comp}. A summary of our findings is provided in Sec.~\ref{sec:summary}.

\section{Effective interactions and current bounds}
\label{sec:eft2exist}

In this section, we first introduce the effective interactions between the ALP and the EW gauge bosons, which set up the theoretical framework adopted in this work. Afterwards, we will review the existing bounds as the starting point of our analysis.    

\subsection{The effective interactions of Axion-Like Particles}
\label{sec:eft}

An axion-like particle (ALP) is a singlet pseudoscalar of the SM gauge group. In the framework of effective field theory (EFT) and under the linear representation, the ALP decay constant, $f_a$, represents the relevant new physics scale.
The most general CP-conserving effective Lagrangian that describes bosonic ALP couplings up to next-to-leading order (NLO) is expressed in terms of dimension-five operators. This work focuses on the effective interactions between the ALP and the SM EW gauge bosons ($\gamma$, $W$, and $Z$). The interaction terms in the effective Lagrangian are given by
\begin{align}\label{eq:EffectiveL}
\mathcal{L}_{\mathrm{eff}} \supset -C_B \frac{a}{f_a} B_{\mu\nu} \tilde{B}^{\mu\nu} - C_W \frac{a}{f_a} \sum_{i=1}^3 W^i_{\mu\nu} \tilde{W}^{i;\mu\nu},
\end{align}
where $B_{\mu\nu}$ and $W^i_{\mu\nu}$ $(i=1,2,3)$ are the field strength tensors of the SM EW gauge bosons, and the dual tensors are defined as $\widetilde{F}^{\mu\nu} = \frac{1}{2} \varepsilon^{\mu\nu\alpha\beta} F_{\alpha\beta}$. The self-interactions of ALPs are suppressed by additional powers of $1/f_a$ and therefore omitted in Eq.~\eqref{eq:EffectiveL}. Similarly, couplings between an ALP and more than two vector bosons are also suppressed by $1/f_a$ and are neglected.
Some earlier studies, such as \cite{Bao:2022onq}, included a term proportional to $B_{\mu\nu} W^{3,\mu\nu}$ , which can arise from dimension-7 operators. These terms are further suppressed by two powers of $v_\text{EW}/f_a$, where $v_\text{EW} = 246$ GeV is the EW vacuum expectation value. Since our analysis focuses on the regime $f_a \gg v_\text{EW}$, this term is irrelevant to our considerations.
After electroweak symmetry breaking (EWSB), the interaction terms in the effective Lagrangian can be written as
\begin{equation}
\begin{aligned}\label{eq:L-after}
    \begin{split}
    \mathcal{L}_{\mathrm{eff}}\supset
    -\frac{g_{a\gamma\gamma}}{4} a F_{\mu\nu}\widetilde{F}^{\mu\nu}
    -\frac{g_{a\gamma Z}}{4} a F_{\mu\nu}\widetilde{Z}^{\mu\nu}
    -\frac{g_{aZZ}}{4} aZ_{\mu\nu}\widetilde{Z}^{\mu\nu}
    -\sum_{i=\pm} \frac{g_{aWW}}{2} a W_{\mu\nu}^i\widetilde{W}^{i;\mu\nu}.
    \end{split}
\end{aligned}    
\end{equation}
The interaction couplings in the above equation depend linearly on $C_W$ and $C_B$ as follows:
\begin{equation}\label{eq:couplings}
\begin{aligned}
    g_{a\gamma\gamma} &= \frac{4}{f_a} (s^2_{W}C_W + c^2_{W}C_B) ,&
    g_{a\gamma Z} &=\frac{ 8}{f_a}s_W c_W (C_W - C_B),\\
    g_{aZZ} &= \frac{4}{f_a} (c^2_{W}C_W + s^2_{W} C_B) , &
    g_{aWW} &=\frac{4}{f_a} C_W, 
\end{aligned}
\end{equation}
with $s_W=\sin\theta_W$ and $c_W=\cos\theta_W$, where $\theta_W$ is the Weinberg mixing angle. 
Throughout this work, our calculations are primarily carried out using \textsc{MadGraph\_aMC@NLO}~\cite{Alwall:2014hca,Frederix:2018nkq}, with results cross-checked via \textsc{Whizard}~\cite{Kilian:2007gr,Moretti:2001zz,Brass:2018xbv}. The BSM framework is implemented as a UFO model file~\cite{Degrande:2011ua,Darme:2023jdn} generated with \textsc{FeynRules}~\cite{Christensen:2008py,Alloul:2013bka}.

The $a\text{-}V_1\text{-}V_2$ vertex contains the Levi-Civita symbol $\varepsilon^{\mu\nu\alpha\beta}$, and the corresponding Feynman rule is in the form
\begin{eqnarray}
    -i g_{aV_1V_2} \varepsilon^{\mu\nu\alpha\beta} p^{V_1}_{\alpha}p^{V_2}_\beta \epsilon_{\mu}^{V_1} \epsilon_{\nu}^{V_2},\label{eq:gaVV}
\end{eqnarray}
where $p^{V_1}$ ($p^{V_2}$) and $\epsilon_{\mu}^{V_1}$ ($\epsilon_{\nu}^{V_2}$) are the momentum and polarization vector of the vector boson $V_1$ ($V_2$).
The contraction of the Levi-Civita symbol requires four independent 4-vectors, for which the ALP couples only to transversely polarized vector bosons, with no contribution from longitudinally polarized ones.
Also, the interaction prefers large momentum transfer, favoring a higher partonic center-of-mass energy.


\subsection{Review of existing collider constraints}
\label{sec:CurrentConstraints}

A massive ALP can decay into photon pairs, with a decay width
\begin{align}\label{eq:Gamma}
    \Gamma_a=\frac{g_{a\gamma\gamma}^2 m_a^3}{64\pi}.
\end{align}
For simplicity, we assume that the light ALP decays exclusively into a pair of photons, \emph{i.e.}, ${\rm Br}(a\to\gamma\gamma)=1$.
If the ALP decays before reaching the electromagnetic calorimeter (ECAL), the signal for BSM effects will manifest as a photon pair. Conversely, if the ALP does not decay and escapes from the ECAL, it contributes to the missing energy signal.
The decay length of the ALP is expressed as
\begin{equation}
    L_D = c \beta_a \gamma_a \tau_a = \frac{c p_a}{m_a \Gamma_a},
\end{equation}
where $p_a$ is the momentum of the ALP, $\tau_a = 1 / \Gamma_a$ is the ALP lifetime, and $\beta_a$ and $\gamma_a$ are its velocity and Lorentz boost factor, respectively.
Assuming the distance from the collision point to the outer layer of the ECAL is $d$, the fraction of ALPs that survive (\emph{i.e.}, do not decay) within the ECAL is
\begin{equation}
    \mathcal{P}_a = \exp\left(-\frac{d}{L_D}\right). \label{eq:P_a}
\end{equation}
At $d=L_D$, $36.8\%$ of the ALPs will survive. 

\begin{figure}[htb]
\begin{center}
\includegraphics[width=0.5\textwidth]{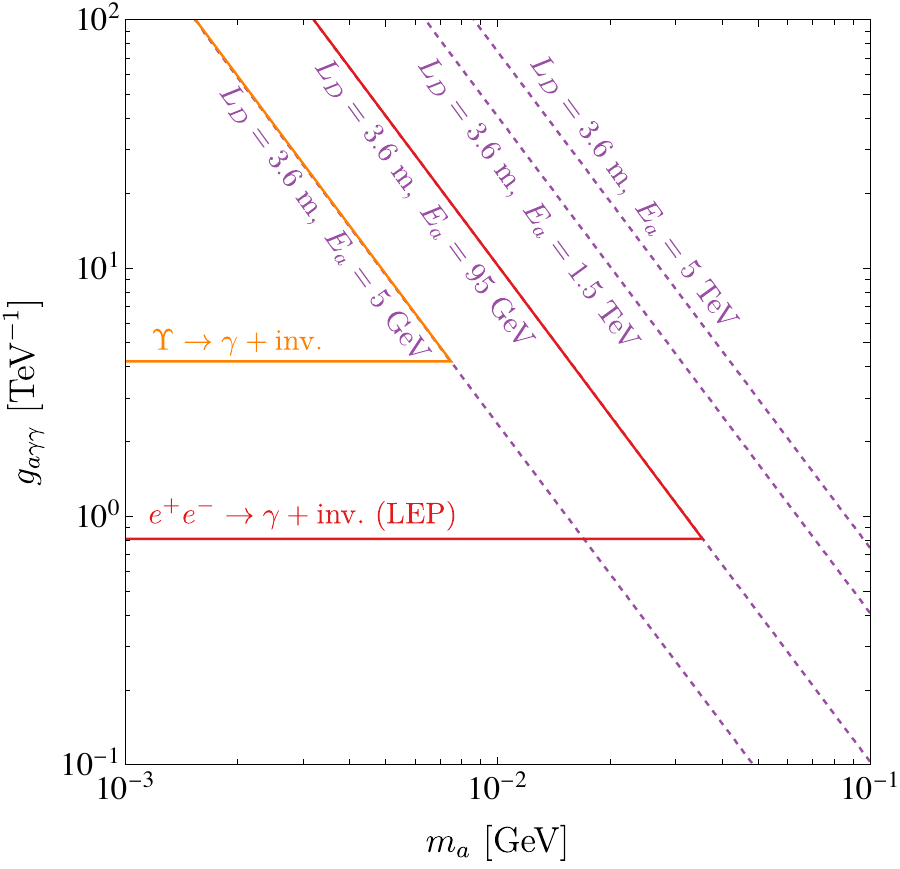}
\caption{The decay length contours of the ALP in the $(m_a,\, g_{a\gamma\gamma})$ plane. Above the purple contours, the ALP can be identified through its decay into photon pairs, while below the contours, the ALP remains invisible. For reference, we also display the constraints on $g_{a\gamma\gamma}$ from $\Upsilon \to \gamma +{\rm inv.}$~\cite{CrystalBall:1990xec,Masso:1995tw} using Crystal Ball data and a recast of the mono-photon search at the LEP 189~GeV~\cite{OPAL:2000puu}. A hard cut at $L_D=3.6\,$ m is applied to ensure the validity of the results.}\label{fig:lifetime}
\end{center}
\end{figure}

For illustration, we present the contours for $L_D=3.6\,$m in the $(m_a,\,g_{a\gamma\gamma})$ plane in Figure~\ref{fig:lifetime}, which assumes a typical distance of $d=3.6\,$m from the collision point to the ECAL exit at future lepton colliders \cite{Linssen:2012hp, Bacchetta:2019fmz}.
Below the purple contours, ALPs are recognized as long-lived and invisible particles. Above the contours, the signal for ALPs arises from photon pairs.
In this study, we focus on the scenario where the light ALPs are long-lived and escape from the ECAL without decaying.

The coefficients $C_W$ and $C_B$ can be extracted by measuring two of the four couplings in Eq.~\eqref{eq:couplings}. Among these, $g_{a\gamma\gamma}$ and $g_{a\gamma Z}$ are the most accessible in current experiments. 
For an ALP lighter than the $Z$ boson, the ALP can be produced in $Z$ decays, with the decay width given by:
\begin{align}
    \Gamma(Z\to \gamma +a)=\frac{g^2_{a\gamma Z}}{384\pi}M_Z^3.
\end{align} 
Experimental measurements of the $Z$ boson decay width can be used to constrain new physics, with $\Gamma(Z\to \mathrm{BSM})\leq 2$ MeV at 95\% confidence level (CL) ~\cite{Brivio:2017ije}. This provides a conservative bound on $|g_{a \gamma Z}|$
\begin{align}
    \vert g_{a\gamma Z}\vert\leq 1.8~\mathrm{TeV}^{-1},
\end{align}
where the branching fraction $\mathrm{Br}(Z\to \gamma + \pi^0) \leq 2.01 \times 10^{-5}$ is used as the upper limit on $\mathrm{Br}(Z\to \gamma +{\rm inv.})$ \cite{Workman:2022ynf}.
%
Hadron rare decays can also be used to constrain $g_{a\gamma\gamma}$ , such as the decay $\Upsilon \to \gamma + {\rm inv.}$ \cite{Masso:1995tw}, where the ALP does not decay before escaping from the ECAL. The branching fraction for this decay is given by~\cite{Wilczek:1977pj,Wilczek:1977zn}
\begin{align}
    \mathrm{Br}(\Upsilon\to\gamma+a) = \frac{ g^2_{a\gamma\gamma} m_b^2}{8\pi\alpha} \mathrm{Br}(\Upsilon\to e^+ e^-), 
\end{align}
where $m_b$ is the bottom quark mass, and $\alpha$ is the fine structure constant. 
Using the measurements of $\Upsilon$ decay from Refs.\cite{CrystalBall:1990xec,Workman:2022ynf}, the corresponding constraint at 95\% CL on $g_{a\gamma\gamma}$ is derived as \cite{Masso:1995tw}. 
\begin{align}
    |g_{a\gamma\gamma}|\leq 4.2~\mathrm{TeV}^{-1}.
\end{align}

\begin{figure}[htb]
\centering
\subfigure[]{\includegraphics[width=0.21\textwidth]{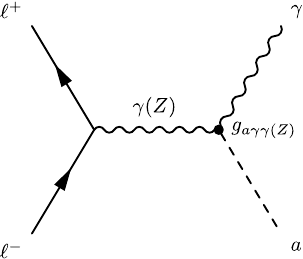}\label{feyn:agm}}
\subfigure[]{\includegraphics[width=0.24\textwidth]{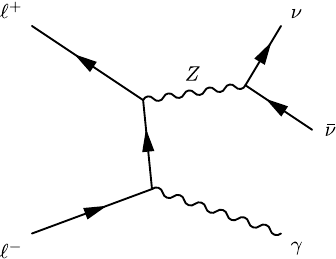}\label{feyn:Zgm}}
\subfigure[]{\includegraphics[width=0.24\textwidth]{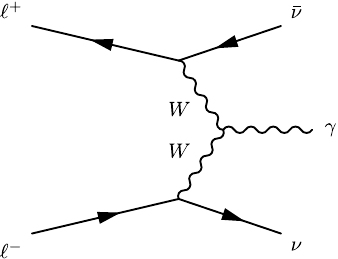}\label{feyn:WW}}
\subfigure[]{\includegraphics[width=0.24\textwidth]{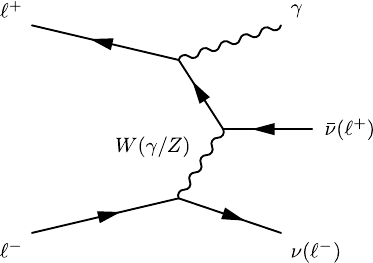}\label{feyn:Wex}}
\caption{Representatives Feynman diagrams for the signal (a) and backgrounds (b--d) for the mono-photon production at lepton colliders.}
\label{feyn:mono-pho}
\end{figure}

At high-energy lepton colliders, the leading production channel for ALPs is the process $\ell^+\ell^- \to \gamma + a$, where the ALP is produced in association with a photon.
For long-lived ALPs, this process manifests as mono-photon production, $\ell^+\ell^-\to \gamma +{\rm inv.}$
The Feynman diagrams for the signal and background processes are shown in Figure~\ref{feyn:mono-pho}.
For light ALPs with $m_a \ll \sqrt{s}$, the theoretical production cross-section for $\ell^+ \ell^- \to \gamma + a$ is given by
\begin{align}
    \sigma_{\ell^+ \ell^- \to \gamma a} = \frac{\alpha}{768} \left[ 32  g_{a\gamma\gamma}^2 + \frac{8 g_{a\gamma\gamma} g_{a\gamma Z} (c_W^2-3 s_W^2) s}{s_W c_W (s-M_Z^2)} + \frac{g_{a\gamma Z}^2 (6 s_W^4 + 2 c_W^4-1)s^2}{s_W^2 c_W^2 (s-M_Z^2)^2}\right],\label{eq:monophoton}
\end{align}
where $s$ is the square of the center-of-mass energy. 
At lower energy machines such as BaBar~\cite{BaBar:2001yhh} or Belle II~\cite{Belle-II:2018jsg}, the second and third terms in Eq.~\eqref{eq:monophoton} are highly suppressed due to the $Z$ boson propagator, leaving the first term dominant. This allows for a direct measurement of $g_{a\gamma\gamma}$. 
For higher energy machines with $\sqrt{s} \geq M_Z$, 
the second and third terms in Eq.~\eqref{eq:monophoton} grow larger, becoming non-negligible compared to the first term. This introduces a dependence on $g_{a\gamma Z}$ in the cross-section, providing an opportunity to simultaneously constrain $C_W$ and $C_B$.

\begin{figure}[!htb]
\centering
\subfigure[]{\includegraphics[width=0.49\textwidth]{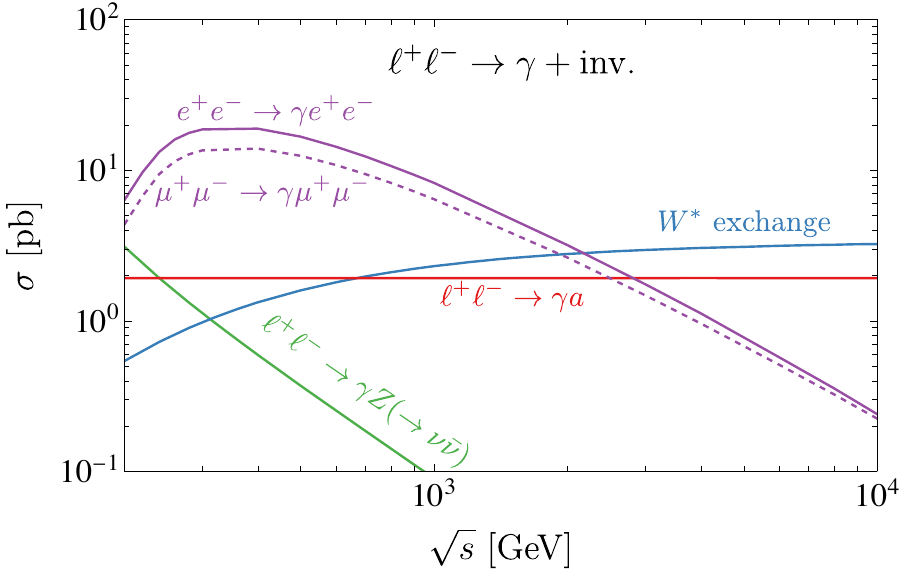}\label{fig:monoAxsec}}
\subfigure[]{\includegraphics[width=0.49\textwidth]{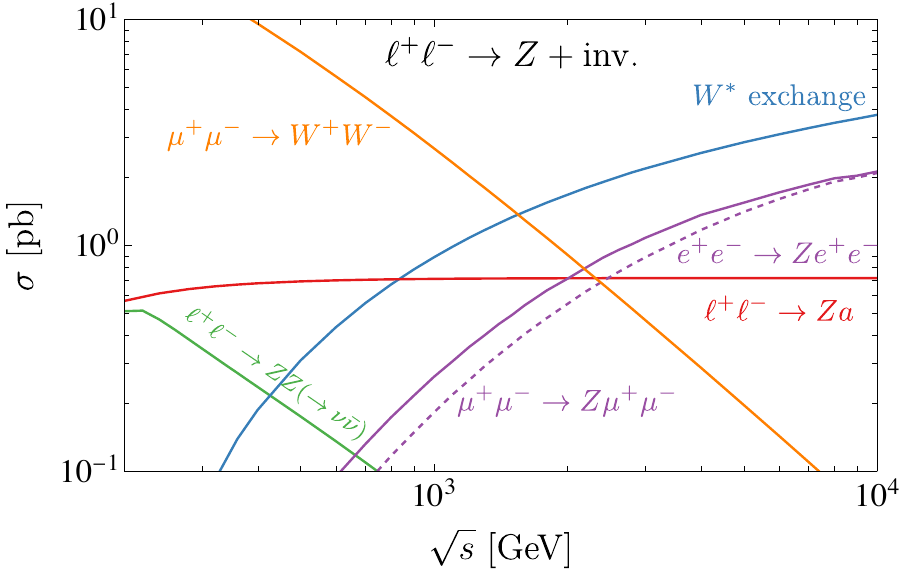}\label{fig:monoZxsec}}
\caption{The cross-sections of the signal and background processes for (a) mono-photon and (b) mono-$Z$ production as functions of the collider energy. The signal cross-sections are computed with $C_W/f_a=C_B/f_a=1\,{\rm TeV}^{-1}$. The mono-photon production is evaluated with universal photon cuts $p_{T,\gamma} > 10$ GeV and $|\eta_\gamma| < 2.5$.
The final-state leptons are required to be outside the detector coverage ($|\eta_\ell| > 2.5$) for the $\gamma/Z$ exchange process $\ell^+ \ell^- \to V + \ell^+ \ell^-$.
Additionally, an invariant mass cut of $M_{\ell\ell,\nu\nu} > 150$ GeV is applied to suppress on-shell $Z \to \ell^+ \ell^- / \nu_\ell \bar{\nu}_\ell$ decays in the $\gamma/Z$ and $W$ exchange channels.}
\label{fig:momoVxsec}
\end{figure}

Assuming $C_W/f_a=C_B/f_a=1\,{\rm TeV}^{-1}$, we present the leading-order theoretical predictions for the ALP signal and the background $\ell^+ \ell^-\to \gamma  \nu\bar{\nu}$ from on-shell $Z$ production $\ell^+ \ell^-\to \gamma Z(\to \nu\bar{\nu})$ in Figure~\ref{feyn:Zgm} as well as $W^*$ exchange, in the charged-current VBS (Figure~\ref{feyn:WW}) and $t$-channel $W\nu_\ell$ scattering (Figure~\ref{feyn:Wex}), with universal cuts $p_{T,\gamma}>10$~GeV\footnote{A basic $p_{T,\gamma}$ cut is necessary to avoid the soft and collinear divergence from the charged- and neutral-current gauge boson exchange, similar to Figure~\ref{feyn:Wex}.
In addition, we require $M_{\ell\ell,\nu\nu}>150~\GeV$ to suppress the on-shell $Z$ contribution.} and $|\eta_\gamma|<2.5$ in Figure~\ref{fig:monoAxsec}. Similarly, we also have another source of the background from the neutral-current $\gamma/Z$ exchange $\ell^+ \ell^-\to \gamma  \ell^+ \ell^-$ process, where final-state leptons move forwardly and escape detectors. 

\begin{figure}[htb]
    \begin{centering}
    \includegraphics[width=0.5\textwidth]{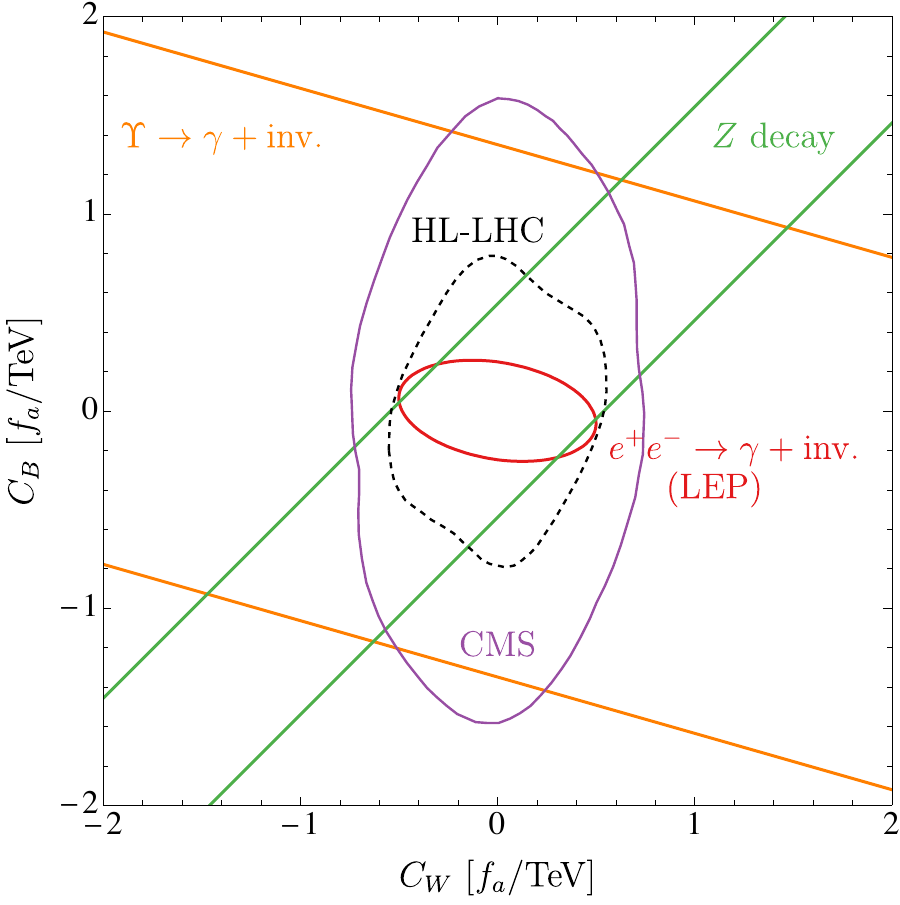}
    \caption{Current constraints on $C_W$ and $C_B$ in units of $f_a / {\rm TeV}$. The red contour is a recast of the mono-photon search at the LEP 189~GeV~\cite{OPAL:2000puu}, the orange and green bounds are derived from the $\Upsilon$~\cite{CrystalBall:1990xec,Masso:1995tw} and $Z$~\cite{Brivio:2017ije} decays, respectively. The purple contour shows constraints from non-resonant VBS at CMS Run 2~\cite{CMS:2020fqz,CMS:2020gfh,CMS:2020ypo,CMS:2021gme}, while the dashed contour denotes the projected sensitivity at the 3~${\rm ab}^{-1}$ HL-LHC~\cite{Bonilla:2022pxu}.
    }\label{fig:alp_limit}
    \end{centering}
\end{figure}

The cross-section of the mono-photon production process $e^+ e^-\to \gamma+{\rm inv.}$ at LEP with $\sqrt{s}= 189$~GeV, as reported by OPAL~\cite{OPAL:2000puu}, is 
\begin{align}
\sigma^{\rm Exp}=4.35 \pm  0.17\pm 0.09~\text{pb}.
\end{align}
The corresponding SM prediction of $e^+ e^-\to \gamma Z(\to\nu\bar{\nu})$ from \textsc{KORALZ} generator~\cite{Jadach:1993yv} is
\begin{align}
\sigma^{\rm SM}=4.66\pm 0.03~\text{pb}.
\end{align}
The consistency between the SM prediction and OPAL measurement can potentially set a strong limit on potential BSM physics. 
In this work, we take the measure
\begin{align}
\mathcal{S}=\left\vert\frac{\sigma^{\rm ALP}+\sigma^{\rm SM}-\sigma^{\rm Exp}}{\sqrt{0.17^2+0.09^2+0.03^2}}\right\vert=2(3),\label{eq:significance}
\end{align}
to project the sensitivity to ALP coupling at the 95\% (98\%) confidence level (CL).
The obtained 95\% CL limit on $C_W$ and $C_B$ is shown as contour plots in Figure~\ref{fig:alp_limit}.
For comparison, the combined constraint at 95\% CL from CMS using LHC Run 2 data~\cite{CMS:2020fqz,CMS:2020gfh,CMS:2020ypo,CMS:2021gme} and the expected sensitivity at the HL-LHC with 3 ab$^{-1}$~\cite{Bonilla:2022pxu} are included. When translating to $g_{a\gamma\gamma}$, the corresponding 95\% CL bounds dependent on $m_a$ are presented in Figure~\ref{fig:lifetime}.

\section{Gauge boson associated ALP productions}
\label{sec:monoV}

Since now on, we begin to explore the ALP phenomenology at future lepton colliders, including electron (CEPC~\cite{CEPCStudyGroup:2018ghi,CEPCStudyGroup:2023quu}/FCC-ee~\cite{FCC:2018evy,Bernardi:2022hny}) and muon colliders~\cite{MuonCollider:2022xlm,Aime:2022flm,Black:2022cth,Accettura:2023ked,Delahaye:2019omf,Bartosik:2020xwr,Schulte:2021hgo,Long:2020wfp,MuonCollider:2022nsa,MuonCollider:2022ded,MuonCollider:2022glg,InternationalMuonCollider:2025sys}.
Our results are equally applicable to ILC~\cite{ILC:2013jhg,ILCInternationalDevelopmentTeam:2022izu} and CLIC~\cite{Aicheler:2012bya,Linssen:2012hp,Lebrun:2012hj,CLIC:2016zwp,CLICdp:2018cto,Brunner:2022usy}, while the corresponding reaches are expected to be lower considering their smaller luminosities.

Apart from the photon-associated ALP production, $\ell^+\ell^- \to \gamma + a$, which has been briefly discussed in Sec.~\ref{sec:CurrentConstraints}, another ALP production process at high-energy lepton colliders is the $Z$-associated production, $\ell^+\ell^- \to Z + a$. The corresponding Feynman diagrams are analogous to those in Figure~\ref{feyn:mono-pho}, with the photon replaced by a $Z$ boson in the final state and/or propagator, and $g_{a\gamma\gamma}$ replaced by $g_{a\gamma Z}$ or $g_{aZZ}$. For long-lived ALPs that escape from the ECAL, the BSM signals manifest mono-$Z$ in addition to the mono-photon production, respectively. In this section, we present a systematic analysis of these two channels.

In our detailed analysis, we use \textsc{MadGraph5\_aMC@NLO}~\cite{Alwall:2014hca,Frederix:2018nkq} to generate the signal and background events, with detector response  simulations carried out in \textsc{Delphes3}~\cite{deFavereau:2013fsa}, utilizing its built-in cards for CEPC~\cite{CEPCStudyGroup:2018ghi,CEPCStudyGroup:2023quu}/FCC-ee~\cite{FCC:2018evy,Bernardi:2022hny} and muon colliders~\cite{InternationalMuonCollider:2024jyv,MuCoL:2024oxj}.   
The signal strength of the BSM effect is expressed as  
\begin{eqnarray}
\mathcal{S} = \sqrt{2(S+B)\log\left(1+\frac{S}{B}\right) - 2S}, \label{eq:signal_strength}
\end{eqnarray}
where $S$ is the number of signal events and $B$ is the number of background events. 
In the limit $S \ll B$, $\mathcal{S}$ can be approximated as  
\begin{eqnarray}
\mathcal{S} \simeq \frac{S}{\sqrt{B}}.
\end{eqnarray}
$\mathcal{S}=1(2)$ corresponds the exclusion at 68\% (95\%) CL.
The signal and background event numbers are given by  
\begin{eqnarray}
S(C_W,C_B) = \mathscr{L} \mathcal{P}_a \sigma_{\ell^+\ell^-\to V a}(C_W,C_B), \quad B = \mathscr{L} \sigma_{\ell^+\ell^-\to V+\rm{inv.}},
\end{eqnarray}
where $\mathscr{L}$ is the collider luminosity, $\mathcal{P}_a$ is the survival fraction of ALPs leaving the ECAL as described in Eq.~\eqref{eq:P_a}, $\sigma_{\ell^+\ell^-\to V a}$ is the ALP production cross-section, and $\sigma_{\ell^+\ell^-\to V+\rm{inv.}}$ is the SM background cross-section.  
In this section, we present the constraints on $C_W$, $C_B$, and the $g_{aVV}$ couplings in the long-lived limit ($L_D > d$), where the survival fraction simplifies to $\mathcal{P}_a \simeq 1$. A discussion of constraints near the mono-$V$ validity threshold ($L_D \simeq d$) will be provided in Sec.~\ref{sec:combine}.

\subsection{Mono-photon production}
\label{sec:monoA}
The mono-photon production process is the most widely used channel to study ALPs at lepton colliders, as photons are well-detected by detectors and the associated background is relatively simple. Notably, the cross-section for $\ell^+\ell^- \to \gamma + a$, given in Eq.~\eqref{eq:monophoton}, is independent of the ALP mass $m_a$ as long as $m_a \ll \sqrt{s}$. Consequently, the constraints on $g_{a\gamma\gamma}$ and $g_{a\gamma Z}$ depend on the ALP mass $m_a$ only through the probability ${\mathcal P}_a$, which determines whether the mono-photon channel is valid.
When the ALP is sufficiently long-lived (i.e., $L_D \gg d$), the constraints become independent of $m_a$. However, when $L_D \sim d$, the lifetime and decay length of the ALP must be carefully considered, as they influence the signal classification and the resulting constraints.
\subsubsection{Mono-photon production at future $Z$ factories}
\label{sec:monoA1}
The Tera-$Z$ phase of future $e^+e^-$ colliders, such as the CEPC and the FCC-ee, offers a powerful tool for precision physics studies. Near the $Z$ resonance peak, i.e. $\sqrt{s}\simeq M_Z$, the FCC-ee is projected to deliver integrated luminosities of 40 ab$^{-1}$ at 87.9 GeV, 125 ab$^{-1}$ at 91.2 GeV, and 40 ab$^{-1}$ at 94.3 GeV~\cite{janot_2024_yr3v6-dgh16}. Similarly, the CEPC expects a comparable luminosity, with $\mathscr{L}\sim 100$ ab$^{-1}$ at the $Z$ peak and lower luminosities off-peak ($\mathscr{L}\sim 1$ ab$^{-1}$)~\cite{CEPCPhysicsStudyGroup:2022uwl}.

Near the $Z$ resonance peak, the cross section in Eq.~\eqref{eq:monophoton} can be written in the Breit-Wigner form:
\begin{equation}
\begin{aligned}\label{eq:monophoton_BW}  
    \sigma_{\ell^+ \ell^- \to \gamma a} = \frac{\alpha}{768}& \left[ 32  g_{a\gamma\gamma}^2 + \frac{8\, g_{a\gamma\gamma}\, g_{a\gamma Z}\, (c_W^2-3 s_W^2)\, s(s-M_Z^2)}{s_W c_W \left(s-M_Z^2+ M_Z^2\Gamma_Z^2\right)}+\right. \\
    &\left. \frac{g_{a\gamma Z}^2\, (6 s_W^4 + 2 c_W^4-1)\, s^2}{s_W^2 c_W^2 \left((s-M_Z^2)^2 + M_Z^2 \Gamma_Z^2\right)}\right].  
\end{aligned}    
\end{equation}
Here, the $g_{a\gamma Z}^2$ term is highly enhanced and the interference ($g_{a\gamma\gamma}\, g_{a\gamma Z}$) term is suppressed when $(s-M_Z^2)\simeq 0$. In particular, the interference term vanishes on the $Z$ peak, leading to
\begin{eqnarray}
    \sigma_{\ell^+ \ell^- \to \gamma a} (s=M_Z^2) = \frac{\alpha}{768} \left[ 32  g_{a\gamma\gamma}^2 + \frac{g_{a\gamma Z}^2\, (6 s_W^4 + 2 c_W^4-1)\, M_Z^2}{s_W^2 c_W^2\, \Gamma_Z^2}\right].\label{eq:monophoton_Z}
\end{eqnarray}
As indicated in Eq.~\eqref{eq:couplings}, when $C_W = C_B$, the coupling $g_{a\gamma Z}$ vanishes, and thus the resonant enhancement from the on-shell $Z$ does not occur in this case.

\begin{figure}[!htb]
\centering
\subfigure[]{\includegraphics[width=0.48\textwidth]{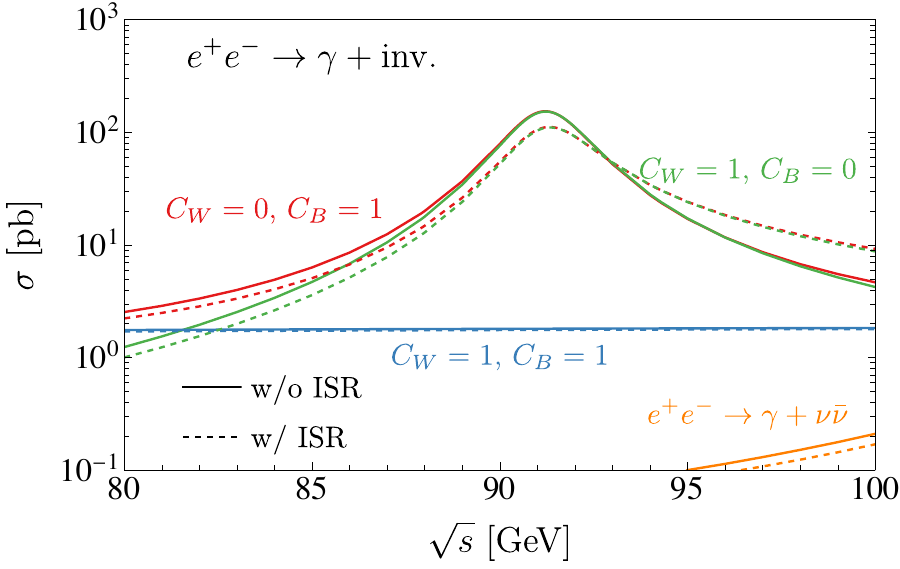}\label{fig:monoAxsec_ZF}}
\subfigure[]{\includegraphics[width=0.475
\textwidth]{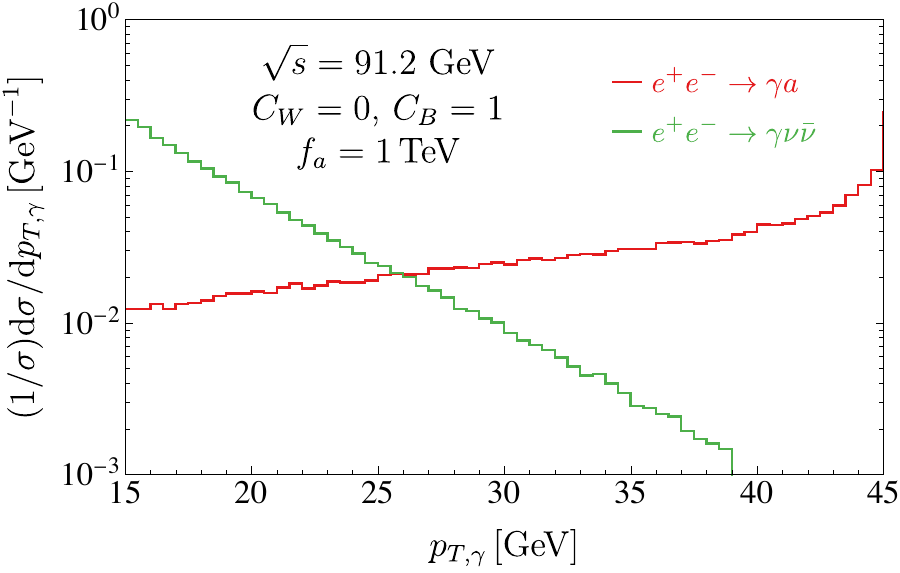}\label{fig:alp_ZF_pt}}
\caption{(a) Cross-sections of the signal and background processes for mono-photon production at $e^+e^-$ colliders with $\sqrt{s} \approx M_Z$, applying a baseline cut of $p_{T,\gamma} > 15$~GeV and $|\eta_\gamma| < 2.5$.  
(b) Normalized $p_{T,\gamma}$ distribution of the photon produced by the signal and background processes at a 91.2~GeV $e^+e^-$ collider.}
\end{figure}

The dominant SM background for the mono-photon process at the Tera-$Z$ phase arises from $e^+e^- \to \gamma Z(\nu \bar{\nu})$, which can be efficiently reduced by applying cuts on the photon energy $E_\gamma$ or its transverse momentum $p_{T,\gamma}$. Since initial-state radiation (ISR) may significantly affect production cross sections near a resonance peak, we include ISR effects in our analysis using the \textsc{Whizard} package ~\cite{Kilian:2007gr,Moretti:2001zz,Brass:2018xbv}. In Figure~\ref{fig:monoAxsec_ZF} the cross section for $e^+e^- \to \gamma + {\rm inv.}$ is displayed as a function of the center-of-mass energy $\sqrt{s}$ under a baseline selection of $p_{T,\gamma} > 15$~GeV and $|\eta_\gamma| < 2.5$. The solid curves show the cross sections without ISR, while the dashed curves include ISR effects. For illustration, three benchmark scenarios in $(C_W,\,C_B)$ are considered: (0,1) in red, (1,0) in green, and (1,1) in blue. The irreducible SM background from the process $e^+e^- \to \gamma \nu_e \bar{\nu}_e$ is represented by the orange curves. 
Notably, the ALP production rate near the $Z$ peak could be resonantly enhanced via the $s$-channel $Z$ diagram, imposing strong constraints on $g_{a\gamma Z}$. Furthermore, the SM background is almost one order of magnitude smaller compared to the case with $g_{a\gamma Z}=0$, allowing for precise measurements of the $g_{a\gamma\gamma}$ coupling. To further elucidate the differences, we present the normalized $p_{T,\gamma}$ distribution at $\sqrt{s}=91.2$~GeV for the (0,1) benchmark (in red) and the SM background (in green) in Figure~\ref{fig:alp_ZF_pt}, demonstrating that an appropriate $p_{T,\gamma}$ cut can further optimize the signal sensitivity.

\begin{figure}[htb]
    \begin{centering}
    \includegraphics[width=0.5\textwidth]{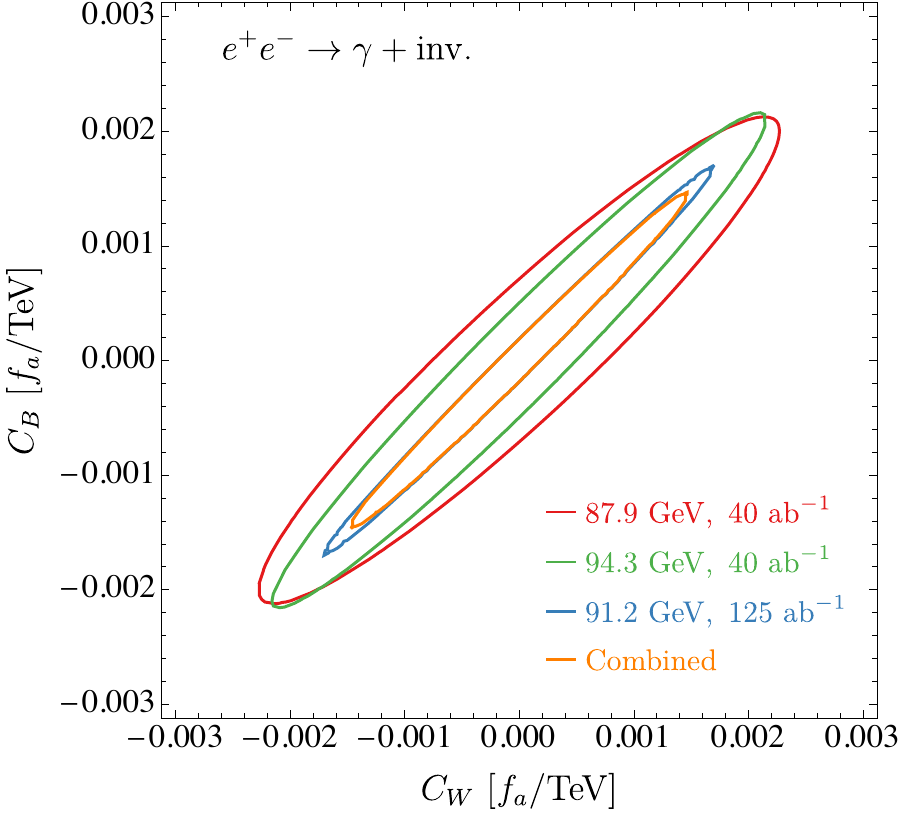}
    \caption{Constraints on $C_W$ and $C_B$ in units of $f_a/\text{TeV}$ for the proposed Tera-$Z$ phase of FCC-ee.}\label{fig:alp_ZF}
    \end{centering}
\end{figure}

\begin{table}[htbp]
\centering
\resizebox{\textwidth}{!}{
\renewcommand{\arraystretch}{1.2}
\begin{tabular}{c|c|c|ccc}
\hline\hline
$e^+e^-$ Collider & 87.9 GeV& 94.3 GeV& \multicolumn{3}{c}{91.2 GeV} \\
\hline
Luminosity [ab$^{-1}$] &40  & 40 & 100 & 125  & 200  \\
\hline
$|g_{a\gamma\gamma}|^{\rm max}$ [TeV$^{-1}$] & $8.54 \times 10^{-3}$  & $8.60 \times 10^{-3}$ & $7.20 \times 10^{-3}$ & $6.81 \times 10^{-3}$ & $6.05 \times 10^{-3}$ \\
$|g_{a\gamma Z}|^{\rm max}$ [TeV$^{-1}$] & $2.54 \times 10^{-3}$  & $1.68\times 10^{-3}$ & $6.80 \times 10^{-4}$ & $6.43 \times 10^{-4}$ & $5.72\times 10^{-4}$ \\
$|g_{aZZ}|^{\rm max}$ [TeV$^{-1}$]& $8.87 \times 10^{-3}$  & $8.59 \times 10^{-3}$ & $7.22 \times 10^{-3}$ & $6.83 \times 10^{-3}$ & $6.07 \times 10^{-3}$ \\
$|g_{aWW}|^{\rm max}$ [TeV$^{-1}$]& $9.09 \times 10^{-3}$  & $8.62 \times 10^{-3}$ & $7.24 \times 10^{-3}$ & $6.84 \times 10^{-3}$ & $6.08 \times 10^{-3}$ \\
\hline\hline
\end{tabular}
}
\caption{The upper limits $|g_{aVV}|^{\rm max}$ from the mono-photon production process at future $Z$ factories.}\label{tab:gconstraints_mono_photon_ZF}
\end{table}

In our detailed analysis, we apply the cuts
\begin{eqnarray}
    |\eta_\gamma|<2.5,~~~p_{T,\gamma} > 30~{\rm GeV},~~~E_\gamma > \frac{\sqrt{s}}{2}-5~{\rm GeV},
\end{eqnarray}
to further optimize the signal significance. In the ALP long-lived limit, i.e. $\mathcal{P}_a \simeq 1$, the 95\% CL constraints are shown as contours in the $(C_W,~C_B)$ plane in Figure~\ref{fig:alp_ZF}, where the red, blue, and green contours correspond to the FCC-ee runs at 
87.9 GeV ($\mathscr{L} = 40 ~{\rm ab}^{-1}$), 91.2 GeV ($\mathscr{L} = 125~{\rm ab}^{-1}$), and 94.3 GeV ($\mathscr{L} = 40~{\rm ab}^{-1}$)~\cite{janot_2024_yr3v6-dgh16}, respectively. The total signal significance is then combined as
\begin{eqnarray}
\mathcal{S}=\sqrt{\mathcal{S}_{87.9~{\rm GeV}}^2+\mathcal{S}_{91.2~{\rm GeV}}^2+\mathcal{S}_{94.3~{\rm GeV}}^2},
\end{eqnarray}
with the resulting 95\% constraint displayed as the orange contour in Figure~\ref{fig:alp_ZF}. Due to the resonance enhancement in the $g_{a\gamma Z}^2$ term, the contours appear as sequential narrow spindles, highlighting a stronger constraint in the $g_{a\gamma Z}$ direction. The 91.2 GeV run yields the most stringent limits because of both the resonance enhancement and its higher luminosity. Moreover, while ISR shifts the effective collision energy of the 94.3 GeV run closer to the $Z$ peak (the so-called ``radiation return''), it slightly shifts the 87.9 GeV run away from the peak. Consequently, under identical luminosity conditions, the 94.3 GeV constraint in the $g_{a\gamma Z}$ direction is marginally better than that at 87.9 GeV. Table~\ref{tab:gconstraints_mono_photon_ZF} lists the corresponding constraints on $|g_{aVV}|$ for various luminosity scenarios, where we also include 100 ab$^{-1}$ and 200 ab$^{-1}$ at the $Z$ peak to illustrate the luminosity dependence. By combining the signal significances from the three FCC-ee runs, we obtain
\begin{eqnarray}
    &&|g_{a\gamma\gamma}| \leq 5.87 \times 10^{-3}~{\rm TeV}^{-1},~~~|g_{a\gamma Z}| \leq 6.39 \times 10^{-4}~{\rm TeV}^{-1},\nonumber \\
    &&|g_{aZZ}| \leq 5.89 \times 10^{-3}~{\rm TeV}^{-1},~~~|g_{aWW}| \leq 5.91 \times 10^{-3}~{\rm TeV}^{-1}.\nonumber
\end{eqnarray}
These combined constraints are comparable to those expected from a $\sqrt{s}=91.2$ GeV machine operating with a luminosity of 200 ab$^{-1}$.

\subsubsection{Mono-photon production at higher collision energies}
\label{sec:monoA2}
For future lepton colliders with $\sqrt{s} \geq 200~{\rm GeV}$, the ALP production cross-section exhibits minimal dependence on the collision energy $\sqrt{s}$. This is due to the $s$- and $s^2$-scaling terms in the numerator of the production cross-section in Eq.~\eqref{eq:monophoton}, ensuring that the signal remains relatively consistent across different collider setups, as illustrated in Figure~\ref{fig:monoAxsec}. Therefore, the primary factor distinguishing detection capabilities at future lepton colliders will be the level of background noise rather than variations in the signal itself.

\begin{figure}
\begin{center}
\includegraphics[width=0.24\textwidth]{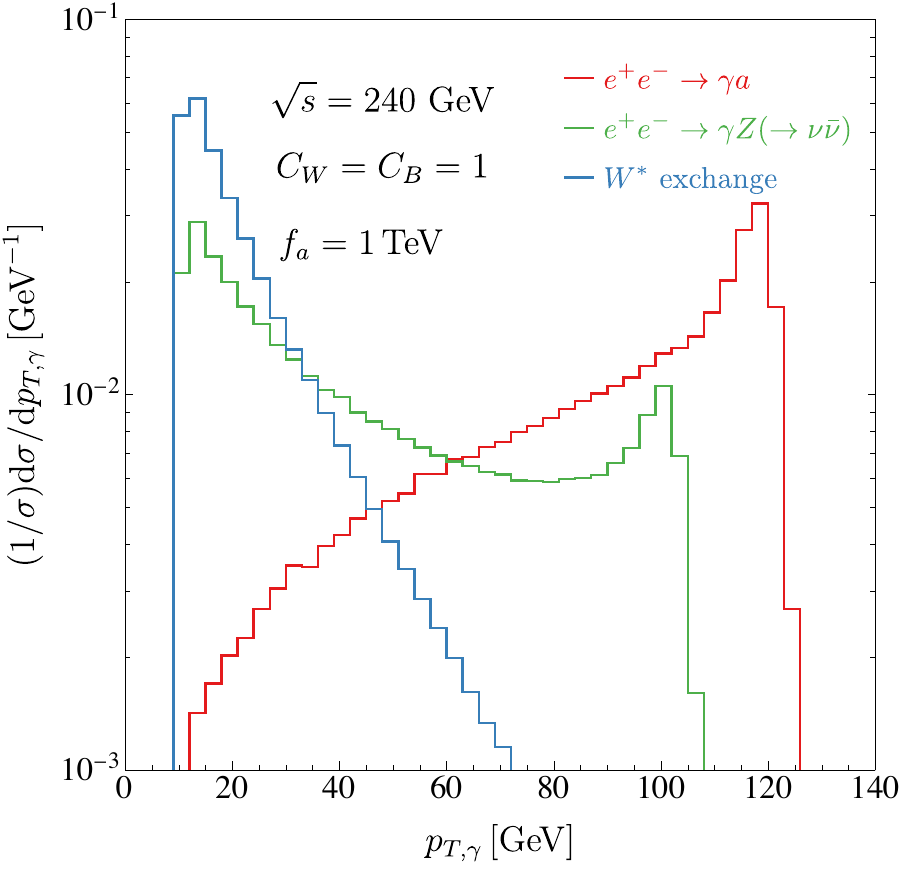}
\includegraphics[width=0.24\textwidth]{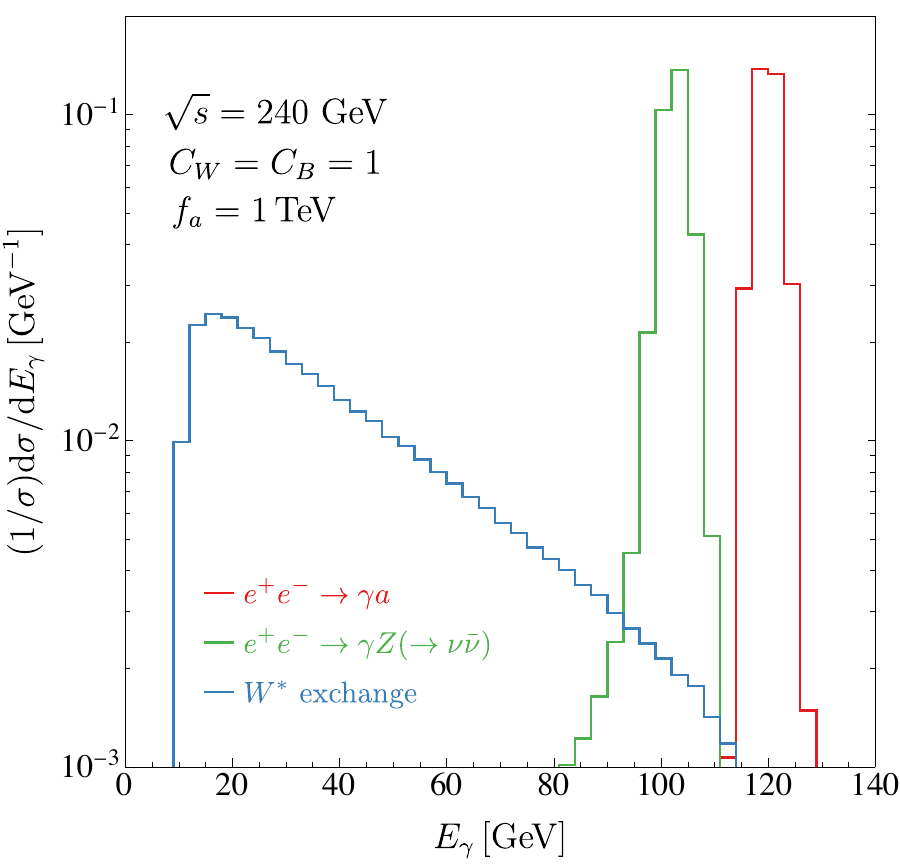}
\includegraphics[width=0.23\textwidth]{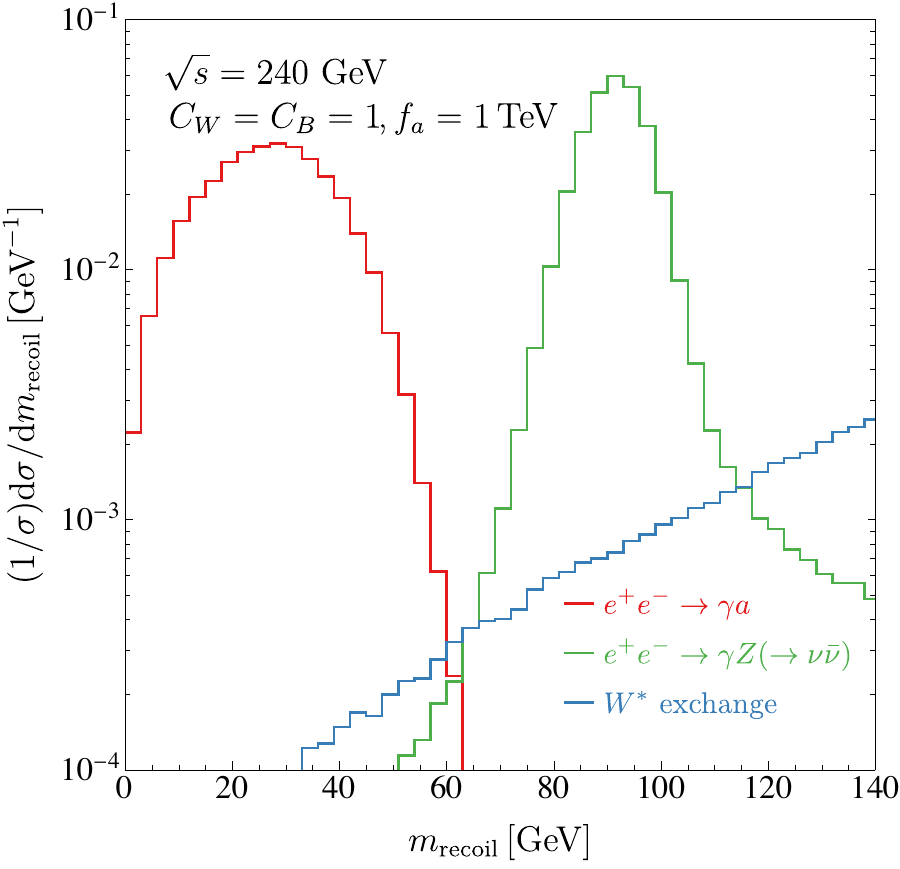}
\includegraphics[width=0.23\textwidth]{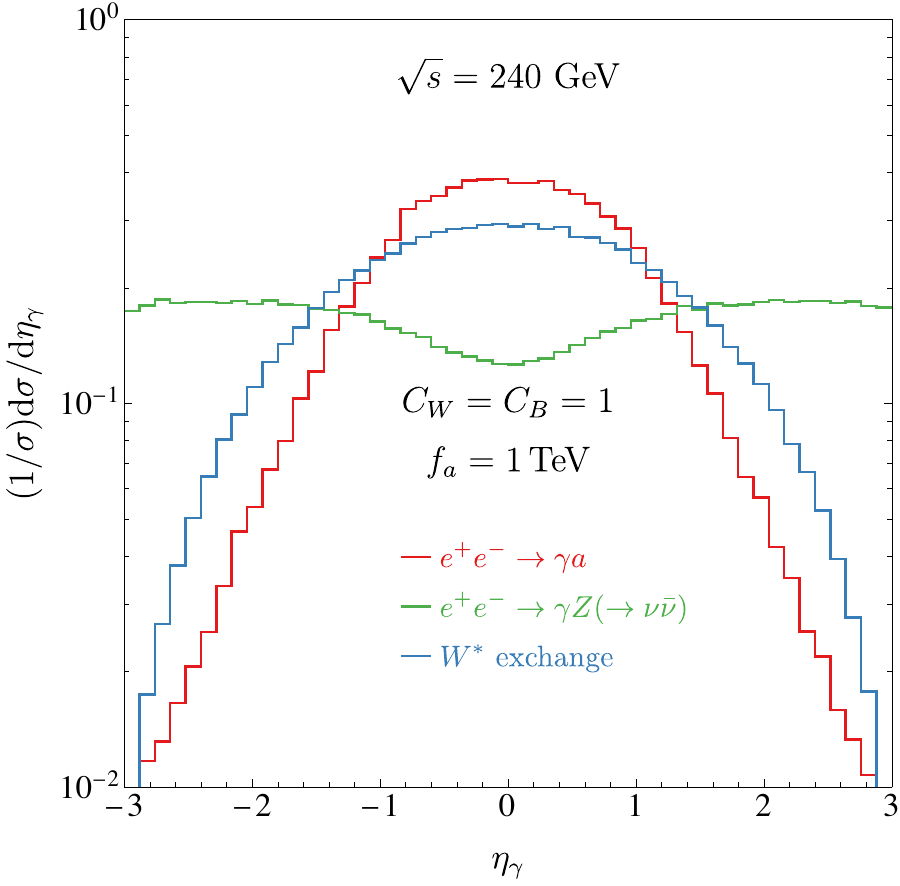}\\
\includegraphics[width=0.24\textwidth]{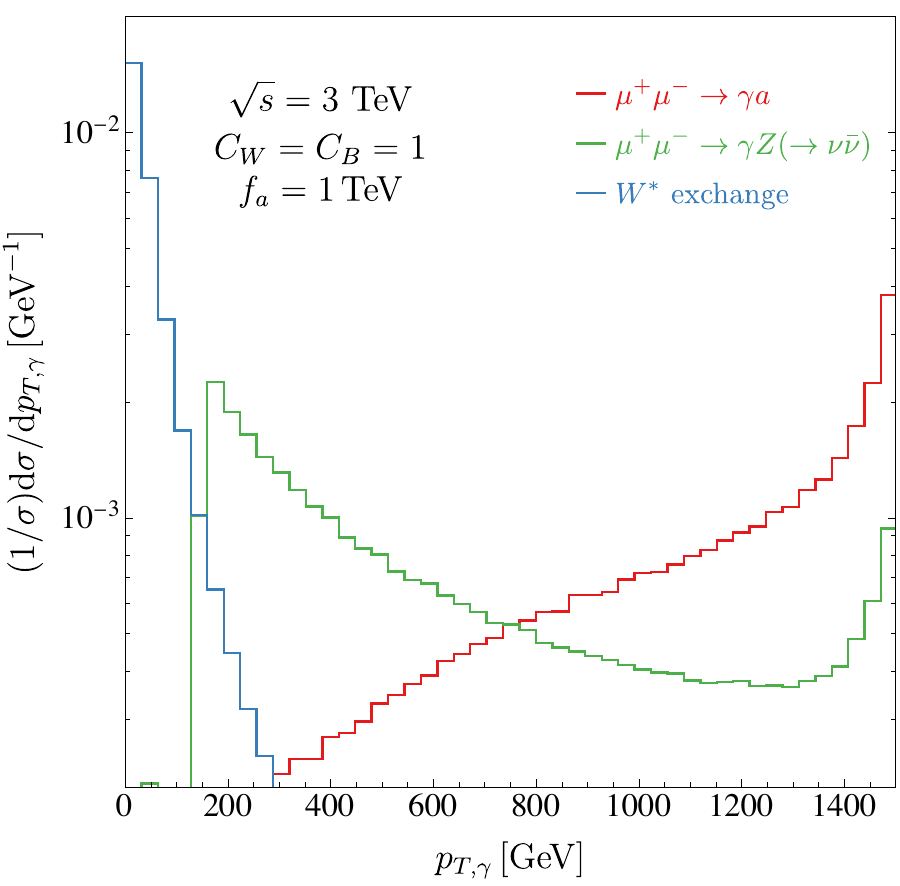}
\includegraphics[width=0.24\textwidth]{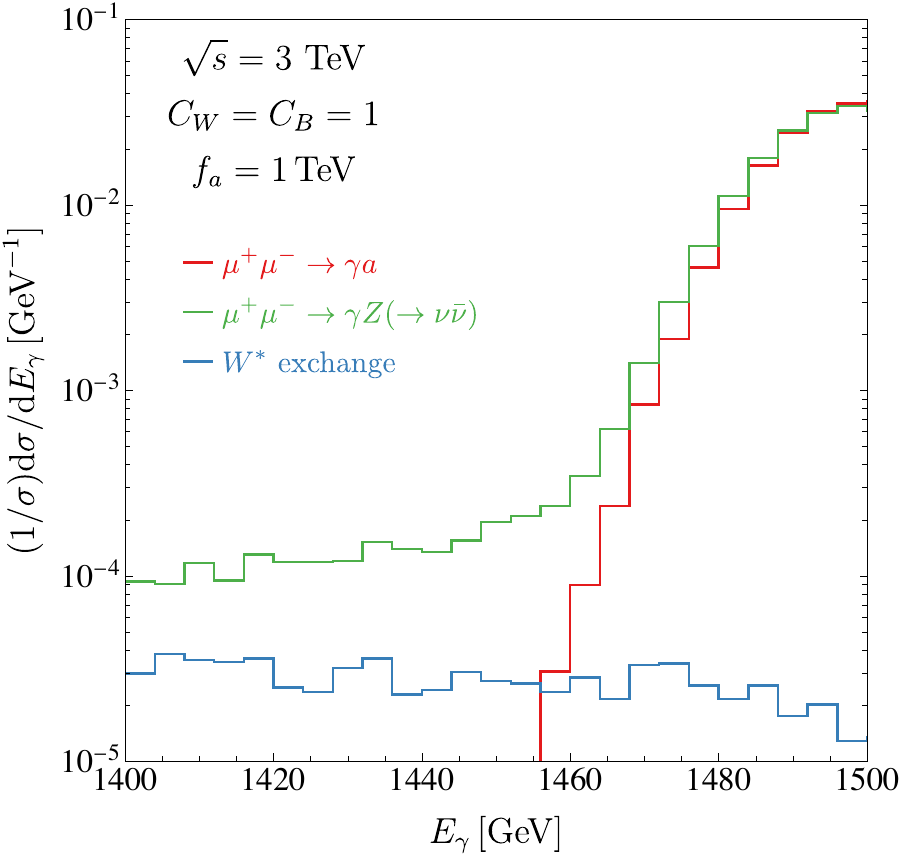}
\includegraphics[width=0.23\textwidth]{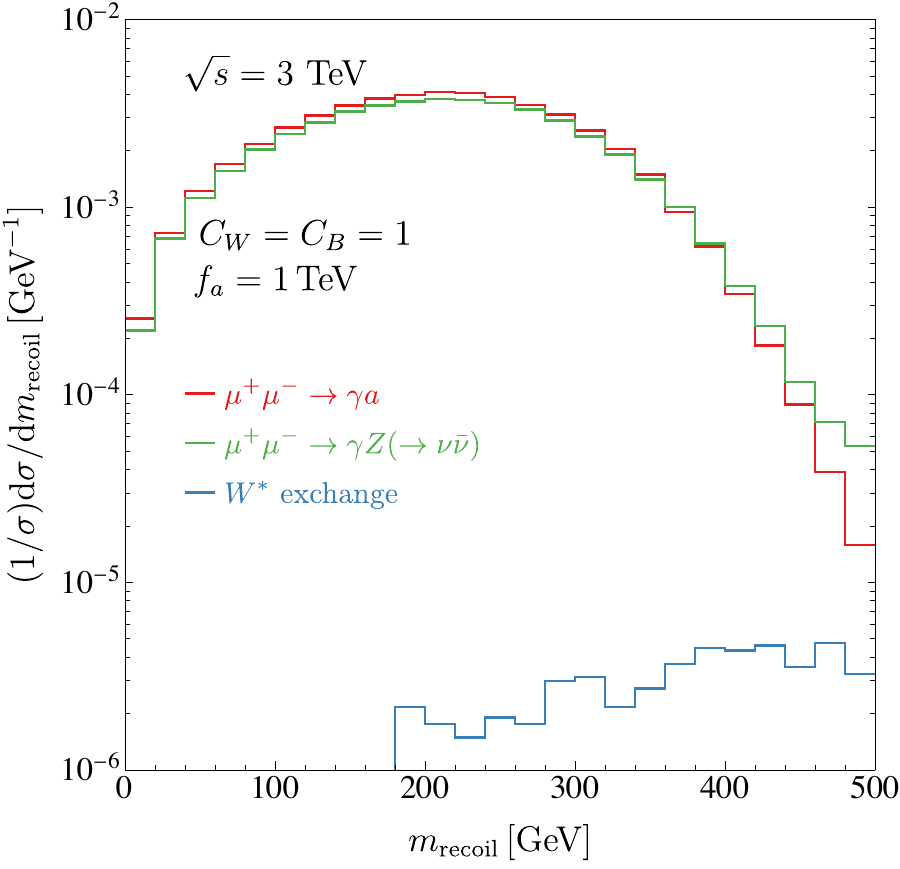}
\includegraphics[width=0.24\textwidth]{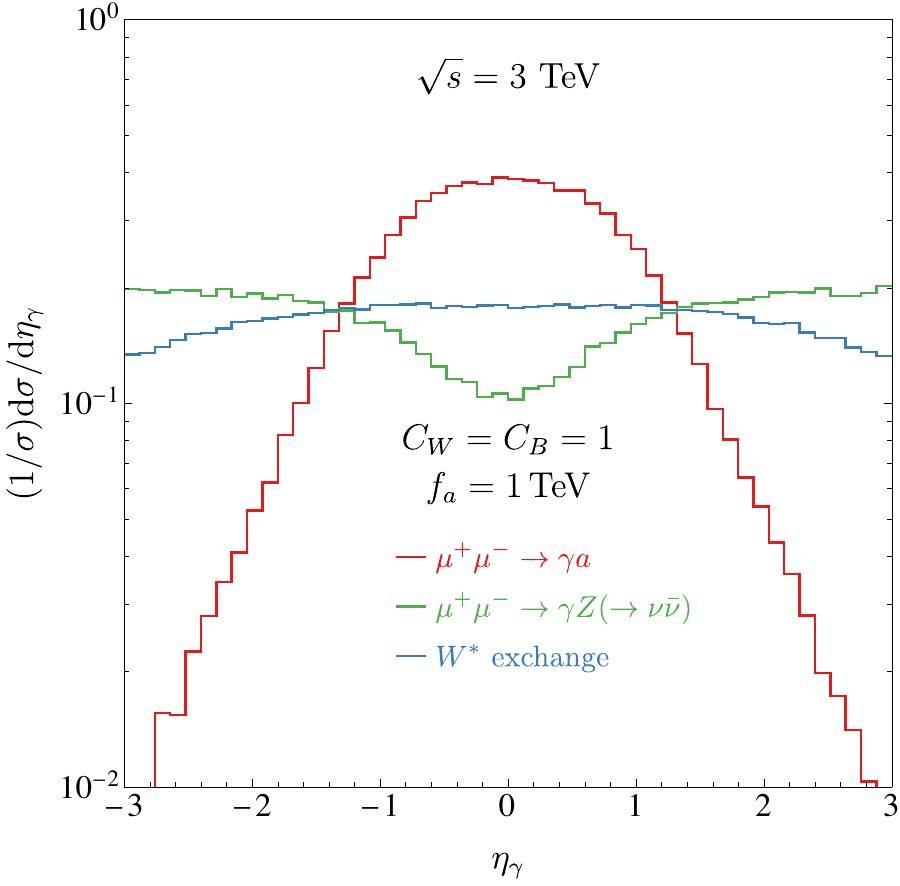}
\caption{The normalized distributions of the photon produced by the signal and background processes at a 240~GeV $e^+e^-$ collider and a 3~TeV muon collider. 
}\label{fig:distMonoPho}
\end{center}
\end{figure}

In Figure~\ref{fig:distMonoPho}, we present the normalized distributions of the photon for the signal and background of $\ell^+\ell^-\to \gamma +{\rm inv.}$ processes at a 240 GeV electron collider and a 3 TeV muon collider as examples. 
Different channels exhibit distinct kinematic features. 
The $W^*$ exchange produces a very soft final-state photon, whereas the background from on-shell $Z$ decay and the signal $\gamma+a$ favor the large photon energy and transverse momentum region. The latter two channels result in monochromatic energy and recoil mass peaks as given by
\begin{equation}\label{eq:mrecoil}
E_{\gamma}\sim\frac{s-m_{\rm inv}^2}{2\sqrt{s}}, \quad
m_{\rm recoil}^2=s-2\sqrt{s}E_\gamma\sim m_{\rm inv}^2,
\end{equation}
due to the 2-to-2 kinematics, where the invisible mass $m_{\rm inv}=m_a$ and $M_Z$, respectively. This distinction can help to reduce the background $\gamma+Z(\nu\bar{\nu})$ at the FCC-ee energies.
At sufficiently high collision energies where $\sqrt{s}\gg M_Z$, the mono-photon energy spectra of the signal $\gamma+a$ and the $\gamma+Z(\nu\bar{\nu})$ background become nearly identical, making $\gamma+Z(\nu\bar{\nu})$ dominate the background.

\begin{table}[h!]
\centering
\begin{tabular}{lcccc}
\hline
\hline
Collider & \multicolumn{2}{c}{$e^+e^-$} & \multicolumn{2}{c}{$\mu^+\mu^-$} \\
$\sqrt{s}$ & 240 GeV & 365 GeV & 3 TeV & 10 TeV \\
\hline
$|\eta_\gamma|^{\max}$ & 2.5 & 2.5 & 2.5 & 2.5 \\
$p_{T,\gamma}^{\min}$ [GeV] & 60 & 90 & 700 & 2500 \\
$E_\gamma^{\min}$ [GeV] & 115 & 177.5 & 1450 & 4900 \\
\hline
\hline
\end{tabular}
\caption{Kinematic cuts on the photon transverse momentum $p_{T,\gamma}$, energy $E_\gamma$, and pseudorapidity $|\eta_\gamma|$ applied in the mono-photon analysis at various center-of-mass energies. }
\label{tab:mono_photon_cuts}
\end{table}

\begin{table}[]
    \centering
    \begin{tabular}{c|ccc}
    \hline
           $e^+e^-$ 240 GeV & Basic & $p_{T,\gamma}>60$~GeV & $E_\gamma>$ 115 GeV \\ 
           \hline
         $\gamma+a$ & 1.93 &  1.58 & 1.56 \\ 
        $\gamma+Z(\nu\bar{\nu})$& 1.83 & 0.704 & $2.12\times10^{-4}$ \\
         $W^*$ exchange &1.04 & 0.0492 & $1.43\times 10^{-3}$\\
    \hline
    \hline
      $\mu^+\mu^-$ 3 TeV  & Basic & $p_{T,\gamma}>700$~GeV & $E_\gamma>1450~\GeV$ \\
           \hline
         $\gamma+a$ & 1.92 & 1.64 & 1.64 \\     
         $\gamma+Z(\nu\bar{\nu})$ & $9.11\times10^{-3}$ & $4.28\times 10^{-3}$ & $4.28\times10^{-3}$\\
         $W^*$ exchange & 2.97 & $7.71\times10^{-3}$ & $4.76\times10^{-4}$ \\
         \hline
    \end{tabular}
    \caption{The cut-flow cross section $\sigma$~[pb] for the mono-photon signal and backgrounds at 240 GeV $e^+ e^-$ and 3 TeV $\mu^+\mu^-$ colliders, starting with basic cuts $p_{T,\gamma}>10~\GeV$ and $|\eta_\gamma|<2.5$. The signal corresponds to a benchmark choice $C_W=C_B=1$ and $f_a=1$~TeV. The background $\gamma+\nu\bar{\nu}$ includes all the contribution in Figure~\ref{feyn:mono-pho}.
    }
    \label{tab:cutflow}
\end{table}

In Table~\ref{tab:mono_photon_cuts}, we summarize the optimized fiducial cuts for 240/365 GeV electron colliders and 3/10 TeV muon colliders. The cut efficiencies for a 240 GeV electron collider and a 3 TeV muon collider are presented in Table~\ref{tab:cutflow}, where the signal cross sections correspond to the benchmark choice $C_B = C_W = 1$ and $f_a = 1~\TeV$. The background includes all the contributions of the SM processes in Figure~\ref{feyn:mono-pho}. The $\gamma+Z(\nu\bar{\nu})$ denotes the neutrino pair is from an on-shell $Z$, while the $W^*$ exchange denotes other contributions to the $\gamma+\nu\bar{\nu}$. From the table, it is evident that the photon energy $E_\gamma$ and transverse momentum $p_{T,\gamma}$ cuts effectively separate the signal from the backgrounds.
After applying these cuts, most of the $\gamma+a$ signal is retained, while the dominant remaining background originates from the residual $W^*$ exchange contribution, since the $\gamma+Z(\nu\bar{\nu})$ can be reduced well at $\sqrt{s}=240$~GeV by the $E_\gamma$ cut as shown in Figure~\ref{fig:distMonoPho}.
However, at multi-TeV muon colliders, while the $E_\gamma$ and $p_{T,\gamma}$ cuts effectively suppress the $W^*$ exchange background, the photon energy cut fails to suppress the $\gamma + Z(\nu\bar{\nu})$ background.
The $p_{T,\gamma}$ cut still works well, which can reduce about 50\% of the $\gamma + Z(\nu\bar{\nu})$ background.
Since the ALP is a pseudoscalar particle, a longitudinally polarized photon produced alongside an ALP favors the transverse direction, whereas a photon produced with an on-shell $Z$ has no preferred direction. As a result, our analysis cannot be directly applied to dark photon searches.

\begin{figure}
\centering
\includegraphics[width=0.45\textwidth]{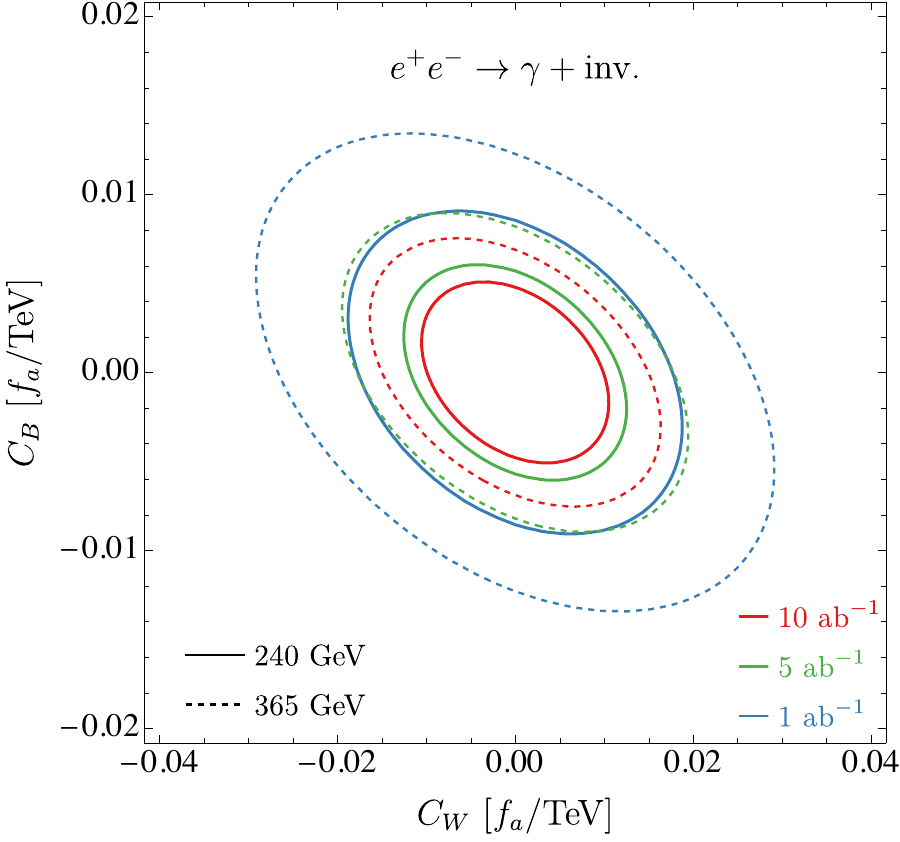}
\includegraphics[width=0.45\textwidth]{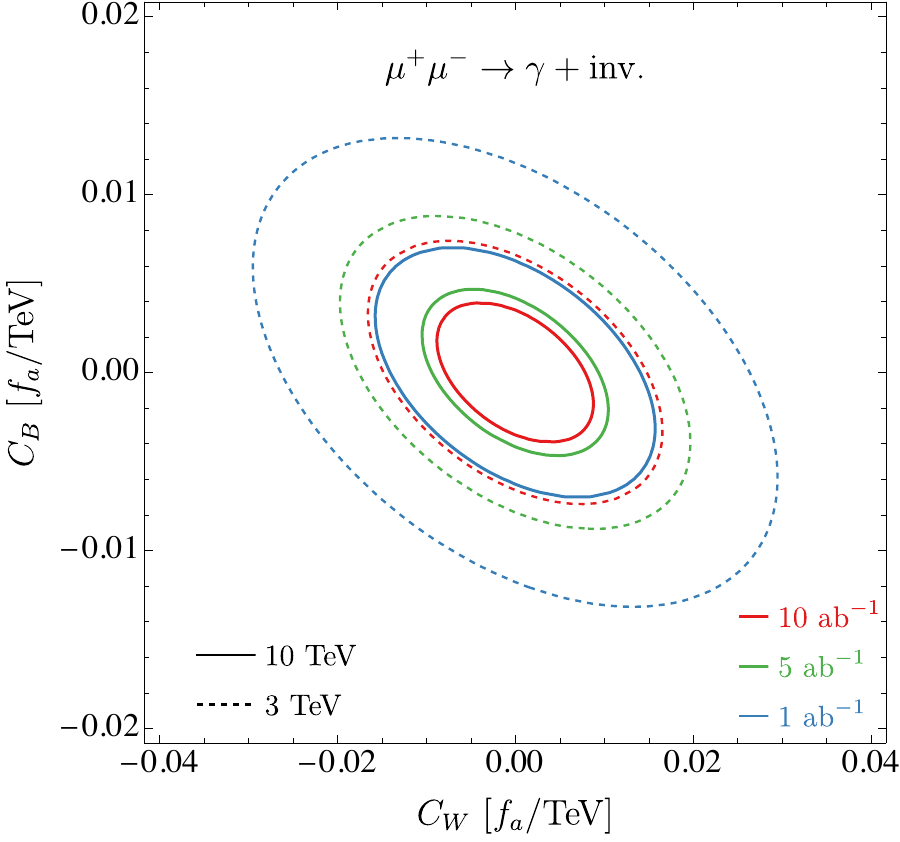}
\caption{Constraints on $C_W$ and $C_B$ from $\ell^+\ell^-\to \gamma +{\rm inv.}$ searches at high-energy lepton colliders with varying machine energies and luminosities at 95\% CL. Red, green, and blue curves correspond to integrated luminosities of 1 ab$^{-1}$, 5 ab$^{-1}$, and 10 ab$^{-1}$, respectively. Solid contours in the left (right) panel represent the 240~GeV $e^+e^-$ collider (10~TeV muon collider), while dashed contours in the left (right) panel correspond to the 365~GeV $e^+e^-$ collider (3~TeV muon collider).}
\label{fig:mono_photon}
\end{figure}

With the significance defined in Eq.~(\ref{eq:signal_strength}), we present the 95\% CL constraints at future lepton colliders on $C_W$ and $C_B$ in the ALP long-lived limit.
The results are shown as contour plots in the $(C_W,\,C_B)$ plane in Figure~\ref{fig:mono_photon}, where the region outside the contours can be probed.
These contours correspond to possible integrated luminosities of $1~{\rm ab}^{-1}$, $5~{\rm ab}^{-1}$, and $10~{\rm ab}^{-1}$, depicted in blue, green, and red, respectively. 
As the center-of-mass energy increases, the $E_\gamma$ cut gradually loses its effectiveness in distinguishing the ALP signal from the $\ell^+\ell^- \to \gamma Z(\nu\bar{\nu})$ background, leading to reduced signal significance at higher energies. Consequently, the 240~GeV Higgs factory can impose stronger constraints than the 365~GeV top factory. 
At TeV-scale lepton colliders, the energy cut becomes entirely ineffective in suppressing the $\ell^+\ell^- \to \gamma Z(\nu\bar{\nu})$ background but remains useful for reducing the background from the $W^*$ exchange channel. As a result, the dominant background at muon colliders comes from on-shell $Z$ production. In this case, the background production rate scales as $1/s$, making the 10~TeV collider more sensitive than the 3~TeV collider.
Interestingly, the 10~TeV muon collider can provide constraints comparable to those of the 240~GeV Higgs factory, while the 3~TeV muon collider performs similarly to the 365~GeV top factory. The corresponding 95\% CL upper limits on the absolute values of $|g_{aVV}|$ are summarized in Table~\ref{tab:gconstraints_mono_photon}.

\begin{table}[htbp]
\centering
\resizebox{\textwidth}{!}{
\renewcommand{\arraystretch}{1.2}
\begin{tabular}{c|ccc|ccc}
\hline\hline
$e^+e^-$ Collider & \multicolumn{3}{c|}{$\sqrt{s}=240$ GeV} & \multicolumn{3}{c}{$\sqrt{s}=365$ GeV} \\
\hline
Luminosity [ab$^{-1}$] & 1  & 5 & 10 & 1 & 5  & 10  \\
\hline
$|g_{a\gamma\gamma}|^{\rm max}$ [TeV$^{-1}$] & $2.76 \times 10^{-2}$ & $1.84 \times 10^{-2}$ & $1.54\times 10^{-2}$ & $3.93 \times 10^{-2}$ & $2.63 \times 10^{-2}$ & $2.21\times 10^{-2}$ \\
$|g_{a\gamma Z}|^{\rm max}$ [TeV$^{-1}$] & $7.80 \times 10^{-2}$ & $5.20 \times 10^{-2}$ & $4.37\times 10^{-2}$ & $1.22 \times 10^{-1}$ & $8.14 \times 10^{-2}$ & $6.85\times 10^{-2}$ \\
$|g_{aZZ}|^{\rm max}$ [TeV$^{-1}$]& $5.62 \times 10^{-2}$ & $3.75 \times 10^{-2}$ & $3.15 \times 10^{-2}$ & $8.65 \times 10^{-2}$ & $5.78 \times 10^{-2}$ & $4.86 \times 10^{-2}$ \\
$|g_{aWW}|^{\rm max}$ [TeV$^{-1}$]& $7.51 \times 10^{-2}$ & $5.01 \times 10^{-2}$ & $4.21 \times 10^{-2}$ & $1.16 \times 10^{-1}$ & $7.78 \times 10^{-2}$ & $6.54 \times 10^{-2}$ \\
\hline
$\mu^+\mu^-$ Collider & \multicolumn{3}{c|}{$\sqrt{s}=3$ TeV} & \multicolumn{3}{c}{$\sqrt{s}=10$ TeV} \\
\hline
Luminosity [ab$^{-1}$] & 1 & 5  & 10 & 1  & 5  & 10  \\
\hline
$|g_{a\gamma\gamma}|^{\rm max}$ [TeV$^{-1}$] & $3.75 \times 10^{-2}$ & $2.50 \times 10^{-2}$ & $2.10 \times 10^{-2}$ & $2.00 \times 10^{-2}$ & $1.33 \times 10^{-2}$ & $1.12\times 10^{-2}$ \\
$|g_{a\gamma Z}|^{\rm max}$ [TeV$^{-1}$] & $1.24 \times 10^{-1}$ & $8.29 \times 10^{-2}$ & $6.97\times 10^{-2}$ & $6.62 \times 10^{-2}$ & $4.41 \times 10^{-2}$ & $3.70 \times 10^{-2}$ \\
$|g_{aZZ}|^{\rm max}$ [TeV$^{-1}$] & $8.71 \times 10^{-2}$ & $5.82 \times 10^{-2}$ & $4.89 \times 10^{-2}$ & $4.65 \times 10^{-2}$ & $3.09 \times 10^{-2}$ & $2.60 \times 10^{-2}$ \\
$|g_{aWW}|^{\rm max}$ [TeV$^{-1}$]& $1.18 \times 10^{-1}$ & $7.88 \times 10^{-2}$ & $6.62 \times 10^{-2}$ & $6.29 \times 10^{-2}$ & $4.19 \times 10^{-2}$ & $3.52 \times 10^{-2}$ \\
\hline\hline
\end{tabular}
}
\caption{The upper limits $|g_{aVV}|^{\rm max}$ from the mono-photon production process at different future lepton colliders.}\label{tab:gconstraints_mono_photon}
\end{table}

\subsection{Mono-$Z$ production}
\label{sec:monoZ}

Similar to the above $\gamma+a$ production in the mono-photon channel, an ALP can also be produced associated with a single $Z$ boson with the ALP flying out of the detector and leaving a mono-$Z$ signal at future lepton colliders. 
In the light ALP mass limit $m_a\ll M_Z$, the corresponding cross section can be written as
\begin{equation} \label{eq:monoZ}
\begin{aligned}
    \sigma_{\ell^+ \ell^- \to Z a} = \frac{\alpha}{192} 
    &\left[\frac{g_{aZZ}^2 (6 s_W^4 + 2 c_W^4 - 1)(s - M_Z^2)}{s_W^2 c_W^2 s}+\right.\\
    &\left.\frac{2 g_{aZZ} g_{a\gamma Z} (c_W^2 - 3 s_W^2)(s - M_Z^2)^2}{s_W c_W s^2} + \frac{2 g_{a\gamma Z}^2 (s - M_Z^2)^3}{s^3} \right],    
\end{aligned}
\end{equation}
where the couplings $g_{a\gamma Z},g_{aZZ}$ are related to $C_B$ and $C_W$ in Eq.~(\ref{eq:couplings}).
Moreover when $\sqrt{s} \gg M_Z$, this cross-section becomes asymptotically independent of the collider energy, as shown in Figure~\ref{fig:monoZxsec}. We expect the corresponding measurements to probe the ALP couplings, especially $g_{aZZ}$.

The primary SM background for the mono-$Z$ production comes from $\ell^+ \ell^- \to Z\nu\bar{\nu}$.
Moreover, similarly as the mono-photon case in Figs.~\ref{feyn:WW} and \ref{feyn:Wex}, we also have the $W^*$ and $\gamma/Z$ exchanges. In addition, there is also a small faked possibility from the $WW$ production, especially in the leptonic channel. 
In Figure~\ref{fig:monoZxsec}, we compare the total cross sections for these backgrounds with the $Z+a$ signal with respect to the collider energy $\sqrt{s}$. We see for the $Z+Z(\nu\bar{\nu})$ and $WW$ decreases with collider energy, while $W^*$ and $\gamma/Z$ exchanges increases due to the enhancement at low $p_T$ region.

In this subsection, we will analyze the signals and backgrounds for the mono-$Z$ production in both leptonic and hadronic modes, respectively.

\subsubsection{Leptonic decay of $Z$ boson}
\label{sec:monoZll}

The easiest way to identify a final-state $Z$ boson is to resolve a di-lepton resonance in the $Z\to\ell^+\ell^-$ mode. Here we consider $\ell=e,\mu$ and do not include the $\tau$-lepton, which is more complicated. 
For the signal of  $Z+a$ production, the final state behaves as $\ell\ell+\slashed{E}$, with the missing energy $\slashed{E}$ carried out by the ALP.
As we discussed above, the background in SM comes from three distinct mechanisms:
\begin{itemize}
    \item di-$Z$: $\ell^+\ell^- \to Z(\to \ell\ell)Z(\to\nu\bar{\nu})$,
    \item di-$W$: $\ell^+\ell^- \to W^+(\to\bar{\ell}\nu) W^-(\to \ell\bar{\nu})$,
    \item $W^*$ and $\gamma/Z$ exchange: $\ell^+\ell^- \to  \bar{\nu}_\ell\nu_\ell Z(\to \ell\ell)$ and $\ell^+\ell^-\to \ell^+\ell^- Z(\to\ell\ell)$.
\end{itemize}

\begin{figure}
    \centering
    \includegraphics[width=0.183\linewidth]{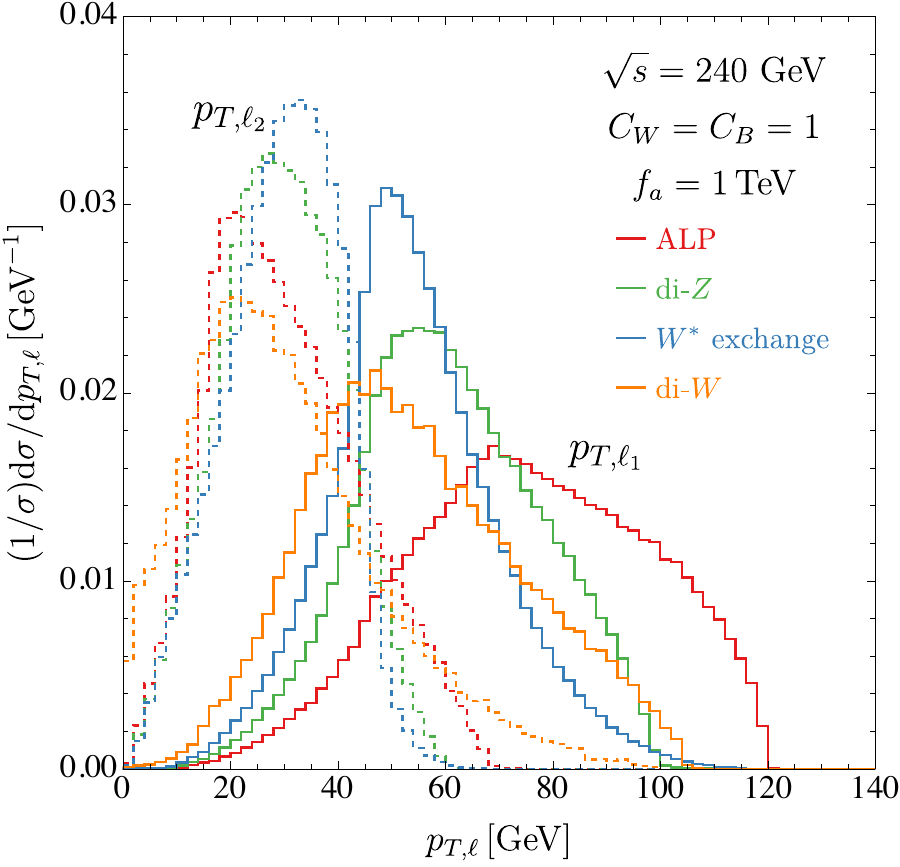}
    \includegraphics[width=0.178\linewidth]{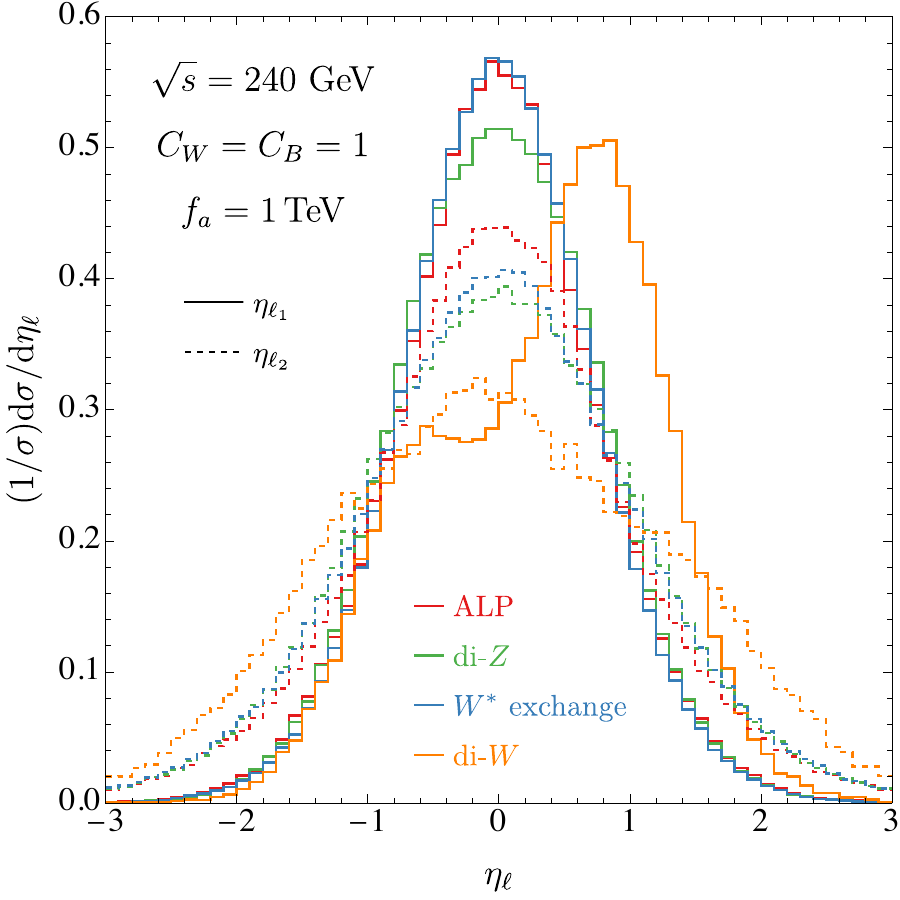}    
    \includegraphics[width=0.183\linewidth]{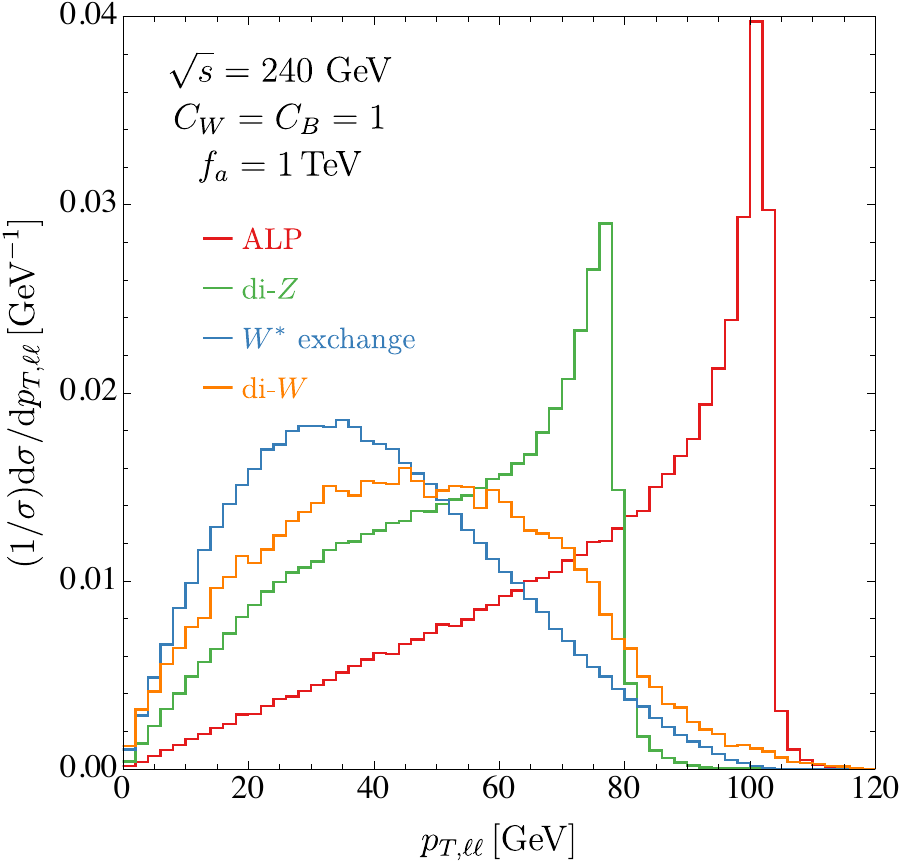}
    \includegraphics[width=0.18\linewidth]{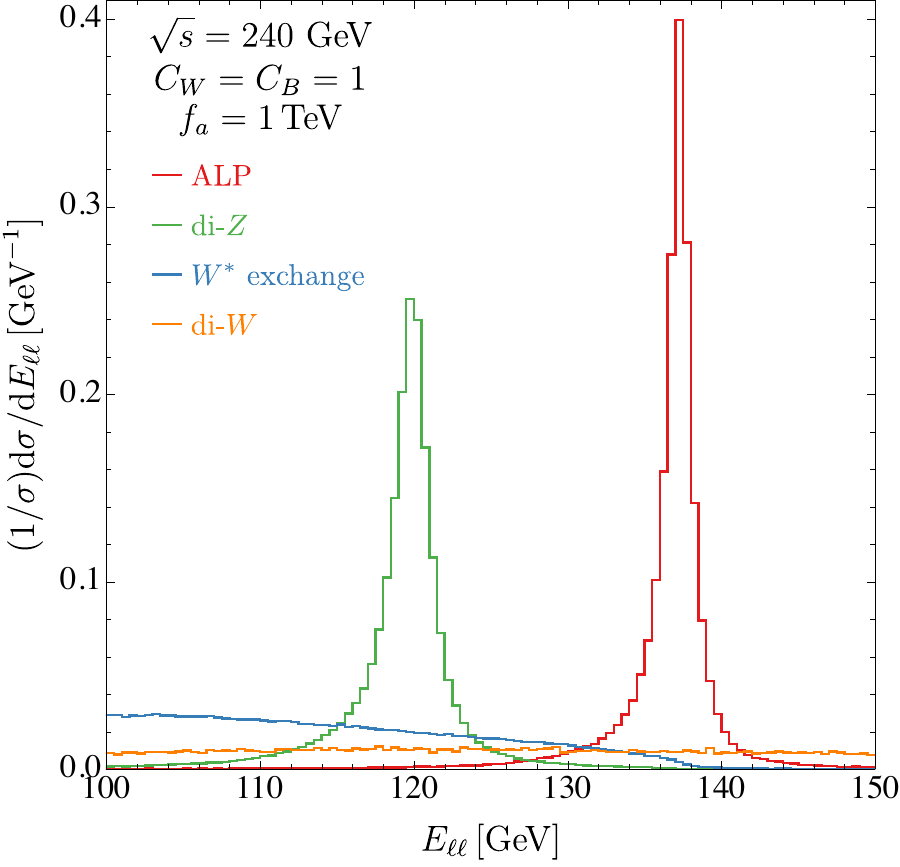}
    \includegraphics[width=0.175\linewidth]{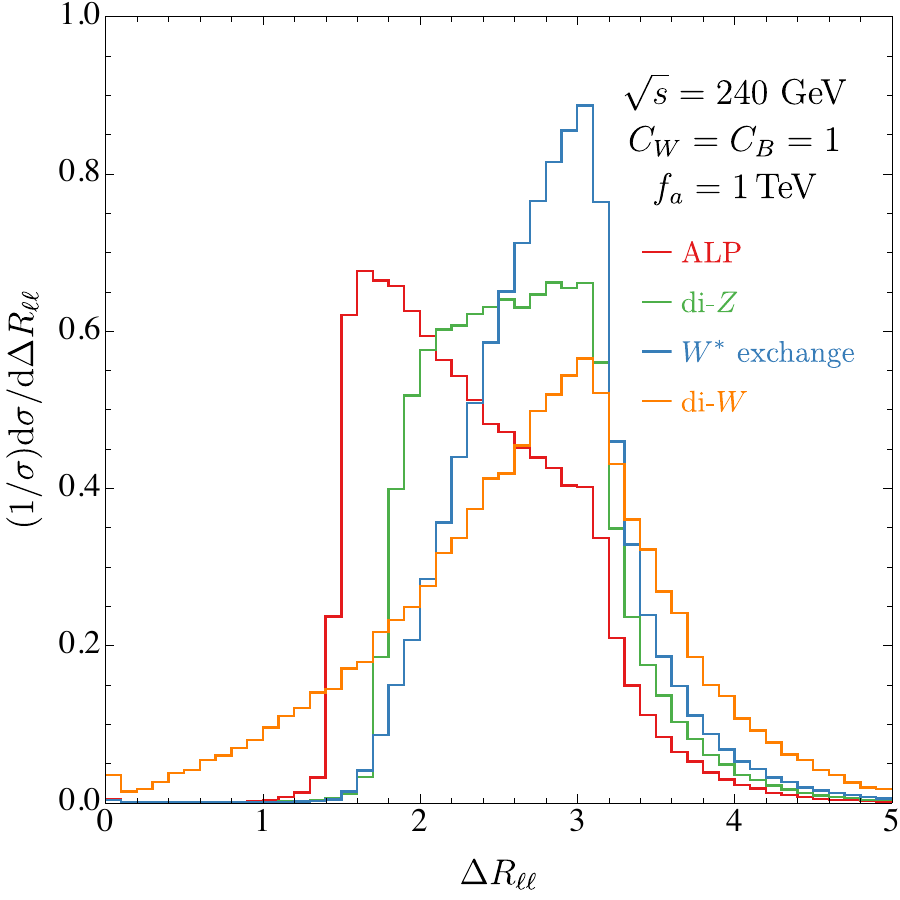}\\
    \includegraphics[width=0.183\linewidth]{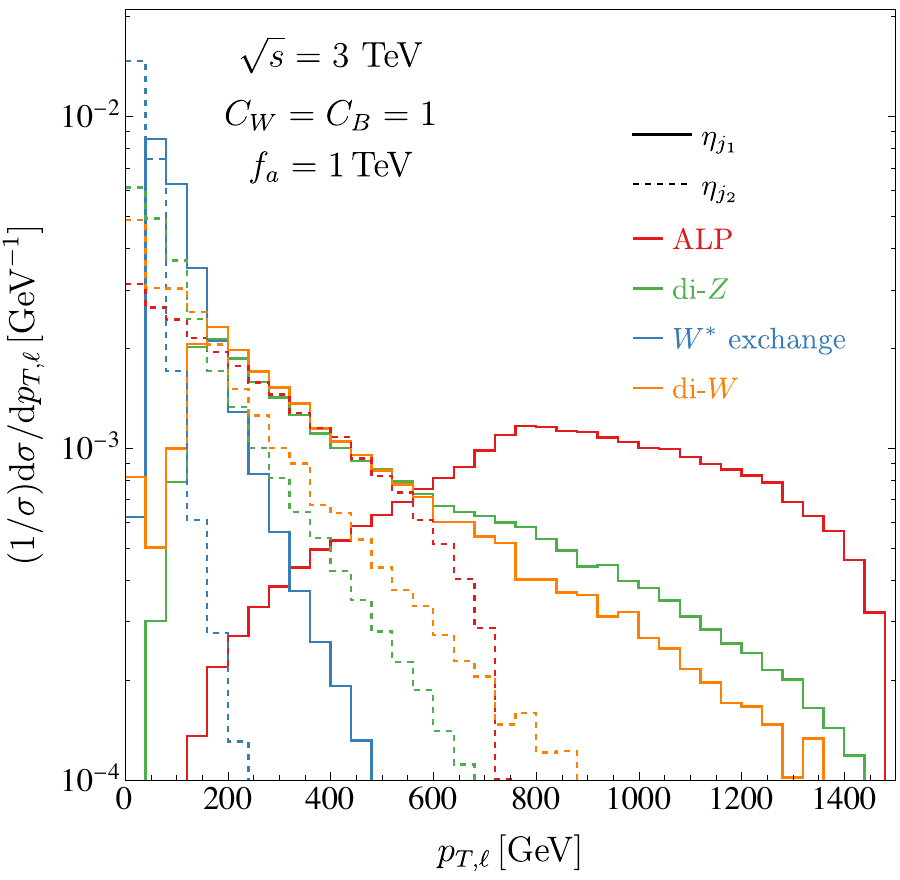}
    \includegraphics[width=0.178\linewidth]{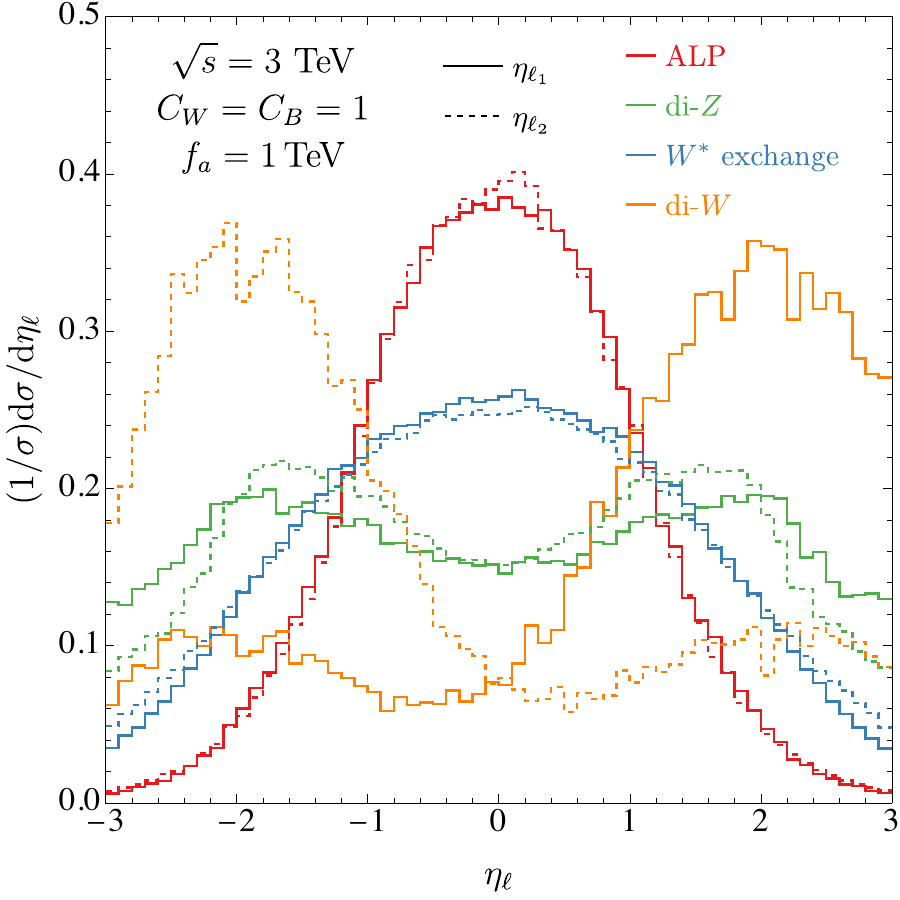}    
    \includegraphics[width=0.18\linewidth]{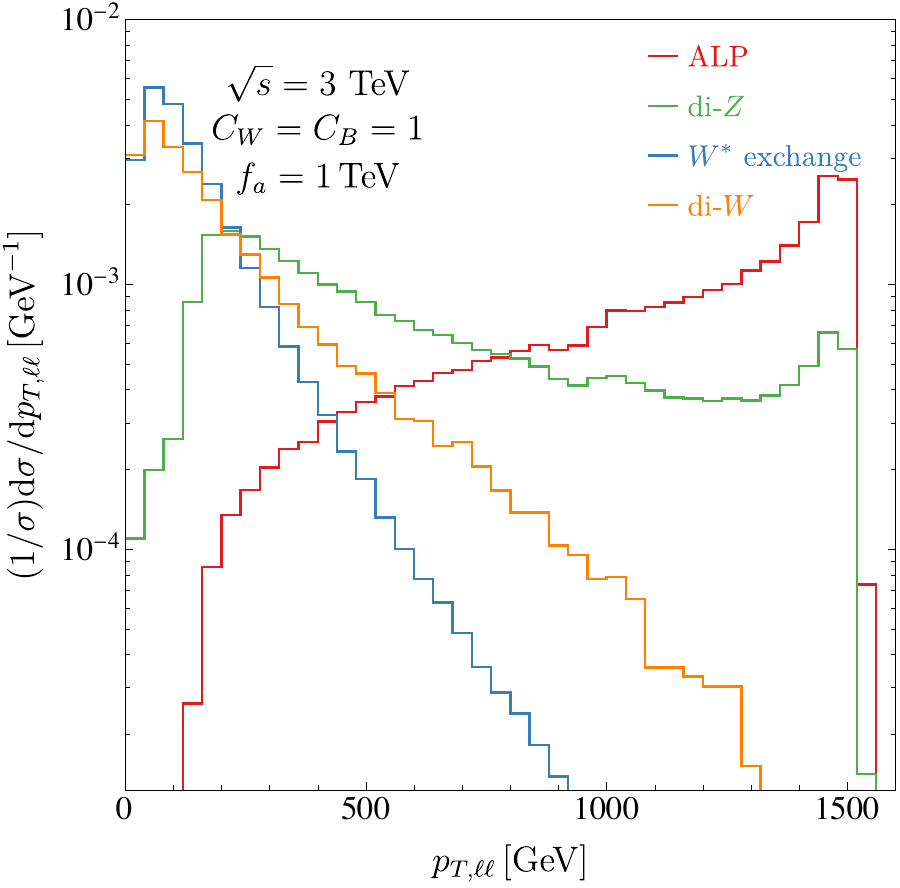}
    \includegraphics[width=0.183\linewidth]{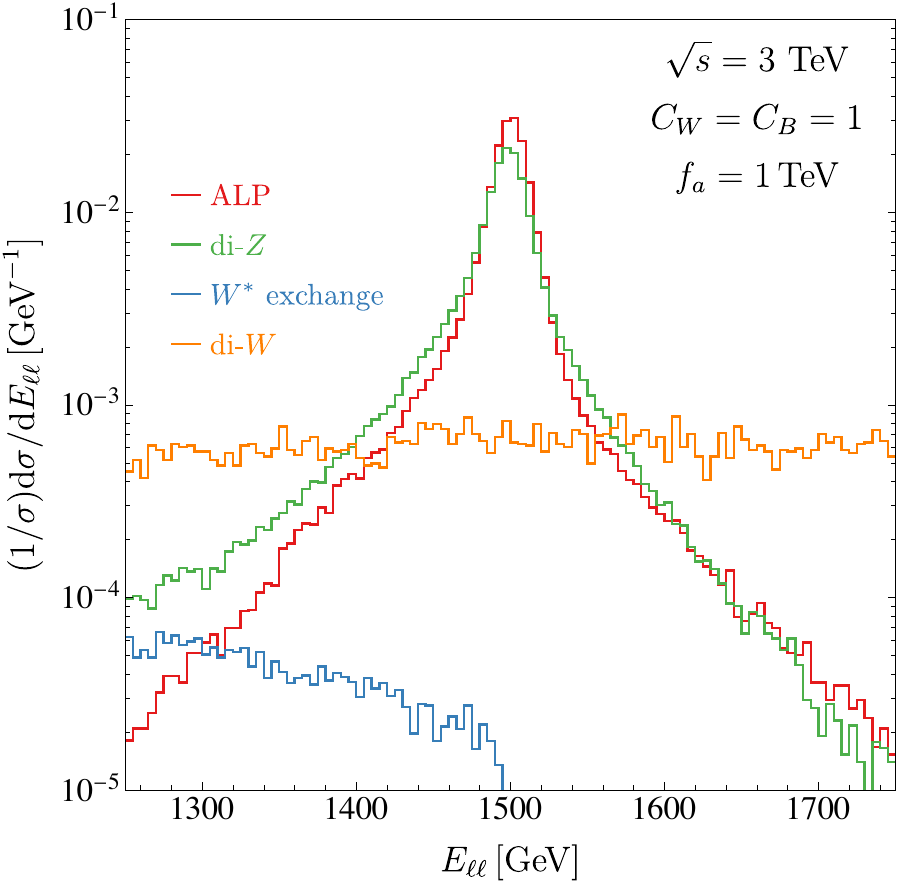}
    \includegraphics[width=0.175\linewidth]{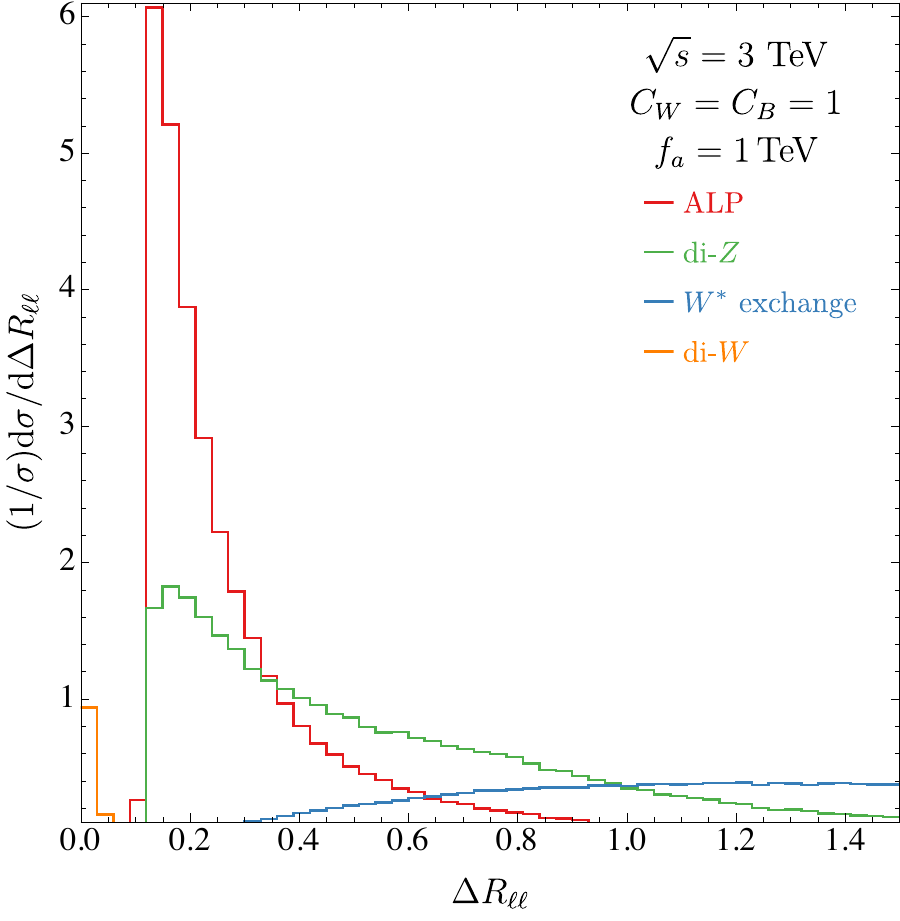} 
    \caption{Normalized distributions for mono-$Z$ production with $Z\to\ell\ell$ at a 240 GeV electron collider and 3 TeV muon collider.}
    \label{fig:distMonoZll}
\end{figure}

In Figure~\ref{fig:distMonoZll}, we present the normalized kinematic distributions for both signal and background processes in di-leptonic mono-$Z$ production at a 240 GeV electron collider and a 3 TeV muon collider. Notably, these distributions exhibit several similarities to those observed in the mono-photon channel shown in Figure~\ref{fig:distMonoPho}.
For example, in the case of $W^*$ and $\gamma/Z$ exchange, the differential cross section ${\rm d}\sigma/{\rm d}p_{T,\ell\ell}$ drops rapidly with increasing $p_{T,\ell\ell}$, allowing the ALP signal to be effectively isolated using a suitable $p_{T,\ell\ell}$ cut.
In comparison, the additional faked $Z$ boson from the di-$W$ production gives a wide spread for the invariant mass $M_{\ell\ell}$. Therefore, we can well exclude this channel with a resonance mass window around $M_{\ell\ell}\sim M_Z$ and an energy cut of lepton pair. Moreover, for the electron collider $\sqrt{s}=240$~ GeV and 365~GeV, the energy cut for the lepton pair is chosen as $E_{\ell\ell}>\sqrt{s}/2$ to suppress the di-$Z$ background. Since $M_Z>m_a$, the missing energy in $Z(\to\ell^+\ell^-)+Z(\to \nu\bar{\nu})$ process is larger that in the signal process $Z(\to\ell^+\ell^-)+a$. However, the energy cut lost its efficiency to exclude the di-$Z$ background at muon colliders where $\sqrt{s}\gg M_Z$. 

\begin{table}[!h]
\centering
\begin{tabular}{lcccc}
\hline
\hline
Collider & \multicolumn{2}{c}{$e^+e^-$} & \multicolumn{2}{c}{$\mu^+\mu^-$} \\
$\sqrt{s}$ & 240 GeV & 365 GeV & 3 TeV & 10 TeV \\
\hline
$p_{T,\ell_1}^{\rm min}$ [GeV] & 60 & 85 & 500 & 1600 \\
$p_{T,\ell_2}^{\rm min}$ [GeV] & 10 & 10 & 150 & 500 \\
$E^{\rm min}_{\ell\ell}$ [GeV] & 125 & 185 & 1450 & 4800 \\
$p_{T,\ell\ell}^{\rm min}$ [GeV] & 80 & 105 & 800 & 2200 \\
$\Delta R_{\ell\ell}^{\max}$ & $2.3$ & $2.0$ & $ 0.4$ & $0.15$ \\
$|\eta_{\ell}|^{\rm max}$ & $2.0$ & $2.0$ & $2.0$ & $2.0$ \\
\hline
\hline
\end{tabular}
\caption{Kinematic selection criteria for mono-$Z$ production events at various machine energies, focusing on reconstructing the $Z$ boson in its leptonic decay mode $Z\to\ell\ell$. The table specifies cuts on the transverse momentum of the leading lepton ($p_{T,\ell_1}^{\rm min}$) and subleading lepton ($p_{T,\ell_2}^{\rm min}$), separation distance $\Delta R_{\ell\ell}$ range between the leptons, and the pseudorapidity $|\eta_\ell|^{\rm max}$ of each lepton.}
\label{tab:cut_ll}
\end{table}

Based on these features, we summarize our final optimized cuts for the di-leptonic mono-$Z$ production in Table~\ref{tab:cut_ll}, with a detailed explanation as follows.
To reconstruct the $Z$ bosons from the lepton pairs, events are selected with more than one isolated leptons satisfying $|\eta_\ell| \leq 2.0$, where the isolation condition from the \textsc{Delphes3} card is applied.  
The leptons are ordered by their transverse momenta $p_{T,\ell}$, with the leading lepton designated as $\ell_1$ and the subleading as $\ell_2$. The $p_{T,\ell}$ thresholds for the two charged leptons, which depend on $\sqrt{s}$, are listed in Table~\ref{tab:cut_ll}.

At high collision energies, the $Z$ boson becomes significantly boosted, causing its decay leptons to be more collimated. To ensure that these leptons are properly resolved in the detector, an angular separation cut, $\Delta R_{\ell\ell}$, is imposed, as specified in Table~\ref{tab:cut_ll}.
The di-lepton invariant mass $M_{\ell\ell}$ is required to be close to the $Z$ boson mass $M_Z$,  
    \begin{equation}
       \vert M_{\ell\ell}-M_Z\vert \leq 10~\mathrm{~GeV},
    \end{equation}

The lepton pair energy $E_{\ell\ell}$ for the signal and di-$Z$ background satisfies a similar peak as Eq.~(\ref{eq:mrecoil}). 
Consequently, the transverse momentum $p_{T,\ell\ell}$ exhibits a Jacobian peak near $E_{\ell\ell}/2$, and cuts on both $p_{T,\ell\ell}$ and $E_{\ell\ell}$ are applied to discriminate the signal from the backgrounds, as summarized in Table~\ref{tab:cut_ll}.
Similar to the $E_\gamma$ cut in the mono-photon analysis, the $E_{\ell\ell}$ cut loses its effectiveness in reducing the di-$Z$ background when the collision energy exceeds 1 TeV. Nonetheless, it remains valuable for suppressing other backgrounds, such as those arising from $W^*$ exchange and di-$W$ processes.

With the application of the above kinematic cuts, we obtain optimized bounds on $(C_W,C_B)$, with the 95\% CL constraints shown as dotted contours in Figure~\ref{fig:monoZ}. At electron colliders, the lower-energy setup yields more stringent constraints since the $E_{\ell\ell}$ cut is more effective at distinguishing the signal from the di-$Z$ background at lower energies. For $\sqrt{s}\gtrsim 1$~TeV, the $E_{\ell\ell}$ distributions for the signal and di-$Z$ background become less distinguishable, as illustrated in Figure~\ref{fig:distMonoZll}; however, the overall di-$Z$ background cross section is suppressed by a factor of $1/s$, as shown in Figure~\ref{fig:momoVxsec}. Consequently, a 10 TeV muon collider can provide stronger constraints than a 3 TeV one.

\subsubsection{Hadronic decay of $Z$ boson}
\label{sec:monoZjj}

Besides the di-lepton channel, we can also resolve a $Z$ boson from its hadronic decay mode $Z\to jj$, with a larger branch fraction but a more complicated hadronic activity. 
In general, the hadronic $Z$ boson is reconstructed from the di-jet events. However, in a high-energy environment, such as at a multi-TeV muon collider, $Z$ bosons can become highly boosted, causing the two final-state partons to be merged into a single jet (referred to as a ``$Z$-jet'').
In our study, we mainly rely on di-jet (or single-jet) events to reconstruct the $Z$ boson for electron (muon) colliders. 
In addition, the $Z$-jet is identified by setting $E_j = E_Z$, without delving into jet substructure in our analysis. Additionally, we treat all jets as generic, without distinguishing between $b$-jets and light jets.

\begin{figure}
    \centering
    \includegraphics[width=0.25\linewidth]{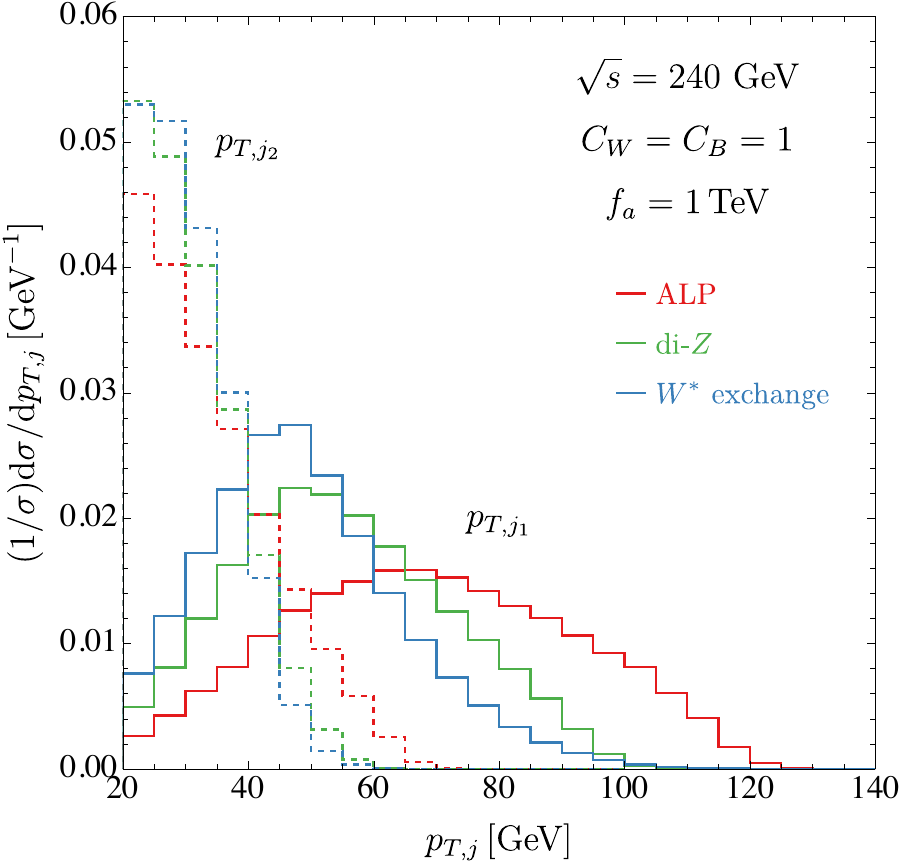}
    \includegraphics[width=0.24\linewidth]{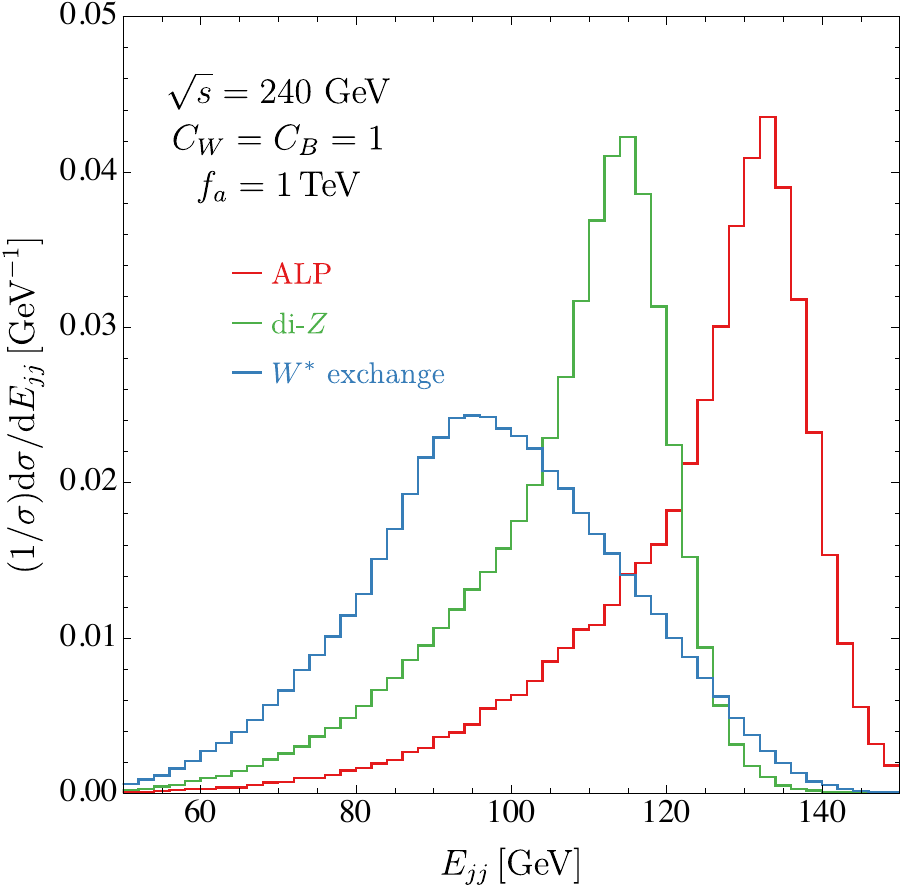}
    \includegraphics[width=0.24\linewidth]{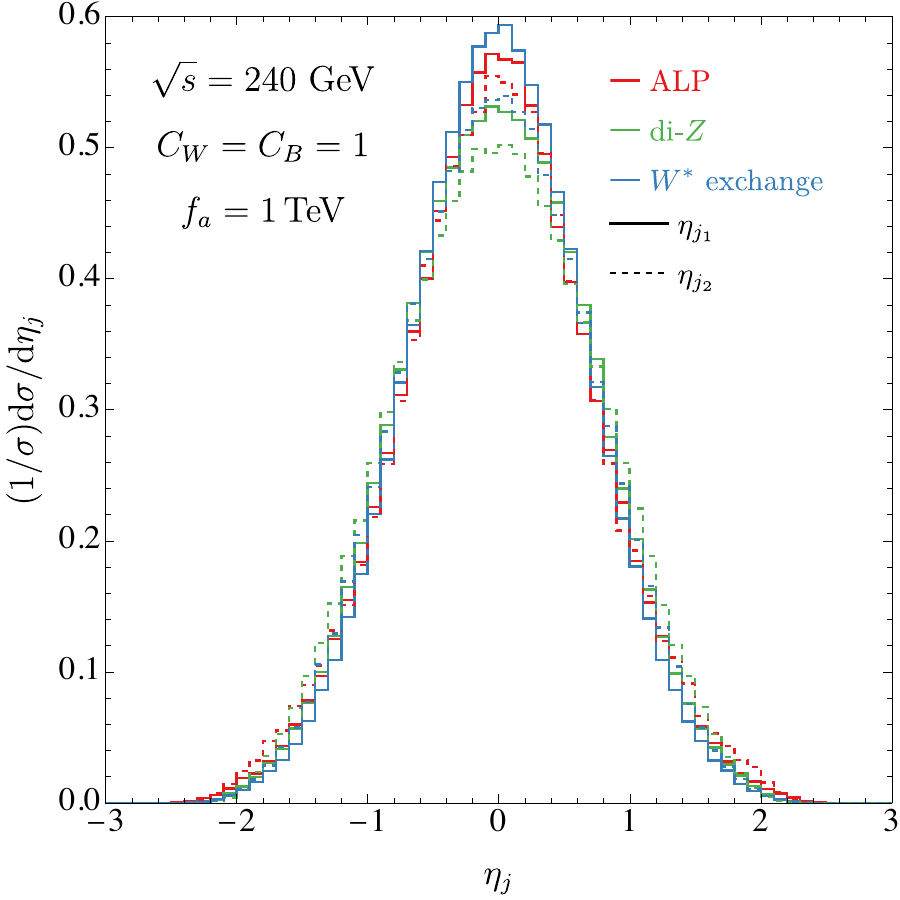}
    \includegraphics[width=0.24\linewidth]{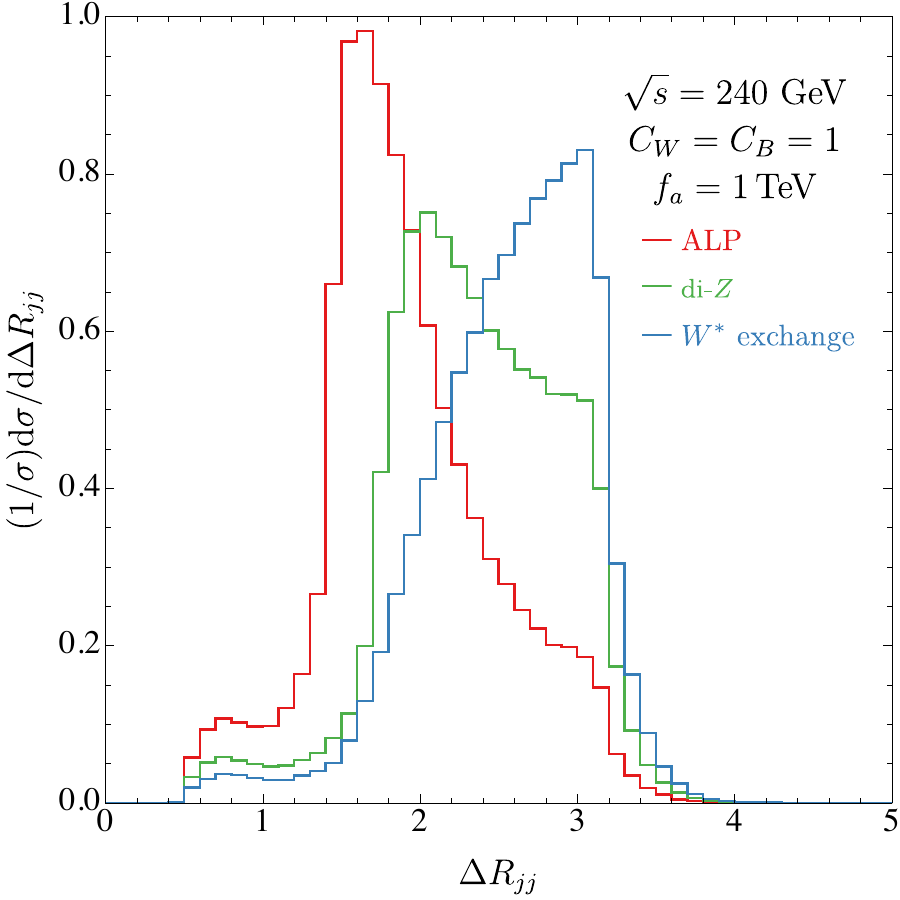}\\
    \includegraphics[width=0.32\linewidth]{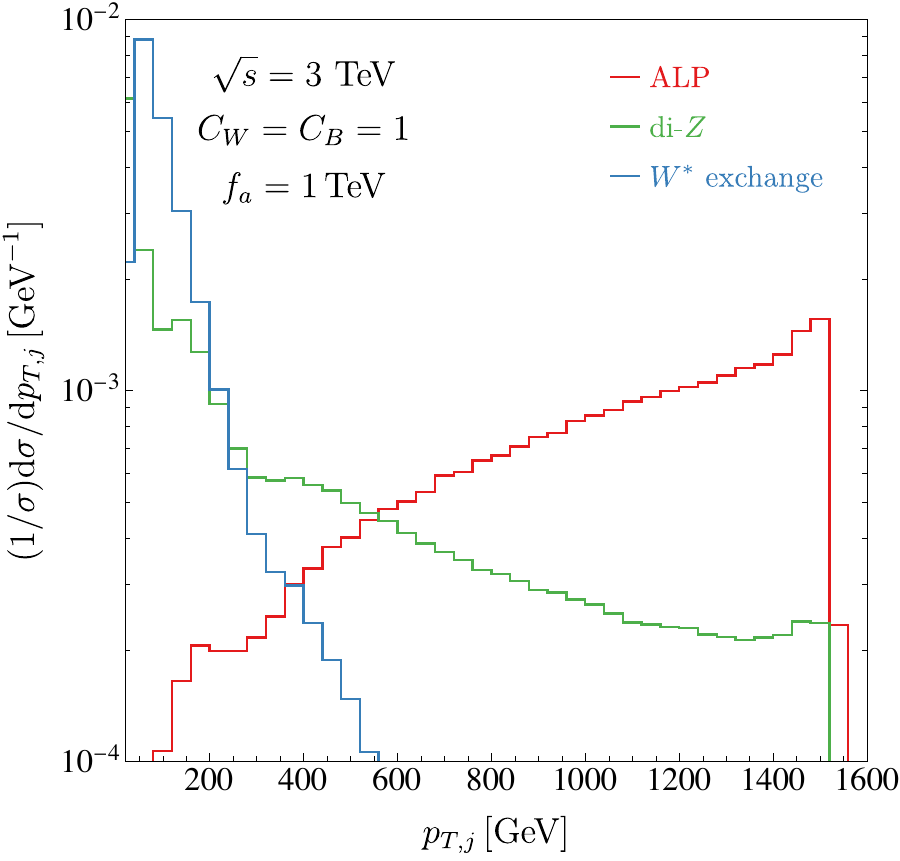}  
    \includegraphics[width=0.32\linewidth]{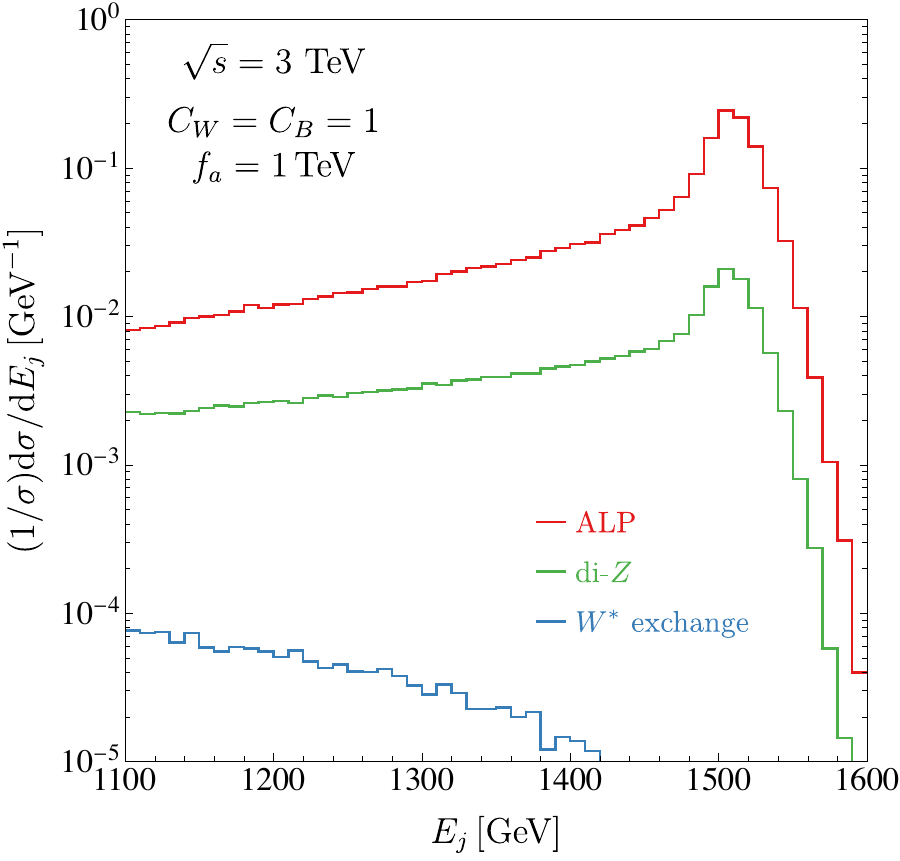}  
    \includegraphics[width=0.315\linewidth]{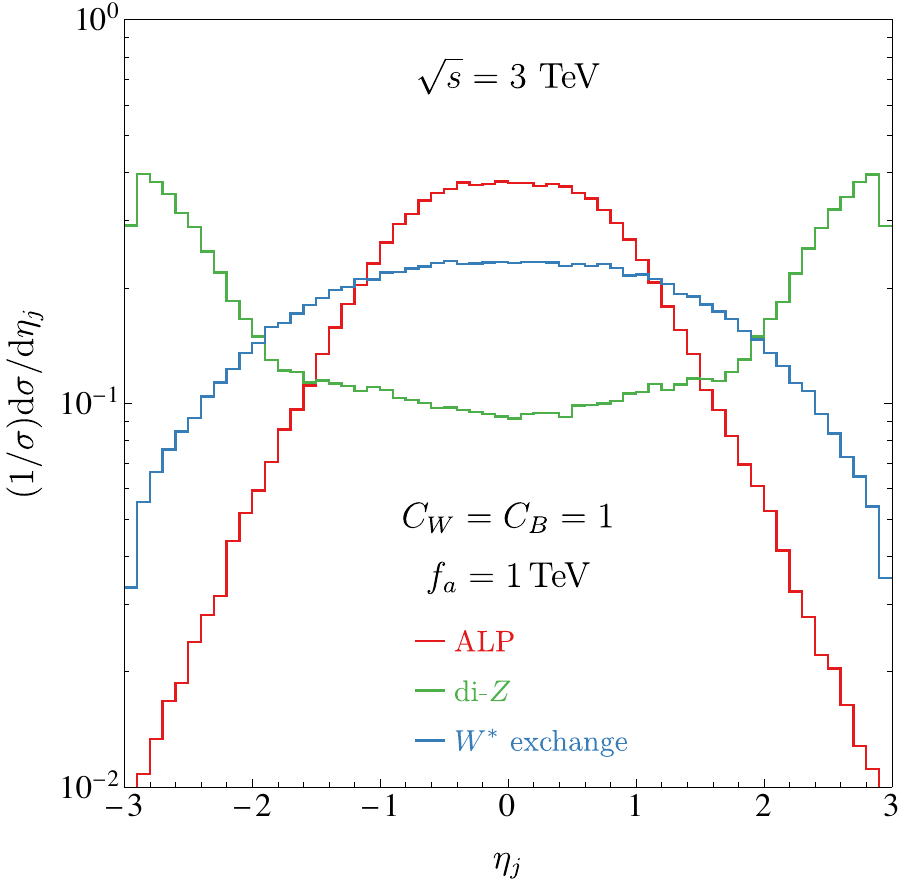}  
    \caption{Similar to Figure~\ref{fig:distMonoZll}, but for the mono-$Z$ production with $Z\to jj$.}
    \label{fig:distMonoZjj}
\end{figure}

Similarly as the leptonic channel, the background can originate from the di-$Z$ production and the $W^*$ exchange scattering. However, we expect that the di-$W$ production can be completely excluded, as it either gives four final-state jets, or two collimated $W$-jets. In both cases, it can be separated with the jet veto or the invariant mass window around $M_{jj}\sim M_Z$.
In Figure~\ref{fig:distMonoZjj}, we present the normalized distributions for the hadronic mono-$Z$ production at a 240 GeV electron collider and a 3 TeV muon collider. Similarly to the di-lepton case, we reconstruct the $Z$-jet from the di-jet or single jet with the invariant mass
\begin{align}
   \vert M_{j(j)}-M_Z\vert\leq 20~{\rm GeV}.
\end{align}
In di-jet events, we rank the jets by their transverse momentum, $p_{T,j}$, and then apply selection cuts on the transverse momenta $p_{T,j_{1,2}}$, the pseudorapidity $\eta_j$, and the angular separation $\Delta R_{jj}$.
For the single jet event, as there is only one jet, we only put the cut on the one jet. The cuts on the energy $E_{jj}$ of the reconstructed $Z$-boson are also applied to further optimize the signal significance.
All the event selection criteria are summarized in Table~\ref{tab:cut_zjj}.

\begin{table}[h!]
\centering
\begin{tabular}{lcccc}
\hline
\hline
Collider & \multicolumn{2}{c}{$e^+e^-$} & \multicolumn{2}{c}{$\mu^+\mu^-$} \\
$\sqrt{s}$ & 240 GeV & 365 GeV & 3 TeV & 10 TeV \\
\hline
$|\eta_j|^{\rm max}$ & 1.0 & 1.0 & 1.75 & 2.0 \\
$p_{T,j_1}^{\rm min}$ [GeV] & 60 & 75 & 550 & 1800 \\
$p_{T,j_2}^{\rm min}$ [GeV] & 40 & 40 & -- & -- \\
$\Delta R^{\rm max}_{jj}$ & 2.0 & 1.8 & -- & -- \\
$E_{Z}^{\rm min}$ [GeV] & 123 & 180 & 1450 & 4800 \\
\hline
\hline
\end{tabular}
\caption{Selection criteria for the $Z$-boson hadronic decay mode $Z\to jj$ in mono-$Z$ events at selected center-of-mass energies. The table specifies pseudorapidity ($|\eta_j|^{\max}$), transverse momentum ($p_T^{\min}$), maximal jet separation ($\Delta R_{jj}^{\max}$), and energy ($E_{Z}^{\min}$) cuts.}
\label{tab:cut_zjj}
\end{table}

\begin{figure}
\centering
\includegraphics[width=0.45\textwidth]{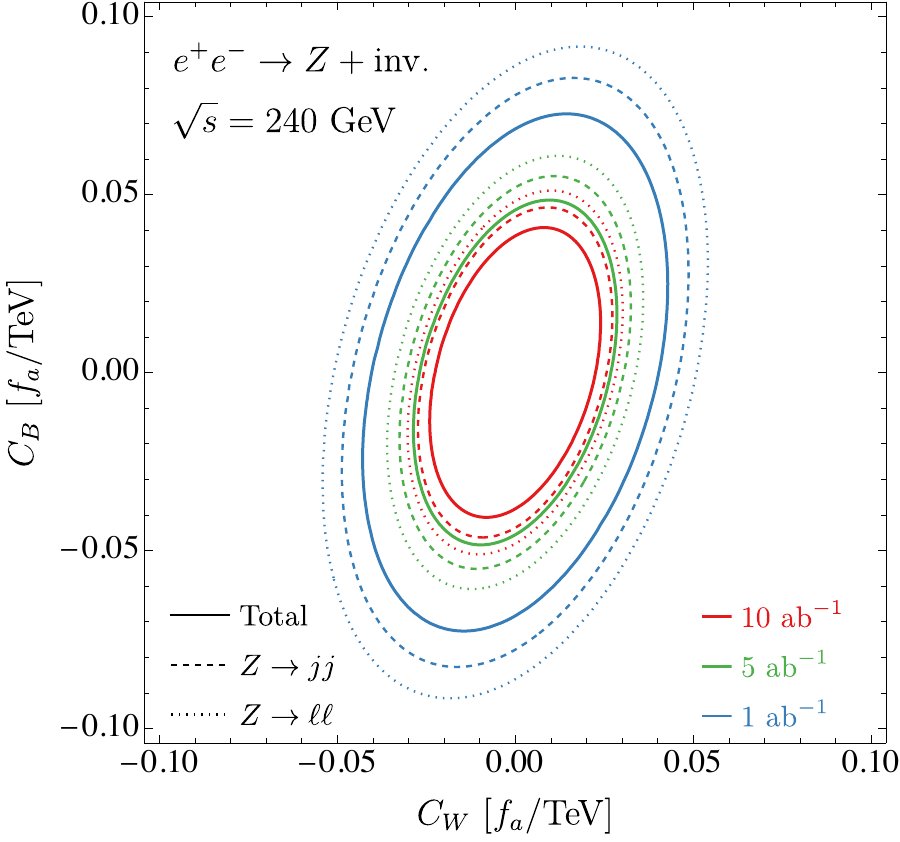}
\includegraphics[width=0.45\textwidth]{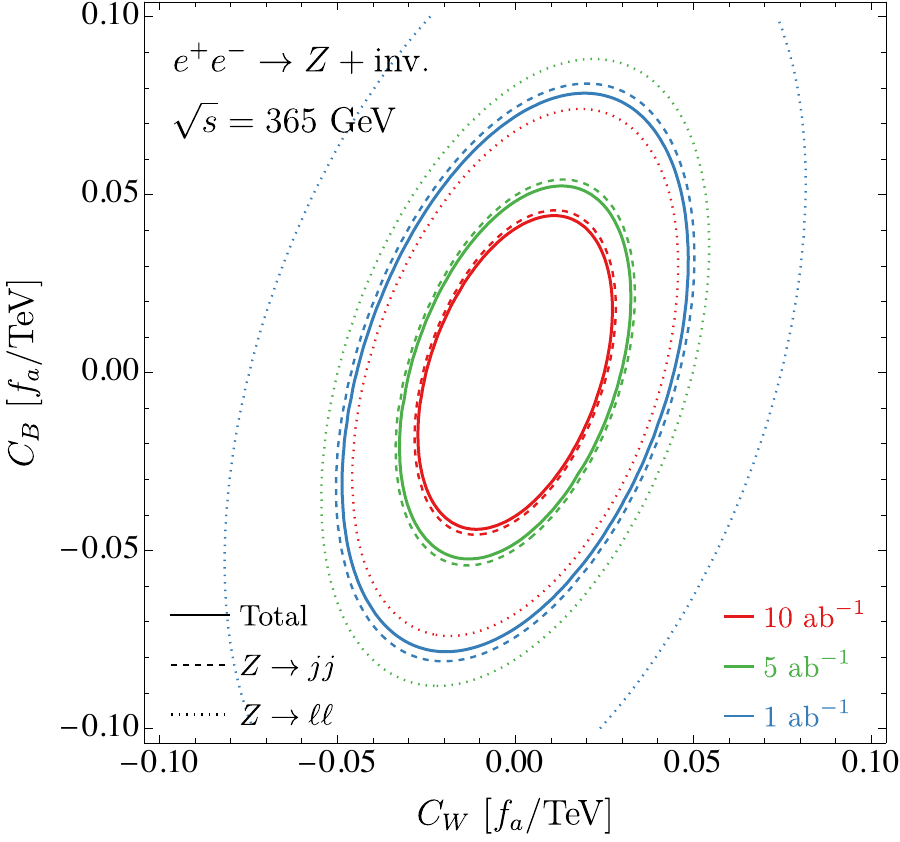}
\includegraphics[width=0.45\textwidth]{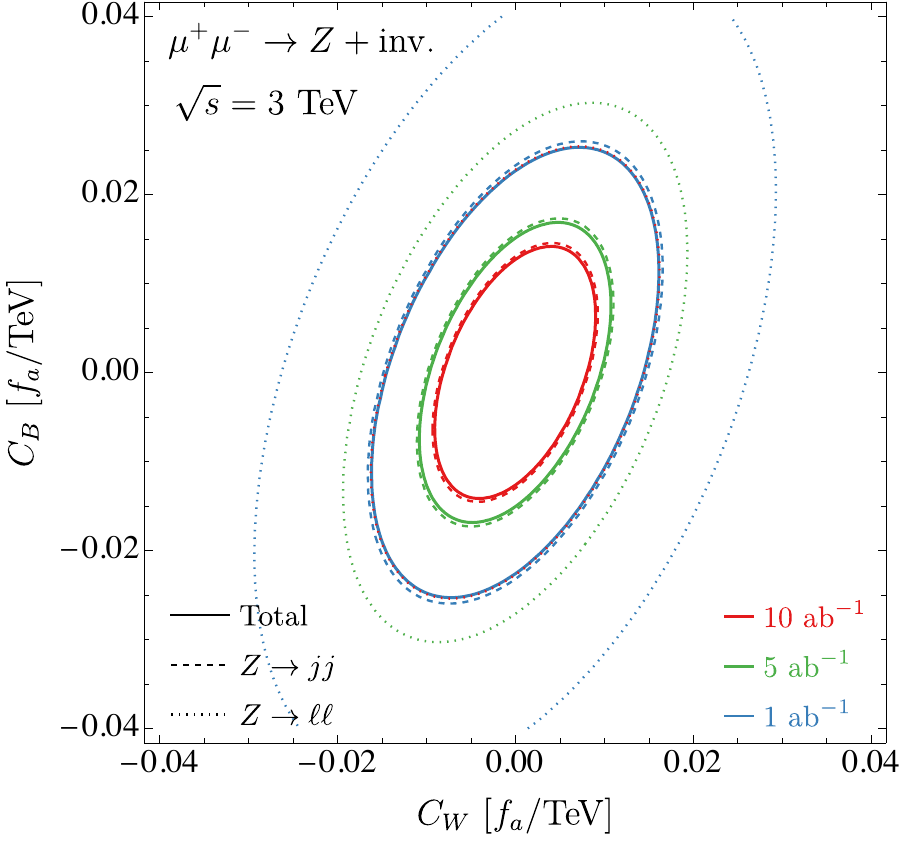}
\includegraphics[width=0.45\textwidth]{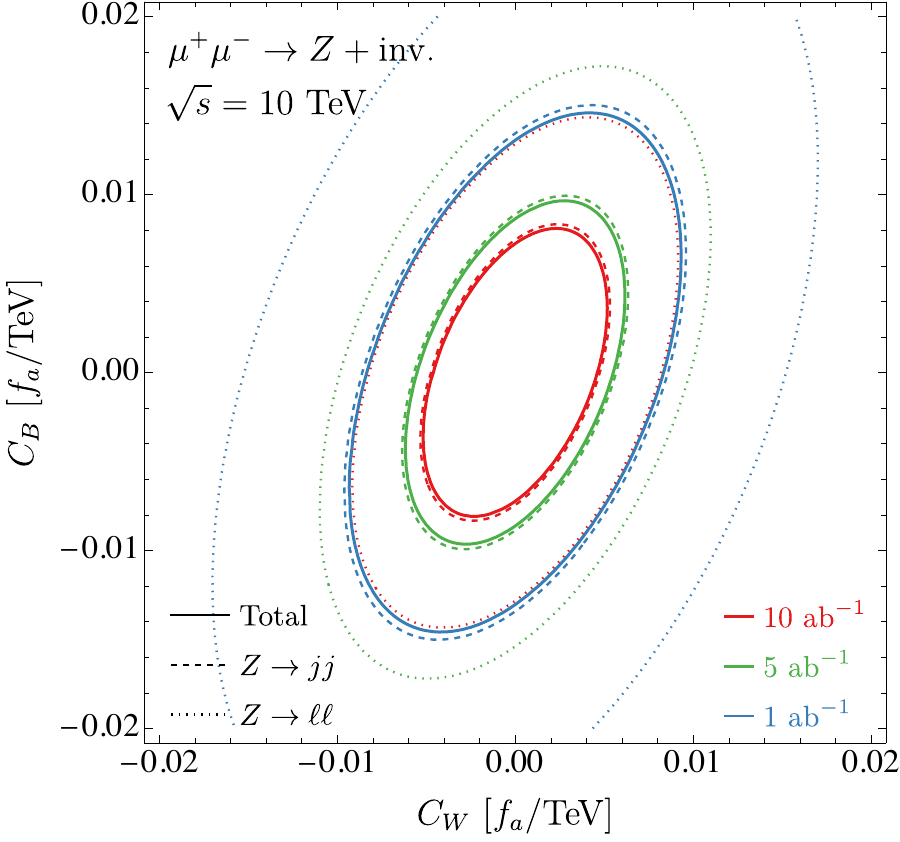}
\caption{Constraints on $C_W$ and $C_B$ from $\ell^+\ell^-\to Z +{\rm inv.}$ searches at high-energy lepton colliders with varying machine energies and luminosities at 95\% CL. Red, green, and blue curves correspond to integrated luminosities of 1 ab$^{-1}$, 5 ab$^{-1}$, and 10 ab$^{-1}$, respectively. Solid contours are based on the total rate, dashed contours use $Z\to jj$, and dotted contours use $Z\to \ell\ell$ for $Z$ boson reconstruction.}\label{fig:monoZ}
\end{figure}

With the above optimized cuts, the 95\% CL constraints on $(C_W,C_B)$ from the hadronic mono-$Z$ measurement are presented as dashed contours in Figure~\ref{fig:monoZ}. We observe that for electron colliders, the 240 GeV configuration yields slightly stronger constraints than the 365 GeV one, while for muon colliders the trend is reversed, with the 10 TeV collider providing tighter constraints than the 3 TeV machine. At electron colliders operating at CEPC/FCC-ee energies, the $Z\to jj$ decay is primarily reconstructed as di-jet events, allowing the 240 GeV setup to employ a more effective $E_Z$ cut compared to the 365 GeV option. However, at higher collision energies, the efficiency of the $E_Z$ cut in separating signal from background diminishes and ultimately fails for $\sqrt{s}\gtrsim 1$~TeV. 
At multi-TeV muon colliders, the hadronic decay of the $Z$ boson typically  results in a highly boosted single jet rather than a resolved di-jet event. The identification efficiency for such boosted jets improves at higher energies, and, combined with a lower irreducible di-$Z$ background, the 10 TeV muon collider imposes stronger constraints than the 3 TeV machine.
Meanwhile, we also see that the hadronic channel gives a stronger bound than the leptonic one, mainly driven by the larger branch fractions.

At last, we combine the constraints from leptonic and hadronic mode of the mono-$Z$ production with the signal significance
\begin{eqnarray}
    \mathcal{S}_{Z}=\sqrt{\mathcal{S}^2_{Z\to \ell \ell}+\mathcal{S}^2_{Z\to jj}}.
\end{eqnarray}
The combined results are presented as solid contours in Figure~\ref{fig:monoZ}. The corresponding 95\% CL constraints on the $g_{aVV}$ couplings are listed in Table~\ref{tab:gconstraints_mono_z}. 
At the multi-TeV muon colliders, the constraints from mono-$Z$ production are comparable to those from the mono-photon production. It's more interesting that the constraint contour in $(C_W,\,C_B$) plane provided by mono-$Z$ production has different direction to provided by mono-photon. And hence mono-$Z$ and mono-photon are complementary at muon colliders.

\begin{table}[htbp]
\centering
\resizebox{\textwidth}{!}{
\renewcommand{\arraystretch}{1.2}
\begin{tabular}{c|ccc|ccc}
\hline\hline
$e^+e^-$ Collider & \multicolumn{3}{c|}{$\sqrt{s}=240$ GeV} & \multicolumn{3}{c}{$\sqrt{s}=365$ GeV} \\
\hline
Luminosity [ab$^{-1}$] & 1  & 5  & 10  & 1  & 5  & 10  \\
\hline
$|g_{a\gamma\gamma}|^{\rm max}$ [TeV$^{-1}$] & $2.42 \times 10^{-1}$ & $1.61 \times 10^{-1}$ & $1.35\times 10^{-1}$ & $2.64 \times 10^{-1}$ & $1.77 \times 10^{-1}$ & $1.48\times 10^{-1}$ \\
$|g_{a\gamma Z}|^{\rm max}$ [TeV$^{-1}$] & $2.36 \times 10^{-1}$ & $1.57 \times 10^{-1}$ & $1.32\times 10^{-1}$ & $2.46 \times 10^{-1}$ & $1.64 \times 10^{-1}$ & $1.38\times 10^{-1}$ \\
$|g_{aZZ}|^{\rm max}$ [TeV$^{-1}$]& $1.67 \times 10^{-1}$ & $1.11 \times 10^{-1}$ & $9.34 \times 10^{-2}$ & $1.90 \times 10^{-1}$ & $1.27 \times 10^{-1}$ & $1.07 \times 10^{-1}$ \\
$|g_{aWW}|^{\rm max}$ [TeV$^{-1}$]& $1.71 \times 10^{-1}$ & $1.14 \times 10^{-1}$ & $9.61 \times 10^{-2}$ & $1.95 \times 10^{-1}$ & $1.30 \times 10^{-1}$ & $1.09 \times 10^{-1}$ \\
\hline
$\mu^+\mu^-$ Collider & \multicolumn{3}{c|}{$\sqrt{s}=3$ TeV} & \multicolumn{3}{c}{$\sqrt{s}=10$ TeV} \\
\hline
Luminosity [ab$^{-1}$] & 1 & 5  & 10 & 1  & 5  & 10  \\
\hline
$|g_{a\gamma\gamma}|^{\rm max}$ [TeV$^{-1}$] & $8.61 \times 10^{-2}$ & $5.74 \times 10^{-2}$ & $4.82 \times 10^{-2}$ & $4.96 \times 10^{-2}$ & $3.28 \times 10^{-2}$ & $2.75\times 10^{-2}$ \\
$|g_{a\gamma Z}|^{\rm max}$ [TeV$^{-1}$] & $7.70 \times 10^{-2}$ & $5.13 \times 10^{-2}$ & $4.31\times 10^{-2}$ & $4.44 \times 10^{-2}$ & $2.94 \times 10^{-2}$ & $2.46 \times 10^{-2}$ \\
$|g_{aZZ}|^{\rm max}$ [TeV$^{-1}$] & $6.36 \times 10^{-2}$ & $4.24 \times 10^{-2}$ & $3.56 \times 10^{-2}$ & $3.67 \times 10^{-2}$ & $2.43 \times 10^{-2}$ & $2.04 \times 10^{-2}$ \\
$|g_{aWW}|^{\rm max}$ [TeV$^{-1}$]& $6.46 \times 10^{-2}$ & $4.31 \times 10^{-2}$ & $3.62 \times 10^{-2}$ & $3.73 \times 10^{-2}$ & $2.47 \times 10^{-2}$ & $2.07 \times 10^{-2}$ \\
\hline\hline
\end{tabular}
}
\caption{The upper limits $|g_{aVV}|^{\rm max}$ from the mono-$Z$ production process at different future lepton colliders.}\label{tab:gconstraints_mono_z}
\end{table}

\section{Non-resonant ALP in vector boson scattering}
\label{sec:VBS}

For future high-energy lepton colliders, the initial state radiation is significant. Gauge bosons ($\gamma$, $W$, and $Z$) can be collinearly radiated off the beam leptons, generating new interactions through these radiation fields. Due to the collinear enhancement, vector boson scattering or fusion (VBS/VBF) processes exhibit double-logarithmic growth, providing additional opportunities for physics studies at future multi-TeV lepton colliders~\cite{Costantini:2020stv,Han:2020uid,Han:2021kes,BuarqueFranzosi:2021wrv,Ruiz:2021tdt}. The gauge bosons radiated off the beams can be treated as the ``partons'' of the beam lepton~\cite{Han:2020uid,Han:2021kes} and shall be described using proper parton distribution functions, namely the EW PDFs~\cite{Han:2020uid,Han:2021kes,Garosi:2023bvq}.

Following the factorization formalism, the cross section for the VBS production of final-state $\mathcal{F}$ can be expressed as
\begin{eqnarray}
    \sigma(V_1 V_2 \to \mathcal{F})\simeq \sum_{V_1 V_2}\int_{\tau_0}^1 dx_1\int_{\tau_0/x_1}^1  dx_2 f_{V_1/\ell} (x_1,Q^2) f_{V_2/\ell} (x_2,Q^2) \hat{\sigma}(V_2 V_2 \to \mathcal{F}),\label{eq:sigma_VBS}
\end{eqnarray}
where the $f_{V_{1,2}/\ell} (x_{1,2},Q^2)$ are the PDFs that describe the probabilities of finding the parton $V_{1,2}$ with an energy fraction $x_{1,2}$ from the beam lepton and $Q$ is the factorization scale. At the leading order, the photon PDF is given by the equivalent photon approximation (EPA) \cite{vonWeizsacker:1934nji,Williams:1934ad} 
\begin{eqnarray}\label{eq:EPA}
    f_{\gamma/\ell} = \frac{\alpha}{2\pi}  \frac{1+(1-x)^2}{x} \log(\frac{Q^2}{m_\ell^2}). 
\end{eqnarray}
Similarly, the weak gauge bosons $V=W,Z$ can be dealt with the effective W approximation (EWA)~\cite{Dawson:1984gx,Kane:1984bb} as
\begin{equation}
\begin{aligned}
      &&f_{V_-/\ell_L}\simeq \frac{g_L^2}{8 \pi^2} \frac{1}{x} \log \frac{Q^2}{M_V^2}, \quad f_{V_+/\ell_L}\simeq \frac{g_L^2}{8 \pi^2} \frac{(1-x)^2}{x} \log \frac{Q^2}{M_V^2}, \quad f_{V_0/\ell_L}\simeq \frac{g_L^2}{4 \pi^2} \frac{1-x}{x},\\
      &&f_{V_+/\ell_R}\simeq \frac{g_R^2}{8 \pi^2} \frac{1}{x} \log \frac{Q^2}{M_V^2}, \quad f_{V_+/\ell_R}\simeq \frac{g_R^2}{8 \pi^2} \frac{(1-x)^2}{x} \log \frac{Q^2}{M_V^2}, \quad f_{V_0/\ell_R}\simeq \frac{g_R^2}{4 \pi^2} \frac{1-x}{x},    
\end{aligned}    
\end{equation}
where $\pm,0$ denote the corresponding polarizations. The couplings of the left- and right-handed leptons to the vector bosons $g_{L,R}$ are
\begin{eqnarray}
    g_L = \frac{g}{\sqrt 2}, ~~~g_R=0,~~~g=\frac{e}{\sin\theta_W},
\end{eqnarray}
for $V=W^\pm$, and 
\begin{eqnarray}
    g_L = \frac{g}{\cos\theta_W} \left( T_3^\ell -Q^\ell \sin^2\theta_W \right), ~~~g_R=-\frac{g}{\cos\theta_W} Q^\ell\sin^2\theta_W,
\end{eqnarray} 
for $V=Z$, with the weak charge $T_3^\ell=-1/2$ and electric charge $Q^\ell=-1$.

To illustrate the potential of future lepton colliders to probe ALP couplings to the SM EW bosons, we study vector boson pair production via VBS,
\begin{eqnarray}
    V_1 V_2 \to V_1' V_2',~~V_{1,2},\,V'_{1,2} \in \{\gamma,\,Z,\,W^\pm\}.
\end{eqnarray}
In our analysis, we adopt the effective vector approximation
(EVA) implementation~\cite{Ruiz:2021tdt} in \textsc{MadGraph5\_aMC@NLO}~\cite{Alwall:2014hca,Frederix:2018nkq} to calculate the VBS processes, with the factorization scale chosen as $Q = \sqrt{\hat{s}}/2$. 
Due to the possible interference, the effect of the ALP can be qualified by the difference between the cross-sections with and without the ALP couplings. 
The signal and background event numbers in Eq.~\eqref{eq:signal_strength} can be expressed as
\begin{equation}
S(C_W,C_B) = \mathscr{L} \left(\sigma_{\rm VBS}(C_W,C_B)-\sigma_{\rm VBS}^{\rm SM} \right), \quad B = \mathscr{L} \sigma_{\rm VBS}^{\rm SM},
\end{equation}
where $\sigma_{\rm VBS}^{\rm SM}\equiv\sigma_{\rm VBS}(C_W=0,C_B=0)$.
Further improvements can be achieved through the fully resummed EW PDFs~\cite{Han:2020uid,Han:2021kes,Garosi:2023bvq} and the next-to-leading-order (NLO) corrections~\cite{Ma:2024ayr,Bredt:2022dmm}, which go beyond the scope of this paper.

\subsection{Light-by-light scattering}

\begin{figure}
    \centering
\subfigure[]{\includegraphics[width=0.24\linewidth]{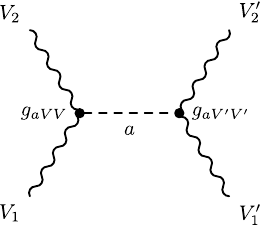}\label{feyn:ALPs}}
\subfigure[]{\includegraphics[width=0.24\linewidth]{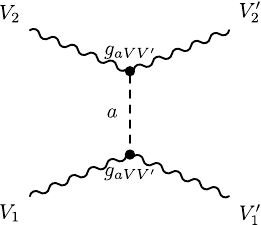}\label{feyn:ALPt}}
\subfigure[]{\includegraphics[width=0.24\linewidth]{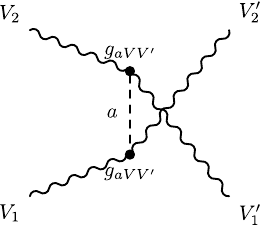}\label{feyn:ALPu}}\\ 
    \caption{Representative Feynman diagrams for the ALP's participation in the vector boson scattering.}
    \label{feyn:VBS}
\end{figure}

\begin{figure}[!htb]
\centering      
\subfigure[]{\includegraphics[width=0.25\linewidth]{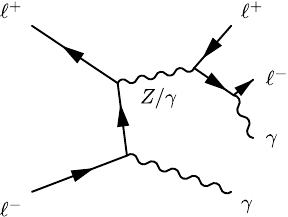}\label{feyn:Zpole}}
\subfigure[]{\includegraphics[width=0.25\linewidth]{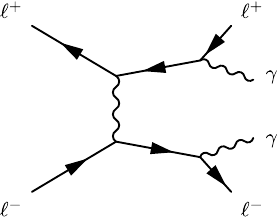}\label{feyn:llaa}}    
\subfigure[]{\includegraphics[width=0.225\linewidth]{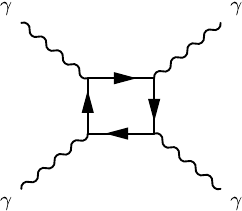}\label{feyn:box}}  
\subfigure[]{\includegraphics[width=0.225\linewidth]{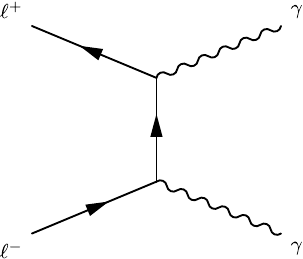}\label{feyn:ll2aa}} 
\caption{The representative Feynman diagrams of SM backgrounds for the photon pair production at lepton colliders.}
\label{fig:SMaa}
\end{figure}

At both future $e^+e^-$ colliders and multi-TeV muon colliders, photon pair production offers a promising channel to probe ALPs. As illustrated in Figure~\ref{feyn:VBS}, ALPs can induce light-by-light scattering through $s$-, $t$-, and $u$-channels. The full process at the colliders is $\ell^+\ell^- \to \ell^+\ell^- \gamma\gamma$, where the ALP-induced light-by-light scattering is logarithmically enhanced. As for the SM background, the leading contribution arises from two categories: (i) initial- and final-state traditions of the $Z$ resonance production as Figure~\ref{feyn:Zpole}; (ii) the bremsstrahlung of the Bhabha scattering as Figure~\ref{feyn:llaa}. In comparison, the SM loop-induced light-by-light scattering as Figure~\ref{feyn:box} is sub-leading.
In addition, we also have background comes from the direct photon pair production process, $\ell^+\ell^- \to \gamma\gamma$, shown in Figure~\ref{feyn:ll2aa}. The initial-state radiation (ISR) can shift the photon pair invariant mass toward lower values, allowing a sizable contribution in the small $m_{\gamma\gamma}$ region. 

The light-by-light scattering process has been employed to probe the $a\text{-}\gamma\text{-}\gamma$ vertex through on-shell ALP production at future $e^+e^-$ colliders, assuming a dominant decay branching fraction Br$(a\to\gamma\gamma) \simeq 1$~\cite{RebelloTeles:2023uig,Zhang:2021sio}.  
In this scenario, the ALP is considered short-lived, so that the resulting constraints on $g_{a\gamma\gamma}$ depend on the ALP decay length, and thus on the ALP mass $m_a$.  
In contrast, we focus on non-resonant ALP production via light-by-light scattering. Although the corresponding constraints are generally weaker than those obtained from resonant ALP production, they are more robust and less sensitive to theoretical assumptions such as the ALP mass or decay width. 

In our analysis, we require $m_{\gamma\gamma} > 100$~GeV for the 240 GeV and 365 GeV FCC-ee, as well as for the multi-TeV muon colliders, ensuring that the derived constraints on the $g_{VV}$ couplings remain applicable for ALPs with masses up to $M_Z$.  
To discriminate the signal from the dominant SM background $\ell^+\ell^- \to \gamma\gamma$, we employ forward lepton tagging by requiring $3.13 < |\eta_\ell| < 6$.  
The following selection cuts are further applied to optimize the signal sensitivity:
\begin{eqnarray}
    |\eta_\gamma| < 1.5,\quad p_{T,\gamma} > 5~\mathrm{GeV},\quad \Delta R(\gamma\gamma) > 0.4,
\end{eqnarray}
where the pseudorapidity cut on photons helps to suppress the SM contribution from FSR in $\ell^+\ell^- \to \ell^+\ell^-$.  
Additionally, we impose $m_{\ell\ell} > 150$~GeV at muon colliders, and $|m_{\ell\ell} - M_Z| > 15\Gamma_Z$ at FCC-ee, to suppress the $Z$-resonance background of Figure~\ref{feyn:Zpole}.  
For illustration, we also study the light-by-light scattering process at the Tera-$Z$ phase of FCC-ee, where we require $m_{\gamma\gamma} > 20$~GeV instead of $m_{\gamma\gamma} > 100$~GeV. In this case, the derived constraints are valid up to $m_a \sim 10$~GeV.

\begin{figure}[htb]
\centering
    \includegraphics[width=0.455\textwidth]{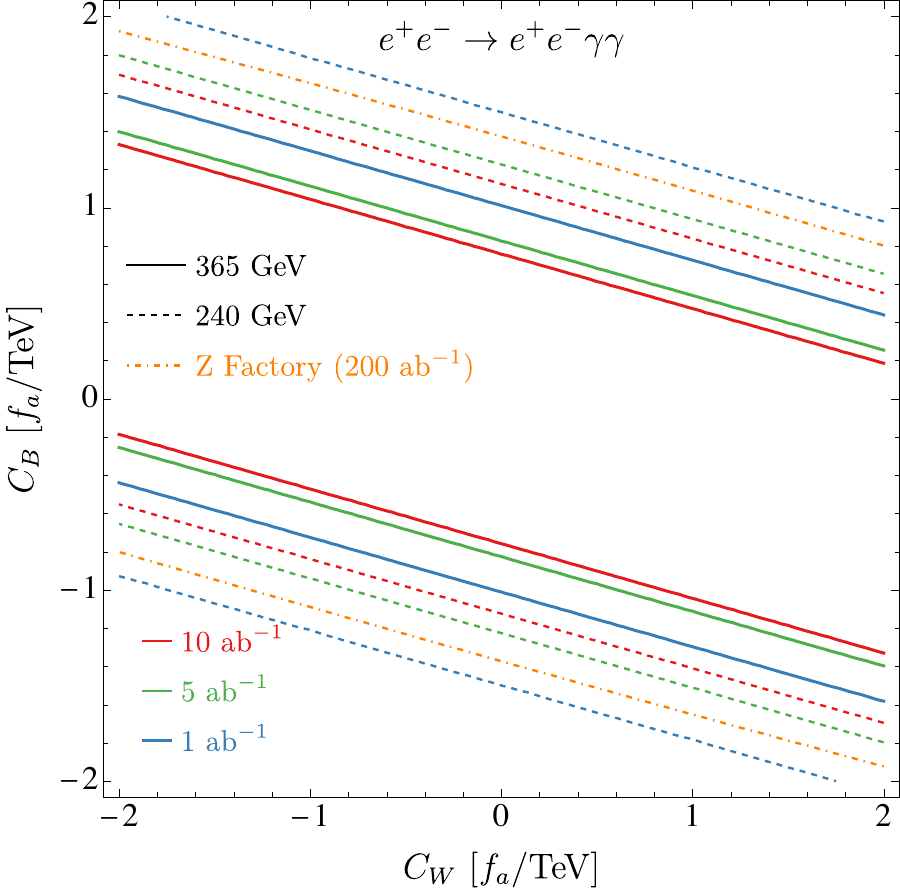}
    \includegraphics[width=0.47\textwidth]{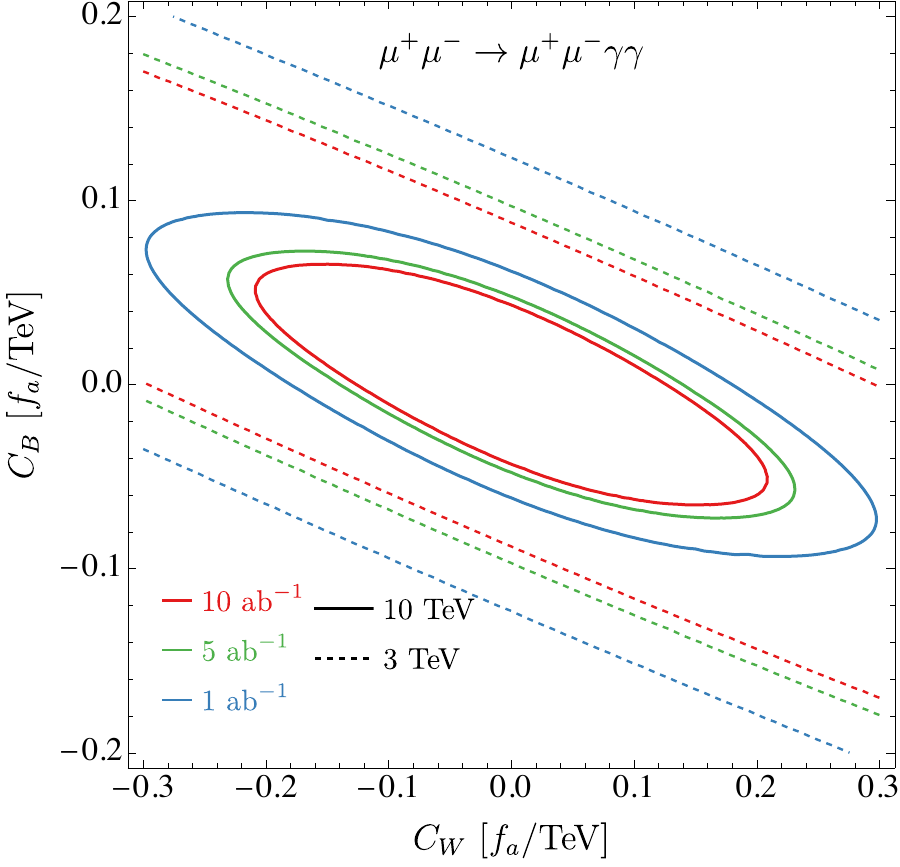}
    \caption{Constraints on $C_W$ and $C_B$ via the light-by-light scattering at (a) future $e^+e^-$ colliders and (b) multi-TeV muon colliders at 95\% CL. Blue, green, and red curves indicate integrated luminosities of 1 ab$^{-1}$, 5 ab$^{-1}$, and 10 ab$^{-1}$, respectively. The Dotdashed orange curve is the 200 ab$^{-1}$ $Z$ factory.}\label{fig:lbl_contour}
\end{figure}

With the kinematic cuts described above, the 95\% CL constraints on $(C_W, C_B)$ from light-by-light scattering are presented as contours in Figure~\ref{fig:lbl_contour}.  
The blue, green, and red curves correspond to integrated luminosities of 1 ab$^{-1}$, 5 ab$^{-1}$, and 10 ab$^{-1}$, respectively.  
For the $e^+e^-\to e^+e^-\gamma\gamma$ process, solid curves represent $\sqrt{s} = 365$~GeV, while dashed curves correspond to $\sqrt{s} = 240$~GeV.  
For the muon colliders, solid curves represent $\sqrt{s} = 10$~TeV and dashed curves represent $\sqrt{s} = 3$~TeV.  
The orange dot-dashed curve shows the result for the Tera-$Z$ phase of the future $e^+e^-$ collider with an integrated luminosity of 200 ab$^{-1}$.  
At future $e^+e^-$ colliders, the BSM effects are dominated by the $a\text{-}\gamma\text{-}\gamma$ interaction, leading to constraints that appear as straight bands in the $(C_W, C_B)$ plane.  
The corresponding limits on the $g_{a\gamma\gamma}$ coupling are summarized in Table~\ref{tab:lbl_gavv}.  
For the multi-TeV muon colliders, where the $a\text{-}\gamma\text{-}Z$ and $a\text{-}Z\text{-}Z$ interactions become non-negligible, the constraints form closed contours.The corresponding limits on the $g_{aVV}$ couplings at future muon colliders are listed in Table~\ref{tab:lbl_gavv2}.

\begin{table}[h!]
\centering
\begin{tabular}{c| c| ccc |ccc}
\hline
\hline
$\sqrt{s}$ [GeV]  & 91.2 & \multicolumn{3}{c|}{240} & \multicolumn{3}{c}{365} \\
\hline
Luminosity [ab$^{-1}$] & 200 & 1 & 5 & 10 & 1 & 5 & 10 \\
$|g_{a\gamma\gamma}|^{\rm max}$ [TeV$^{-1}$] & 4.28 & 4.67 & 3.82 & 3.50 & 3.15 & 2.57 & 2.36 \\
\hline
\hline
\end{tabular}
\caption{Projected 95\% CL upper limits on $|g_{a\gamma\gamma}|$ from light-by-light scattering at future $e^+e^-$ colliders with different center-of-mass energies and integrated luminosities.}
\label{tab:lbl_gavv}
\end{table}

\begin{table}[htbp]
\centering
\resizebox{\textwidth}{!}{
\renewcommand{\arraystretch}{1.2}
\begin{tabular}{c|ccc|ccc}
\hline\hline
$\sqrt{s}$ & \multicolumn{3}{c|}{3 TeV} & \multicolumn{3}{c}{10 TeV} \\
\hline
Luminosity [ab$^{-1}$] & 1  & 5 & 10  & 1  & 5 & 10  \\
\hline
$|g_{a\gamma\gamma}|^{\rm max}$ [TeV$^{-1}$] & $3.84 \times 10^{-1}$ & $3.02 \times 10^{-1}$ & $2.74\times 10^{-1}$ & $1.93 \times 10^{-1}$ & $1.50 \times 10^{-1}$ & $1.35\times 10^{-1}$ \\
$|g_{a\gamma Z}|^{\rm max}$ [TeV$^{-1}$] & $6.58 \times 10^{0}$ & $5.19 \times 10^{0}$ & $4.71\times 10^{0}$ & $1.25 \times 10^{0}$ & $9.68 \times 10^{-1}$ & $8.74\times 10^{-1}$ \\
$|g_{aZZ}|^{\rm max}$ [TeV$^{-1}$]& $4.40 \times 10^{0}$ & $3.47 \times 10^{0}$ & $3.15 \times 10^{0}$ & $8.62 \times 10^{-1}$ & $6.70 \times 10^{-1}$ & $6.05 \times 10^{-1}$ \\
$|g_{aWW}|^{\rm max}$ [TeV$^{-1}$]& $6.16 \times 10^{0}$ & $4.85 \times 10^{0}$ & $4.40 \times 10^{0}$ & $1.19 \times 10^{0}$ & $9.25 \times 10^{-1}$ & $8.35 \times 10^{-1}$ \\
\hline
\hline
\end{tabular}
}
\caption{The upper limits on $|g_{aVV}|^{\rm max}$ from the light-by-light scattering at a 3 TeV muon collider (left) and a 10 TeV muon collider (right).}\label{tab:lbl_gavv2}
\end{table}

\subsection{Electroweak vector boson scattering}
At multi-TeV lepton colliders, the EW VBS process features large production cross sections, offering a promising opportunity to search heavy ALPs via on-shell production processes~\cite{Han:2022mzp,Bao:2022onq,Inan:2022rcr}. Conversely, light ALPs can also be produced non-resonantly through VBS. Along this line, the process $V_1 V_2 \to V_1' V_2'$, whose Feynman diagrams are shown in Figure~\ref{feyn:VBS}, serves as a probe for light ALP interactions with SM EW gauge bosons.  
In these non-resonant VBS processes, the partonic scattering energy is typically well above the ALP mass, \emph{i.e.}, $\hat{s} \gg m_a^2$, causing the ALP propagator to be off-shell. As a result, the BSM contributions become insensitive to the ALP mass and decay width.  
These processes thus provide constraints on the coupling constants $g_{aVV}$, independent of the ALP mass $m_a$.

\begin{figure}[htb]
\centering
    \includegraphics[width=0.49\textwidth]{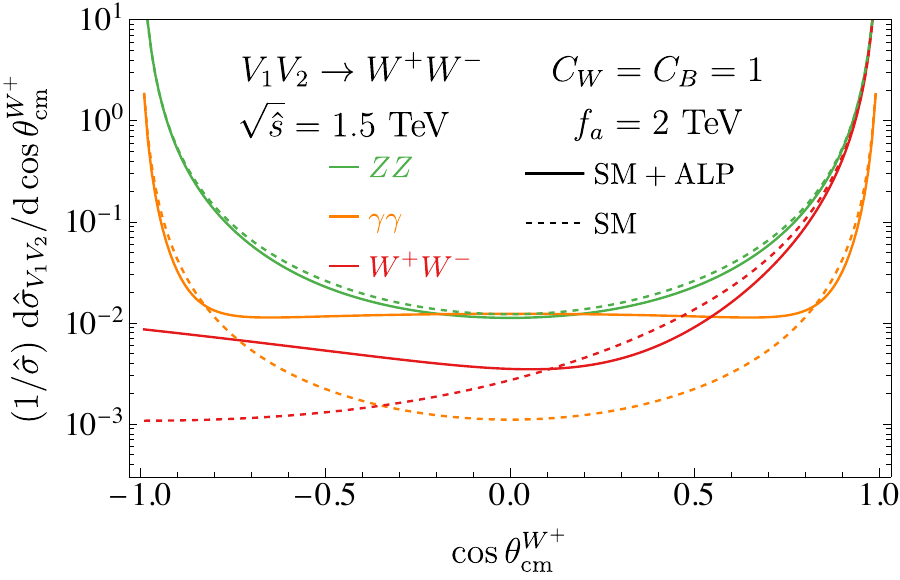}
    \includegraphics[width=0.49\textwidth]{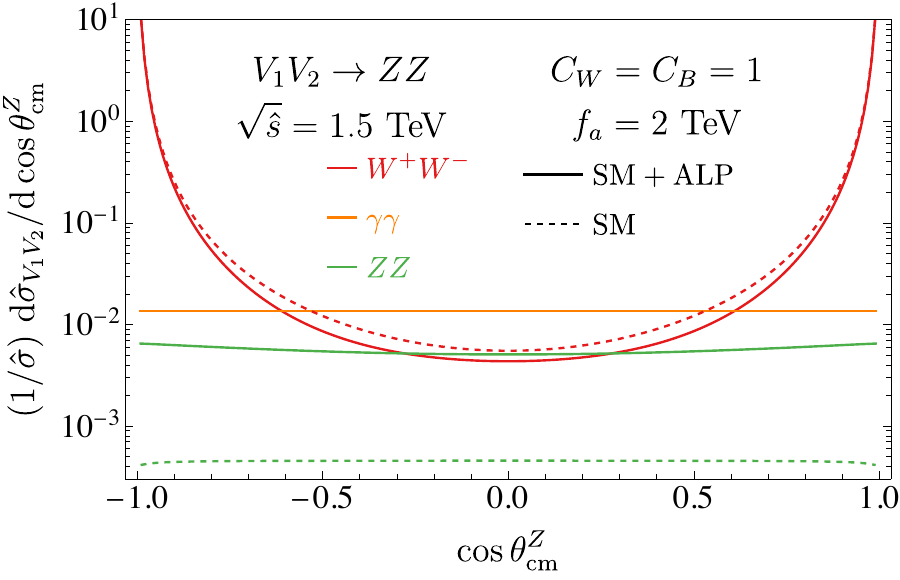}
    \caption{The normalized angular distributions of the partonic cross sections for (a) $V_1 V_2\to W^+W^-$ and (b) $V_1 V_2\to ZZ$. The dashed curves are for the SM diagrams only while the solid curves count both the SM and the ALP contributions. The differential cross sections are normalized using the sum of all partonic channels, ${\hat \sigma}=\sum_{V_1V_2}{\hat \sigma_{V_1V_2\to V_1'V_2'}}$, where $V_1V_2=\gamma\gamma, W^+W^-, ZZ$.}\label{fig:VBS_partonic}
\end{figure}

To explore ALP's impact on EW VBS processes, we take two representative production $V_1V_2 \to W^+W^-$ and $V_1V_2 \to ZZ$ for demonstration. For simplicity, we fix $f_a = 2$ TeV and $C_W = C_B = 1$ as an illustration, which gives $g_{a\gamma Z} = 0$.
We define $\theta_{\rm cm}^{V}$ ($V= W^+, Z$ ) as the angle of the final-state vector boson with respect to the beam axis in the partonic center-of-mass frame, and present the normalized $\cos\theta_{\rm cm}^{V}$ distribution of partonic cross section 
in Figure~\ref{fig:VBS_partonic}. 
We see that the inclusion of the ALP enhances the production cross section in the small $|\cos\theta_{\rm cm}^V|$ region, corresponding to significant signals in the large $p_T$ range. Moreover, the ALP introduces the a new scattering channel $\gamma\gamma \to ZZ$, which is absent in the SM at tree level. 

\begin{figure}[htb]
\centering
    \includegraphics[width=0.49\textwidth]{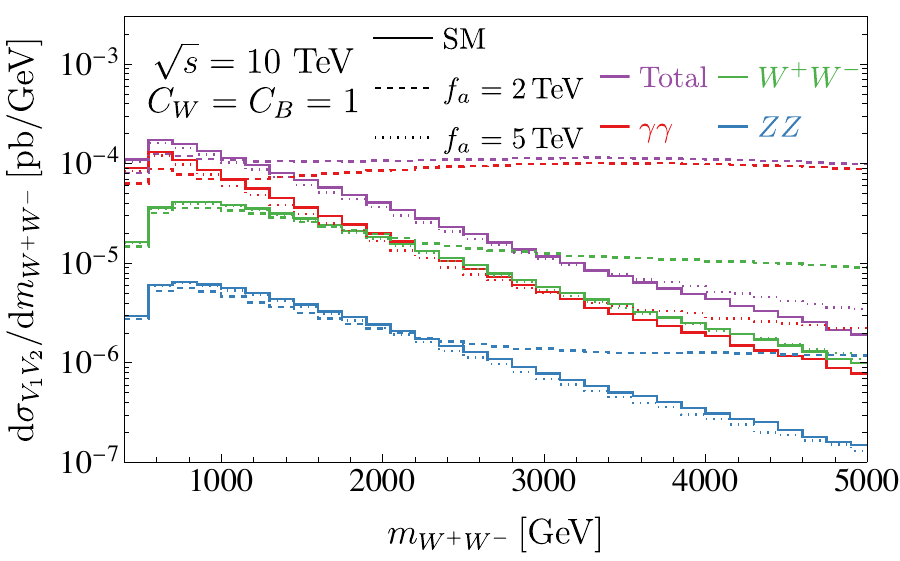}
    \includegraphics[width=0.49\textwidth]{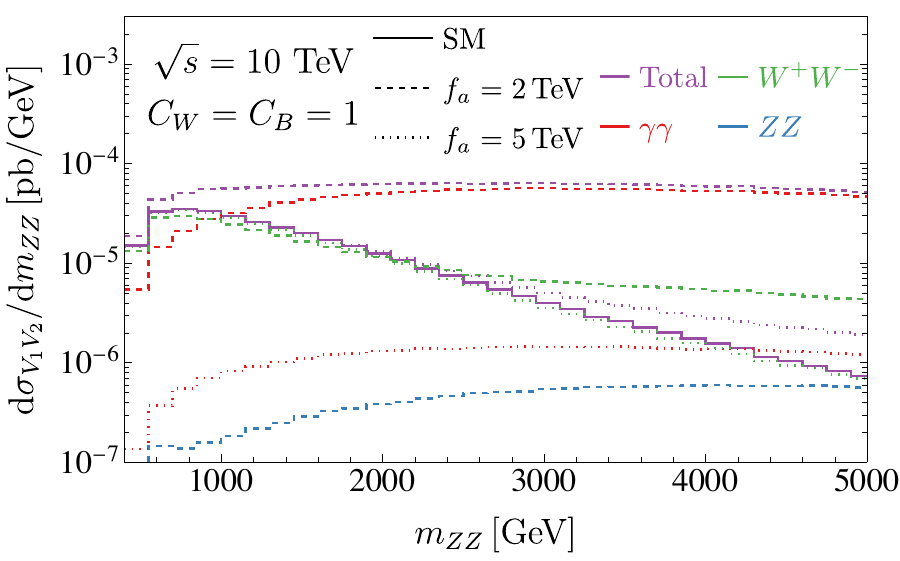}
    \includegraphics[width=0.49\textwidth]{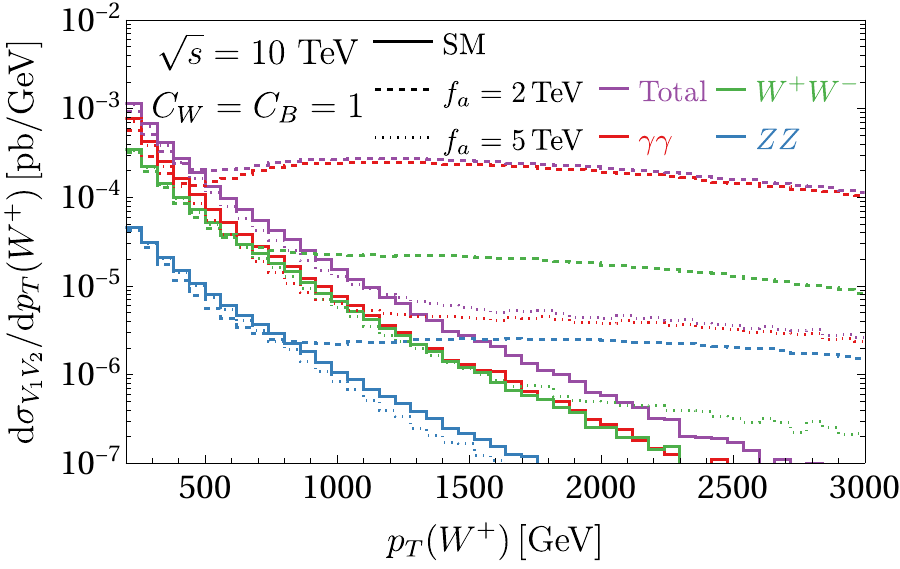}
    \includegraphics[width=0.49\textwidth]{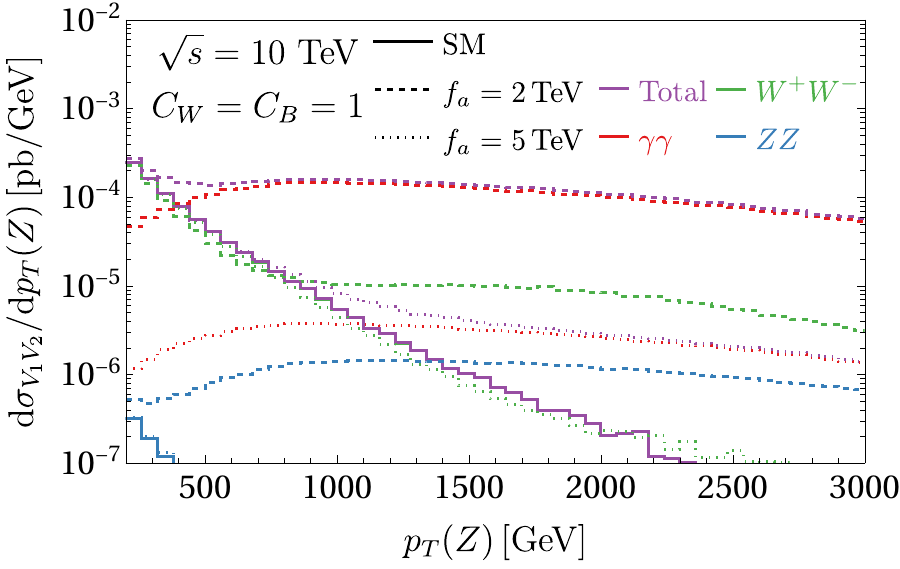}    
    \caption{Distributions of  invariant mass $m_{V_1'V_2'}$ (upper) and transverse momentum $p_T(V)$ (lower) in the $W^+W^-$ and $ZZ$ pair production through VBS at a $\sqrt{s} = 10$ TeV muon collider in the SM and ALP model ($C_B = C_W = 1$). Solid curves show SM predictions, while dashed (dotted) curves represent the ALP model with $f_a = 2$ (5) TeV. Red, green, blue, and violet curves correspond to $\gamma\gamma$, $W^+W^-$, $ZZ$, and their sum, respectively. For the EFT to be valid, the dashed (dotted) curves should be interpolated only up to $\sim 2$ (5) TeV.}
    \label{fig:distVBS}
\end{figure}

Following Eq.~\eqref{eq:sigma_VBS}, the total cross section can be obtained with a convolution of the partonic cross sections ${\hat \sigma}_{V_1V_2\to V_1'V_2'}$ with the corresponding EW PDFs. Similarly as before with $C_W = C_B = 1$, $f_a = 2(5)$ TeV, we present the distributions of invariant mass $m_{V_1'V_2'}$ and transverse momentum $p_T(V')$ for final-state vector boson at a 10 TeV muon collider in Figure~\ref{fig:distVBS}. The numerical simulation is performed with the EVA option~\cite{Ruiz:2021tdt} in \textsc{MadGraph5\_aMC@NLO}~\cite{Alwall:2014hca,Frederix:2018nkq} with a factorization scale $Q=\sqrt{\hat s}/2$. Following our previous work~\cite{Han:2020uid,Han:2021kes}, a baseline kinematic cut 
$|\cos\theta^{V'}_{\rm cm}|<{\rm min}(1-M_W^2/{\hat s},~0.99)$ is imposed to ensure the validity of the EWA and to remove the collinear divergence.
The detector coverage is chosen as $|\eta_{V'}|<2.44$ where $V'=W^\pm,Z$, which corresponds to a $10^\circ$ angle from the beam axis, to mitigate the impact of beam-induced background (BIB) of the proposed muon collider~\cite{MuonCollider:2022ded}.
In our analysis, the following kinematic cuts are applied to enhance the sensitivity to the ALP signal:
\begin{eqnarray}
    p_T(V'_{1,2}) > 150~\text{GeV}, \quad m_{V_1' V_2'} > 500~\text{GeV}, \quad |\eta_{V'_{1,2}}| < 2.44.
\end{eqnarray}

As shown in Figure~\ref{fig:distVBS}, the BSM contribution exhibits a negative interference with the SM diagrams in the low $p_T(V)$ region, leading to a total cross section smaller than the SM prediction. In contrast, at high $p_T(V)$, the BSM cross section exceeds the SM value. Based on this behavior, we divide the analysis into three kinematic regions:
\begin{itemize}
    \item \textbf{Near-threshold region:} $p_T(V'_{1,2}) \in [150,\,300]~\text{GeV}$
    \item \textbf{Intermediate region:} $p_T(V'_{1,2}) \in [300,\,600]~\text{GeV}$
    \item \textbf{High-$p_T$ tail:} $p_T(V'_{1,2}) > 600~\text{GeV}$
\end{itemize}
The total signal significance is then obtained by combining the contributions from each region:
\begin{eqnarray}
    \mathcal{S}_{\text{VBS}} = \sqrt{\mathcal{S}_\textrm{Near-threshold}^2 + \mathcal{S}_\textrm{Intermediate}^2 + \mathcal{S}_{\textrm{High-}p_T}^2}.
\end{eqnarray}
Furthermore, for neutral final states such as $ZZ$, $WW$, and $\gamma Z$ production, we apply an additional $m_{V_1' V_2'} < 0.8 \sqrt{s}$ cut to suppress SM background from $\mu^+ \mu^-$ annihilation.

\begin{figure}[!tb]
    \centering
    \includegraphics[width=0.32\textwidth]{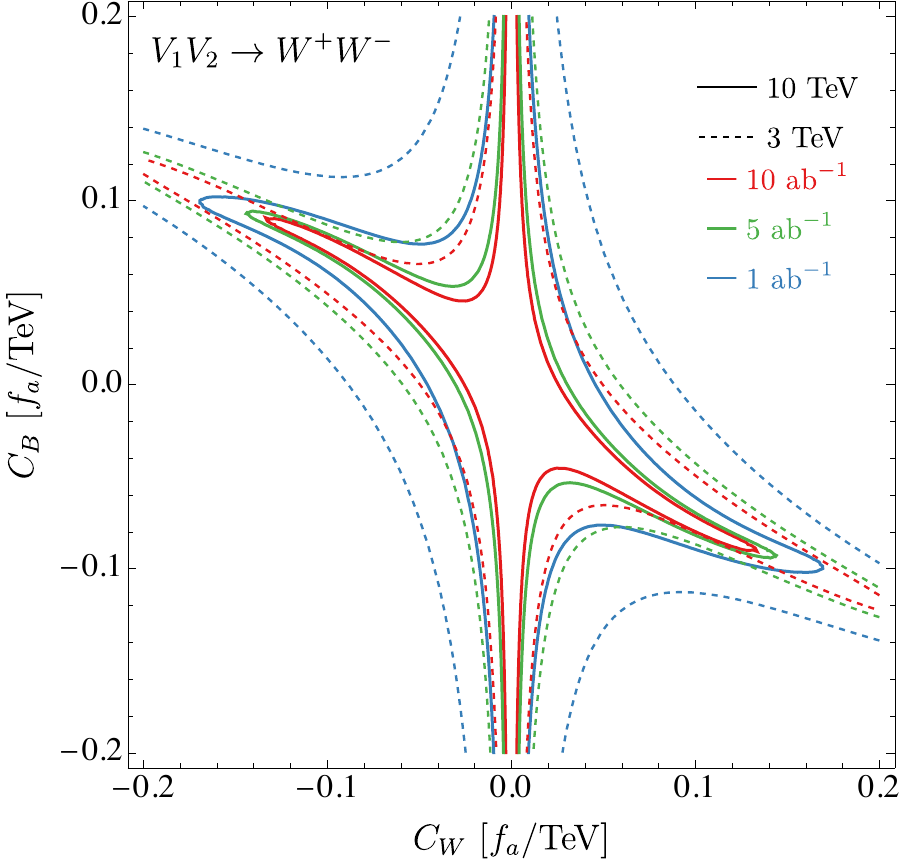}
    \includegraphics[width=0.32\textwidth]{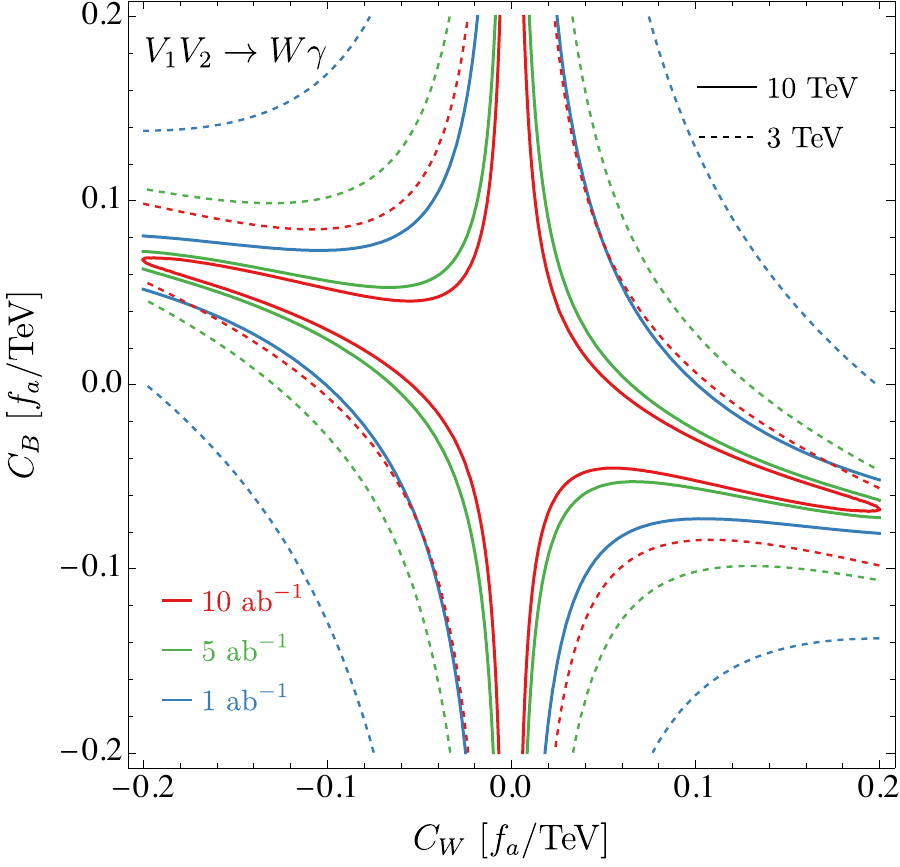}
    \includegraphics[width=0.32\textwidth]{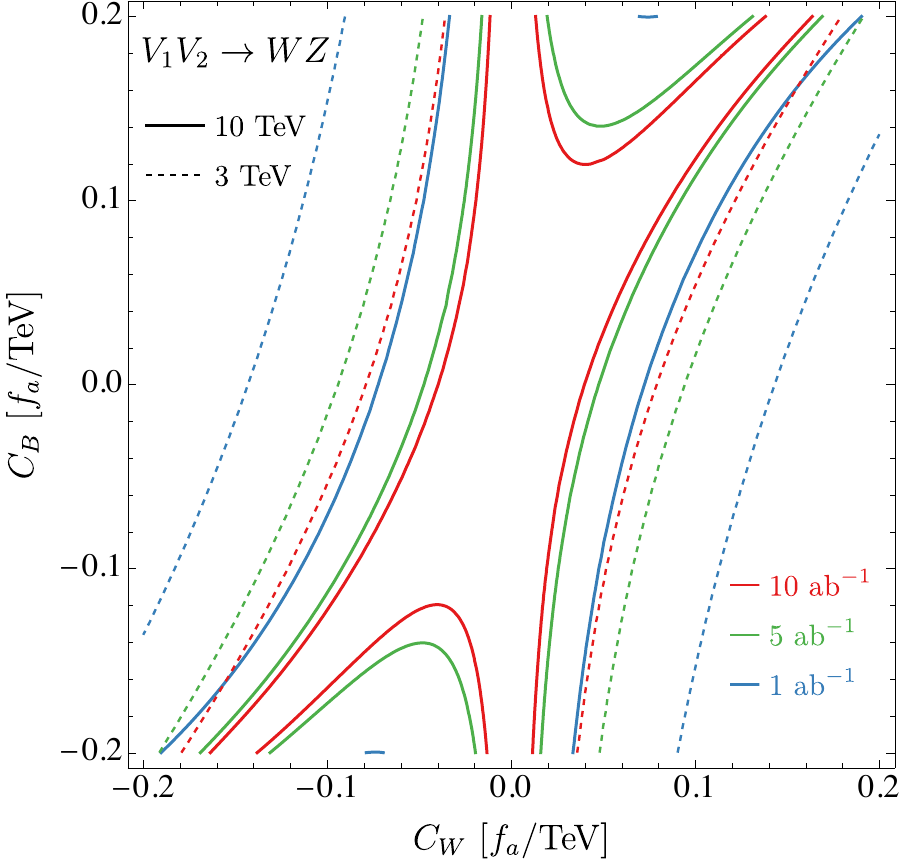}
    \includegraphics[width=0.32\textwidth]{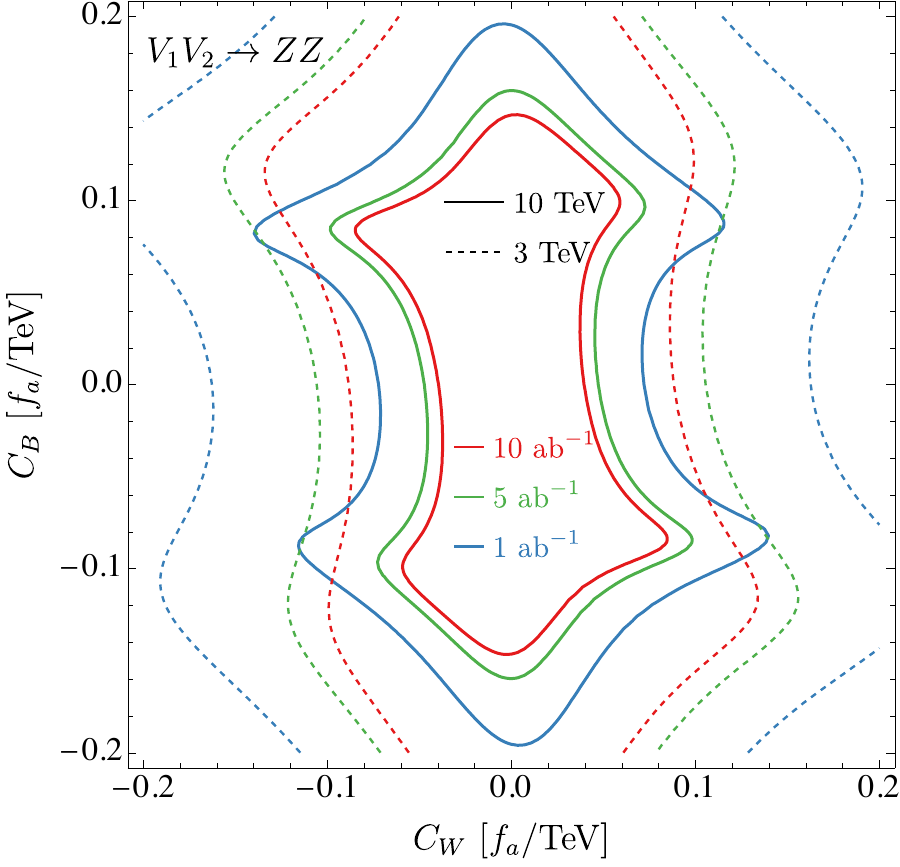}
    \includegraphics[width=0.32\textwidth]{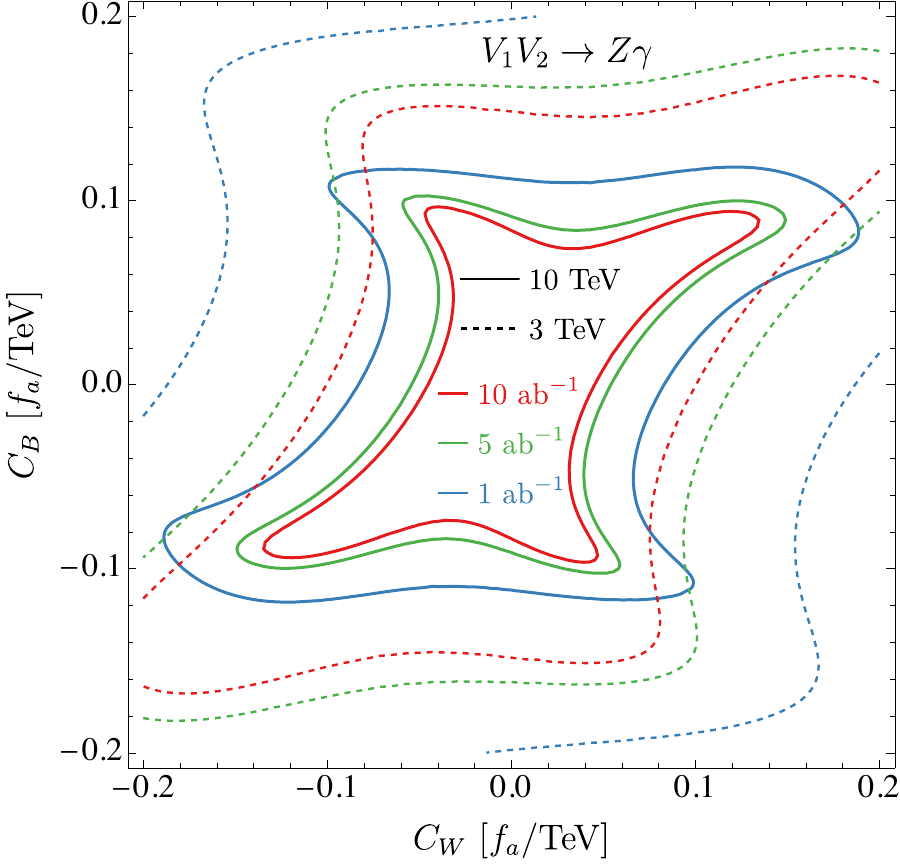}
    \includegraphics[width=0.32\textwidth]{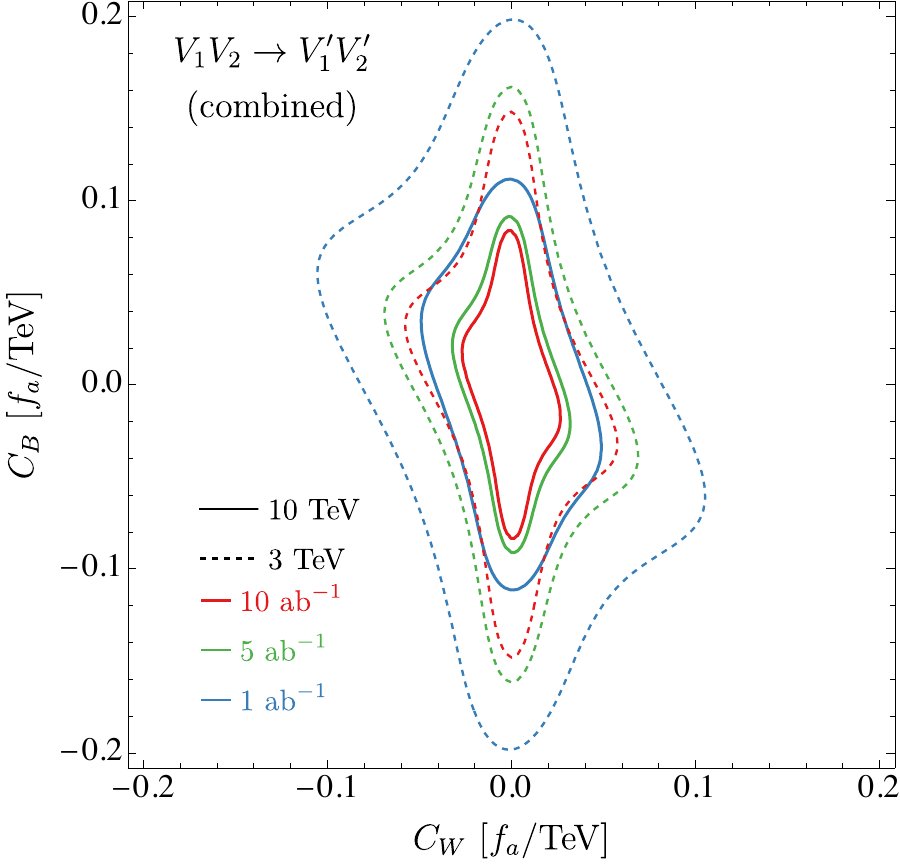}
    \caption{Constraints on $C_W$ and $C_B$ at future muon colliders at 95\% CL. Solid contours represent the results for a 10 TeV collider, while dashed contours correspond to a 3 TeV collider. Blue, green, and red curves indicate integrated luminosities of 1 ab$^{-1}$, 5 ab$^{-1}$, and 10 ab$^{-1}$, respectively.}
    \label{fig:VBS_contours}
\end{figure}

By relaxing $C_W = C_B$ and including all possible EW VBS $V_1V_2 \to V_1' V_2'$ processes, we present the 95\% CL constraints on $C_W$ and $C_B$ from individual processes at future multi-TeV muon colliders in Figure~\ref{fig:VBS_contours}.
The dashed contours represent results for the 3 TeV muon collider, while the solid contours correspond to the 10 TeV muon collider. The blue, green, and red curves indicate scenarios with integrated luminosities of 1~ab$^{-1}$, 5~ab$^{-1}$, and 10~ab$^{-1}$, respectively. 
By translating the $(C_W, C_B)$ parameters into the couplings $g_{aVV}$ through Eq.~(\ref{eq:couplings}), we obtain the corresponding upper limits in Table~\ref{tab:gconstraints_VBS}.
Together with the light-by-light scattering, we obtain the combined bounds also listed in the right columns in Table~\ref{tab:gconstraints_VBS}.
As shown, the light-by-light scattering provides the strongest constraints on $g_{a\gamma\gamma}$ and significantly improves those on $g_{a\gamma Z}$ from the EW VBS processes. 
On the other hand, EW VBS remains the dominant channel for constraining $g_{aZZ}$ and $g_{aWW}$, with only a modest additional contribution from light-by-light scattering.  
Although EW VBS has limited sensitivity to $g_{a\gamma\gamma}$, it provides an effective means to constrain $g_{aZZ}$, $g_{a\gamma Z}$, and $g_{aWW}$.

\begin{table}[htbp]
\centering
\resizebox{\textwidth}{!}{
\renewcommand{\arraystretch}{1.2}
\begin{tabular}{c|ccc|ccc}
\hline\hline
& \multicolumn{3}{c|}{EW VBS} & \multicolumn{3}{c}{Combined with light-by-light} \\
\hline
Luminosity [ab$^{-1}$] & 1 & 5  & 10  & 1  & 5  & 10  \\
\hline
&\multicolumn{6}{c}{3 TeV muon collider} \\
\hline
$|g_{a\gamma\gamma}|^{\rm max}$ [TeV$^{-1}$] & $6.20 \times 10^{-1}$ & $5.04 \times 10^{-1}$ & $4.61\times 10^{-1}$ & $3.82 \times 10^{-1}$ & $3.01 \times 10^{-1}$ & $2.73\times 10^{-1}$ \\
$|g_{a\gamma Z}|^{\rm max}$ [TeV$^{-1}$] & $6.77 \times 10^{-1}$ & $5.46 \times 10^{-1}$ & $4.99\times 10^{-1}$ & $5.90 \times 10^{-1}$ & $3.89 \times 10^{-1}$ & $3.27\times 10^{-1}$ \\
$|g_{aZZ}|^{\rm max}$ [TeV$^{-1}$]& $2.80 \times 10^{-1}$ & $1.84 \times 10^{-1}$ & $1.54 \times 10^{-1}$ & $2.80 \times 10^{-1}$ & $1.84 \times 10^{-1}$ & $1.54 \times 10^{-1}$ \\
$|g_{aWW}|^{\rm max}$ [TeV$^{-1}$]& $4.21 \times 10^{-1}$ & $2.76 \times 10^{-1}$ & $2.31 \times 10^{-1}$ & $4.21 \times 10^{-1}$ & $2.76 \times 10^{-1}$ & $2.31 \times 10^{-1}$ \\
\hline
&\multicolumn{6}{c}{10 TeV muon collider} \\
\hline
$|g_{a\gamma\gamma}|^{\rm max}$ [TeV$^{-1}$] & $3.48 \times 10^{-1}$ & $2.84 \times 10^{-1}$ & $2.60\times 10^{-1}$& $1.92 \times 10^{-1}$ & $1.49 \times 10^{-1}$ & $1.35\times 10^{-1}$ \\
$|g_{a\gamma Z}|^{\rm max}$ [TeV$^{-1}$] & $3.88 \times 10^{-1}$ & $3.12 \times 10^{-1}$ & $2.85\times 10^{-1}$ & $3.13 \times 10^{-1}$ & $2.10 \times 10^{-1}$ & $1.78 \times 10^{-1}$ \\
$|g_{aZZ}|^{\rm max}$ [TeV$^{-1}$] & $1.30 \times 10^{-1}$ & $9.14 \times 10^{-2}$ & $8.09 \times 10^{-2}$ & $1.30 \times 10^{-1}$ & $8.61 \times 10^{-2}$ & $7.23 \times 10^{-2}$ \\
$|g_{aWW}|^{\rm max}$ [TeV$^{-1}$]& $1.96 \times 10^{-1}$ & $1.28 \times 10^{-1}$ & $1.07 \times 10^{-1}$ & $1.96 \times 10^{-1}$ & $1.28 \times 10^{-1}$ & $1.07 \times 10^{-1}$ \\
\hline\hline
\end{tabular}
}
\caption{The upper limits on $|g_{aVV}|^{\rm max}$ from the EW VBS alone (left) and combined with the light-by-light scattering (right) at 3 TeV and 10 TeV muon colliders.}\label{tab:gconstraints_VBS}
\end{table}

\section{Combination and comparison}
\label{sec:comb2comp}
In this section, we perform a combined analysis of the constraints on ALP interactions explored in Sec.~\ref{sec:monoV} and Sec.~\ref{sec:VBS}. Following this, we will compare our results with the existing bounds in the literature.

\subsection{Combined analysis for each collider}
\label{sec:combine}

Similarly as the combination procedure in the mono-$Z$ leptonic and hadronic channels in Sec.~\ref{sec:monoZ}, we take the combined significance measure as
\begin{equation}
\mathcal{S}=\sqrt{\mathcal{S}_{\gamma}^2+\mathcal{S}_{Z}^2+\mathcal{S}_{\rm VBS}^2}  
\end{equation}
where $\mathcal{S}_{\gamma(Z)}$ correspond to the mono-photon($Z$) significance in Sec.~\ref{sec:monoV}, and $\mathcal{S}_{\rm VBS}$ refers to the VBS one in Sec.~\ref{sec:VBS}, including both light-by-light and EW VBS.

\begin{figure}[!htb]
    \centering
    \includegraphics[width=0.45\textwidth]{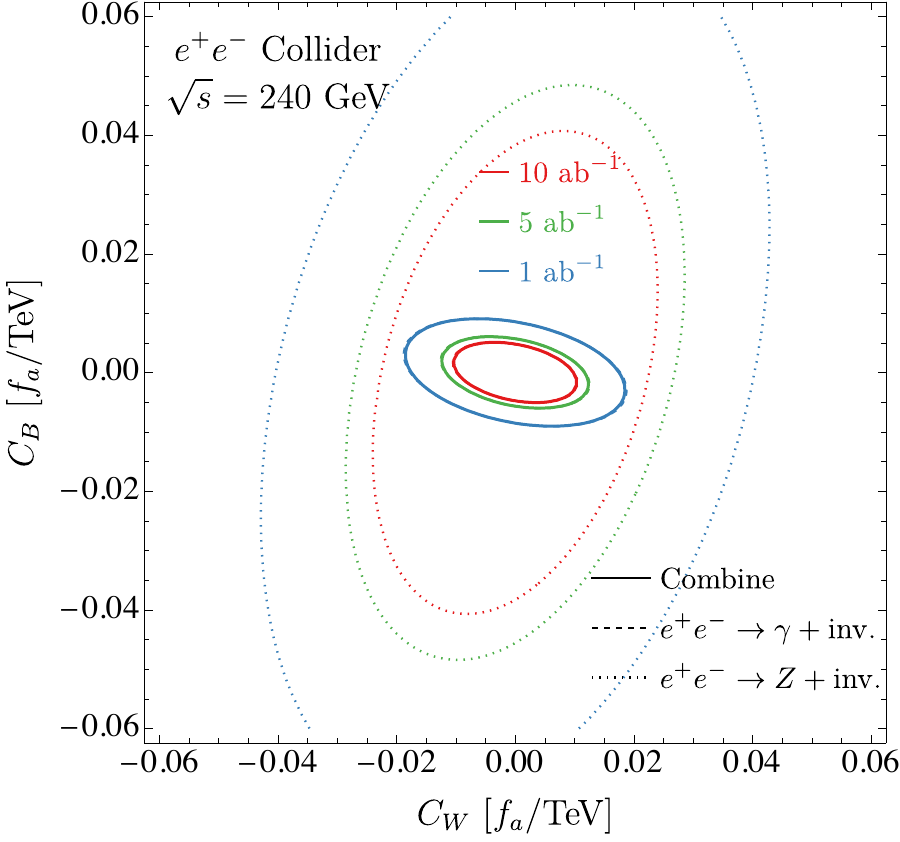}
    \includegraphics[width=0.45\textwidth]{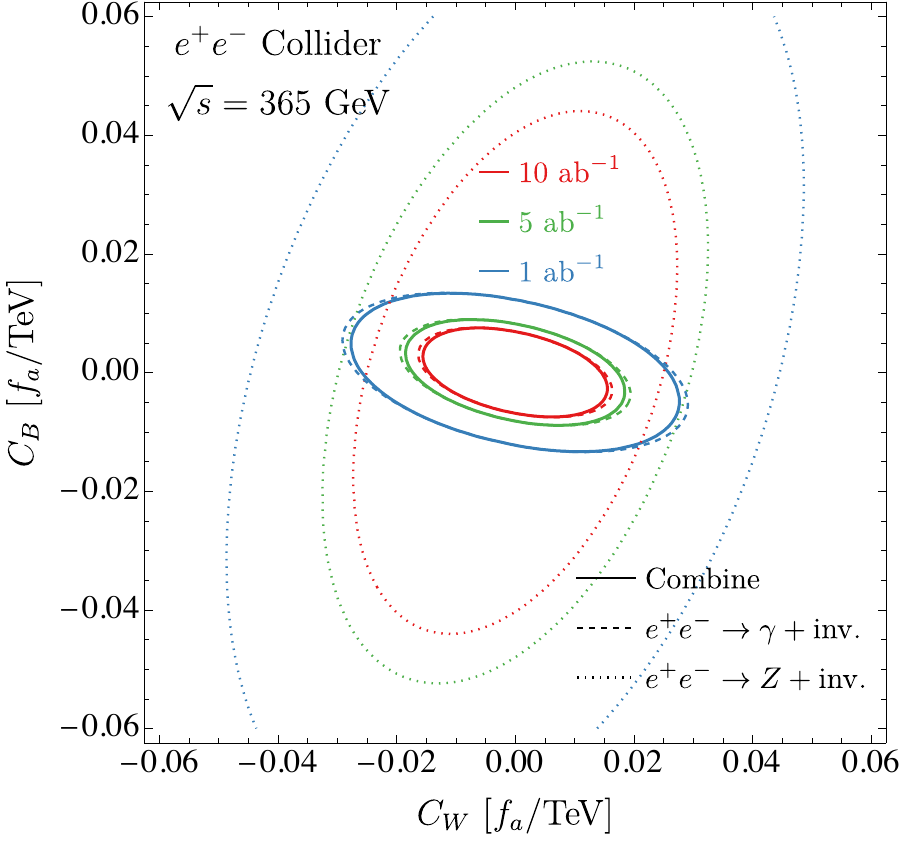}
    \includegraphics[width=0.45\textwidth]{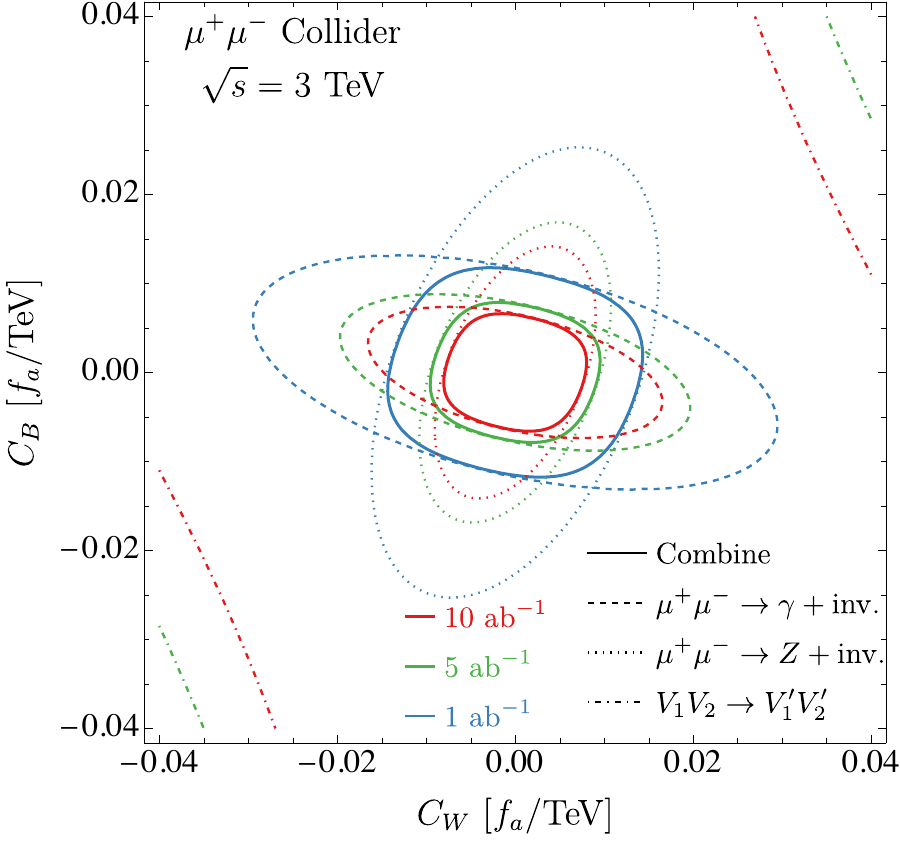}
    \includegraphics[width=0.45\textwidth]{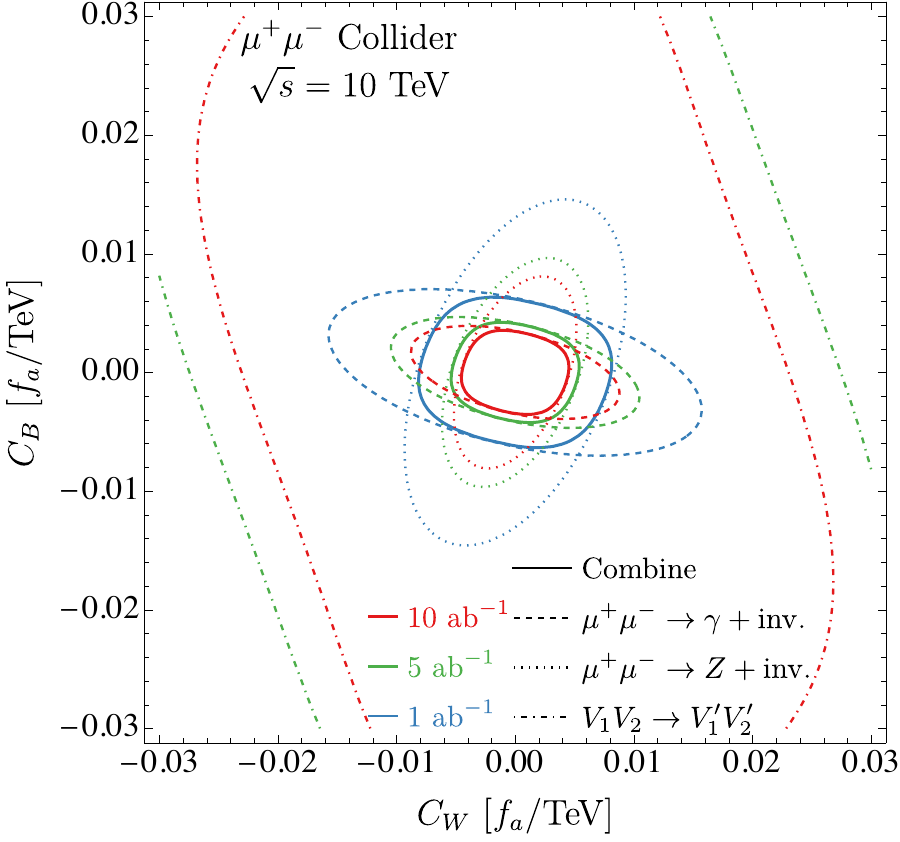}
    \caption{Constraints in the long-lived ALP limit on $C_W$ and $C_B$ from $\ell^+\ell^- \to \gamma/Z + \text{inv.}$ and $V_1V_2 \to V_1'V_2'$ processes at high-energy lepton colliders, evaluated at varying machine energies and luminosities at 95\% CL. Dashed contours correspond to mono-photon production, dotted contours represent mono-$Z$ production, dot-dashed contours depict the $V_1V_2 \to V_1'V_2'$ process, and solid contours show the combined constraints from all these processes. Blue, green, and red curves indicate integrated luminosities of 1 ab$^{-1}$, 5 ab$^{-1}$, and 10 ab$^{-1}$, respectively.}
    \label{fig:contours_combine}
\end{figure}

The final combined constraints on $(C_W, C_B)$, compared with those from individual channels, are presented in Figure~\ref{fig:contours_combine}. The mono-photon and mono-$Z$ bounds are shown as dashed and dotted contours, respectively, while EW VBS is represented by dot-dashed contours, and the final combined result by solid contours.
Due to their relatively low collision energies, electron colliders are limited to light-by-light scattering in VBS, while multi-TeV muon colliders also accommodate EW VBS processes.  
At future $e^+e^-$ colliders such as FCC-ee~\cite{FCC:2018evy,Bernardi:2022hny} and CEPC~\cite{CEPCStudyGroup:2018rmc,CEPCStudyGroup:2018ghi,An:2018dwb,CEPCAcceleratorStudyGroup:2019myu,CEPCPhysicsStudyGroup:2022uwl,CEPCStudyGroup:2023quu}, mono-photon production dominates in sensitivity, while mono-$Z$ provides complementary constraints on $g_{a\gamma Z}$, $g_{aZZ}$, and $g_{aWW}$ at multi-TeV muon colliders ($3~{\rm TeV}$ and $10~{\rm TeV}$).  
Nevertheless, mono-photon remains the most sensitive channel for constraining $g_{a\gamma\gamma}$, even at high-energy muon colliders.
On the other hand, constraints from non-resonant VBS processes are generally weaker than those from mono-$V$ production but remain robust regardless of the ALP's lifetime.

The constraints in the $(C_W,\,C_B)$ space can be translated into limits on the $g_{aVV}$ couplings using Eq.~\eqref{eq:couplings}. We summarize the combined constraints in the ALP long-lived limit ($L_D \gg d$) in Table~\ref{tab:gconstraints_combine}, which applies to low ALP masses.  
For higher ALP masses, e.g., $m_a \sim 1~\GeV$, the ALP becomes short-lived, rendering the mono-$V$ bounds invalid. In this case, the VBS constraints take over, with the strongest limits on $g_{aVV}$ coinciding as those in Table~\ref{tab:gconstraints_VBS}.  

\begin{table}[htbp]
\centering
\resizebox{\textwidth}{!}{
\renewcommand{\arraystretch}{1.2}
\begin{tabular}{l|ccc|ccc}
\hline\hline
$e^+e^-$ Collider & \multicolumn{3}{c|}{$\sqrt{s}=240$ GeV} & \multicolumn{3}{c}{$\sqrt{s}=365$ GeV} \\
\hline
Luminosity [ab$^{-1}$ ] & 1 & 5  & 10  & 1 & 5  & 10  \\
\hline
$|g_{a\gamma\gamma}|^{\rm max}$ [TeV$^{-1}$] & $2.76 \times 10^{-2}$ & $1.84 \times 10^{-2}$ & $1.54\times 10^{-2}$ & $3.93 \times 10^{-2}$ & $2.63 \times 10^{-2}$ & $2.21\times 10^{-2}$ \\
$|g_{a\gamma Z}|^{\rm max}$ [TeV$^{-1}$] & $7.70 \times 10^{-2}$ & $5.14 \times 10^{-2}$ & $4.32\times 10^{-2}$ & $1.16 \times 10^{-1}$ & $7.77 \times 10^{-2}$ & $6.53\times 10^{-2}$ \\
$|g_{aZZ}|^{\rm max}$ [TeV$^{-1}$]& $5.55 \times 10^{-2}$ & $3.70 \times 10^{-2}$ & $3.11 \times 10^{-2}$ & $8.26 \times 10^{-2}$ & $5.52 \times 10^{-2}$ & $4.64 \times 10^{-2}$ \\
$|g_{aWW}|^{\rm max}$ [TeV$^{-1}$]& $7.41 \times 10^{-2}$ & $4.94 \times 10^{-2}$ & $4.15 \times 10^{-2}$ & $1.11 \times 10^{-1}$ & $7.40 \times 10^{-2}$ & $6.22 \times 10^{-2}$ \\
\hline
$\mu^+\mu^-$ Collider & \multicolumn{3}{c|}{$\sqrt{s}=3$ TeV} & \multicolumn{3}{c}{$\sqrt{s}=10$ TeV} \\
\hline
Luminosity [ab$^{-1}$] & 1  & 5  & 10  & 1  & 5 & 10  \\
\hline
$|g_{a\gamma\gamma}|^{\rm max}$ [TeV$^{-1}$] & $3.71 \times 10^{-2}$ & $2.48 \times 10^{-2}$ & $2.08 \times 10^{-2}$ & $1.98 \times 10^{-2}$ & $1.32 \times 10^{-2}$ & $1.11\times 10^{-2}$ \\
$|g_{a\gamma Z}|^{\rm max}$ [TeV$^{-1}$] & $6.32 \times 10^{-2}$ & $4.22 \times 10^{-2}$ & $3.54\times 10^{-2}$ & $3.57 \times 10^{-2}$ & $2.37 \times 10^{-2}$ & $1.99 \times 10^{-2}$ \\
$|g_{aZZ}|^{\rm max}$ [TeV$^{-1}$] & $4.73 \times 10^{-2}$ & $3.16 \times 10^{-2}$ & $2.65 \times 10^{-2}$ & $2.65 \times 10^{-2}$ & $1.76 \times 10^{-2}$ & $1.48 \times 10^{-2}$ \\
$|g_{aWW}|^{\rm max}$ [TeV$^{-1}$]& $5.73 \times 10^{-2}$ & $3.82 \times 10^{-2}$ & $3.21 \times 10^{-2}$ & $3.26 \times 10^{-2}$ & $2.16 \times 10^{-2}$ & $1.81 \times 10^{-2}$ \\
\hline\hline
\end{tabular}
}
\caption{The upper limits $|g_{aVV}|^{\rm max}$ from the combination of the mono-$V$ production and the non-resonant VBS processes at different future lepton colliders. }\label{tab:gconstraints_combine}
\end{table}

\subsection{Comparison with existing bounds}
\label{sec:compare}
This subsection compares our projected bounds at future lepton colliders with existing upper limits on the $g_{aVV}$ couplings, assuming nominal integrated luminosities.
We consider the most recent FCC-ee splitting scheme for the Tera-$Z$ phase~\cite{janot_2024_yr3v6-dgh16}.
For $e^+e^-$ colliders operating at 240 GeV and 365 GeV, we adopt integrated luminosities of 5 ab$^{-1}$ and 1 ab$^{-1}$, respectively \cite{CEPCStudyGroup:2018ghi, CEPCStudyGroup:2023quu, FCC:2018evy, Bernardi:2022hny}. For high-energy muon colliders at 3 TeV and 10 TeV, we assume integrated luminosities of 1 ab$^{-1}$ and 10 ab$^{-1}$, respectively \cite{MuonCollider:2022xlm, Aime:2022flm, Black:2022cth, Accettura:2023ked, Delahaye:2019omf, Bartosik:2020xwr, Schulte:2021hgo, Long:2020wfp, MuonCollider:2022nsa, MuonCollider:2022ded, MuonCollider:2022glg,InternationalMuonCollider:2025sys}.

\begin{figure}[!htb]
    \centering
    \includegraphics[width=0.45\textwidth]{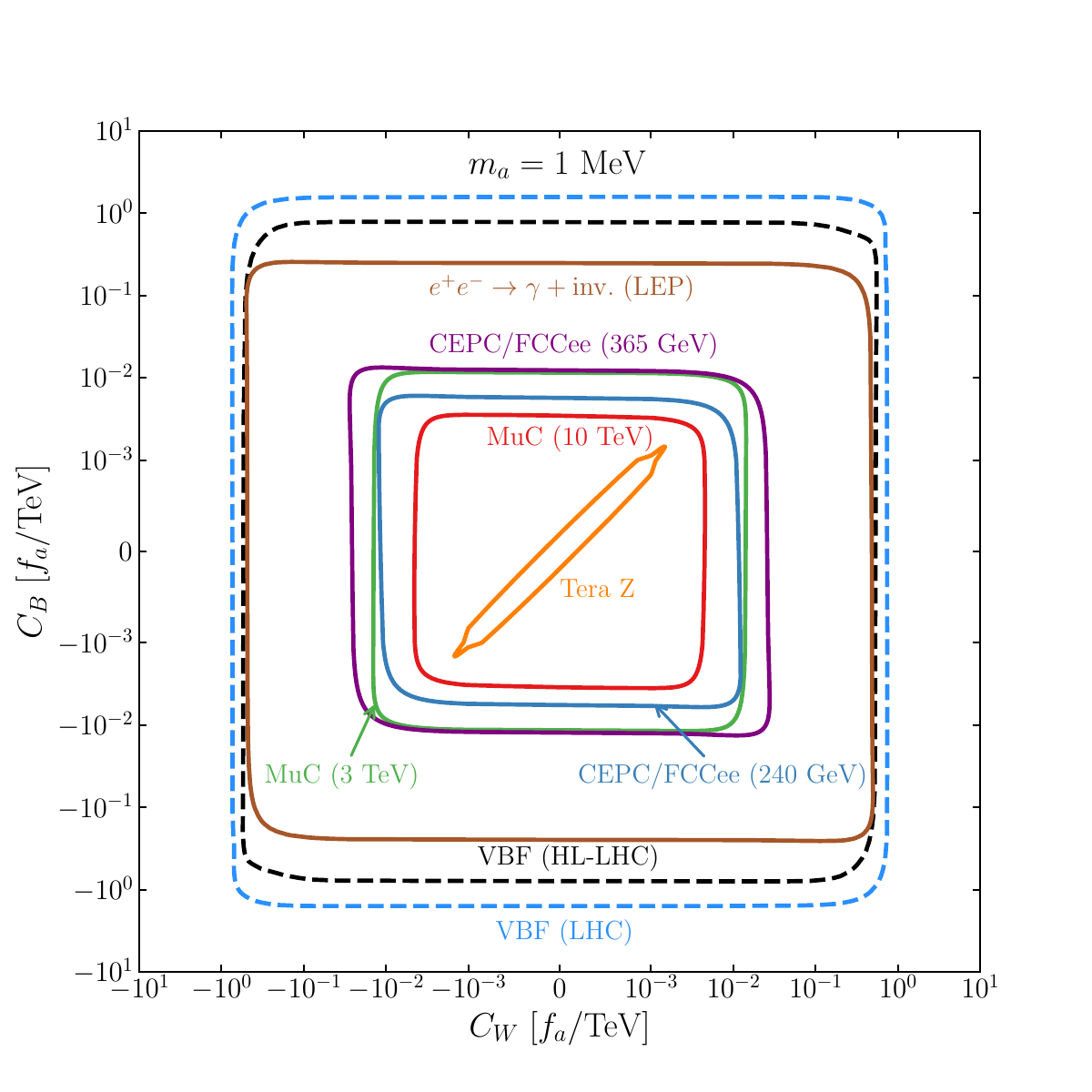}
    \includegraphics[width=0.45\textwidth]{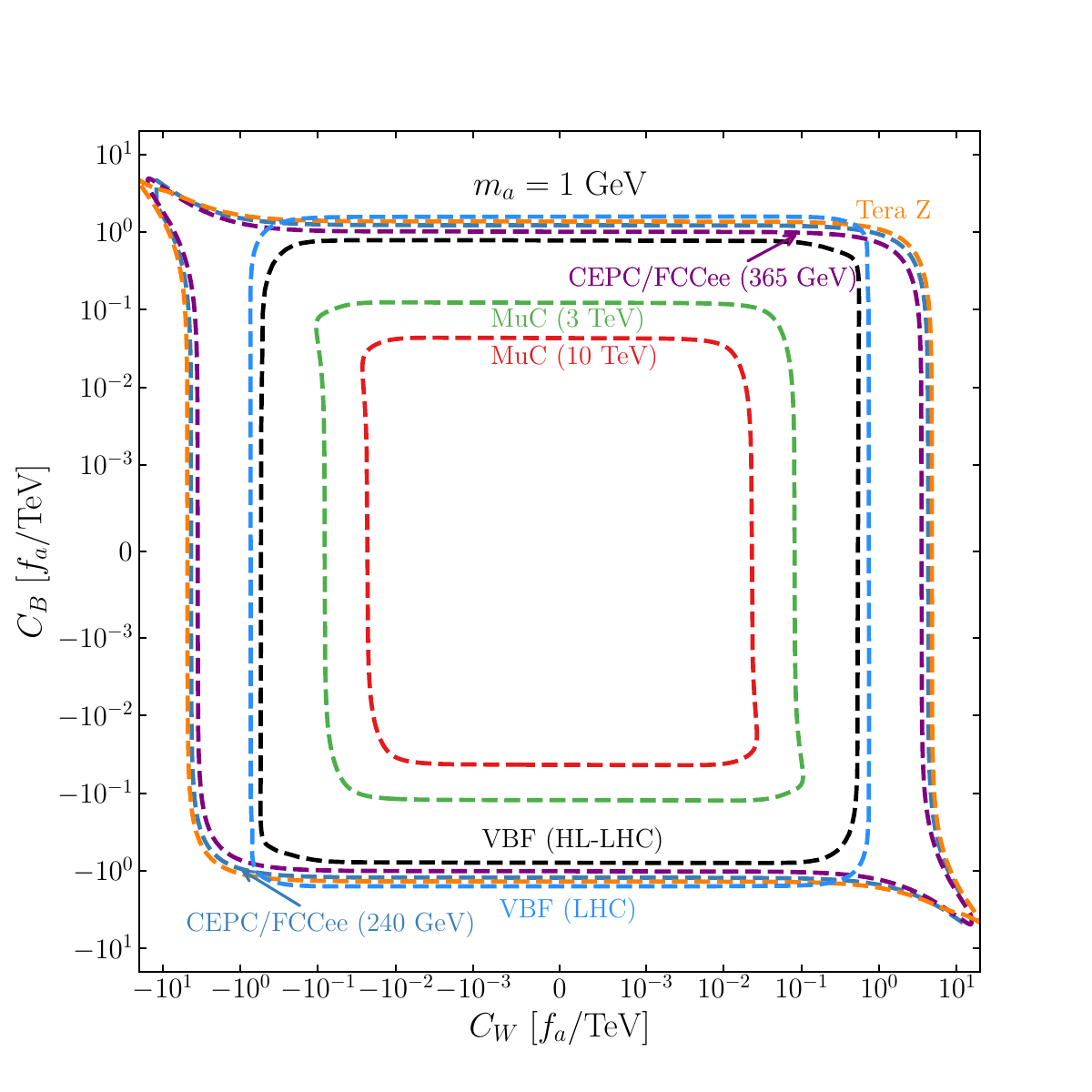}
    \caption{Constraints at 95\% CL for $m_a=1$ MeV (left) and $m_a=1$ GeV (right) at proposed $e^+e^-$ colliders (labed as ``CEPC/FCC-ee'') operating at 240 GeV (luminosity 5 ab$^{-1}$) and 365 GeV (luminosity 1 ab$^{-1}$), as well as muon colliders (labeled as ``MuC'') at 3 TeV (luminosity 1 ab$^{-1}$) and 10 TeV (luminosity 10 ab$^{-1}$). Also shown are constraints from mono-photon production at LEP (189 GeV)~\cite{OPAL:2000puu} and non-resonant VBS constraints from CMS Run 2~\cite{CMS:2020fqz,CMS:2020gfh,CMS:2020ypo,CMS:2021gme} and the HL-LHC with 3 ab$^{-1}$ luminosity~\cite{Bonilla:2022pxu}, labeled as VBS (LHC) and VBS (HL-LHC), respectively.}\label{fig:CompareContours}
\end{figure}

In Figure~\ref{fig:CompareContours}, we compare our final combined constraints in the ($C_W, C_B$) plane against the existing limits from mono-photon production from LEP at 189 GeV~\cite{OPAL:2000puu} and non-resonant VBS constraints from CMS Run 2~\cite{CMS:2020fqz,CMS:2020gfh,CMS:2020ypo,CMS:2021gme} and the 3 ab$^{-1}$ HL-LHC~\cite{Bonilla:2022pxu}. As discussed above, here we take two benchmark ALP masses $m_a=1$~MeV and 1 GeV to represent the long- and short-lived scenarios, where mono-$V$ production and non-resonant VBS processes dominate the constraints, respectively. The solid contours indicate constraints from mono-$V$ production or a combination of mono-$V$ and VBS, while the dashed curves represent constraints from non-resonant VBS alone.

As shown in Figure~\ref{fig:CompareContours}, future lepton colliders significantly improve the constraints on $a$-$V$-$V$ interactions, especially in the ALP long-lived regime where mono-$V$ production dominates.
In the ALP long-lived limit, the most stringent constraint arises from the mono-photon channel at the Tera-$Z$ phase of future $e^+e^-$ colliders, benefiting from the $> 100$ ab$^{-1}$ integrated luminosity, the resonant enhancement to the ALP signal from the on-shell $Z$ boson, and the small SM background. At higher-energy lepton colliders, the improvement over LEP’s mono-photon limits is mainly driven by the increased luminosity. As a result, the 10 TeV muon collider and the 240 GeV CEPC/FCC-ee options yield the stronger constraints due to their high luminosities. In comparison, the 3 TeV muon collider offers sensitivities similar to those of the 365 GeV CEPC/FCC-ee, assuming equal integrated luminosities.
Outside the long-lived ALP regime, mono-$V$ production is no longer effective, and non-resonant VBS processes take over. Notably, due to the higher partonic luminosities in VBS processes at the 10 TeV muon collider, the constraints from its non-resonant VBS alone are comparable to those from the combined mono-$V$ and VBS constraints at the 365 GeV FCC-ee.

\begin{figure}[!htb]
    \centering
    \includegraphics[width=0.49\textwidth]{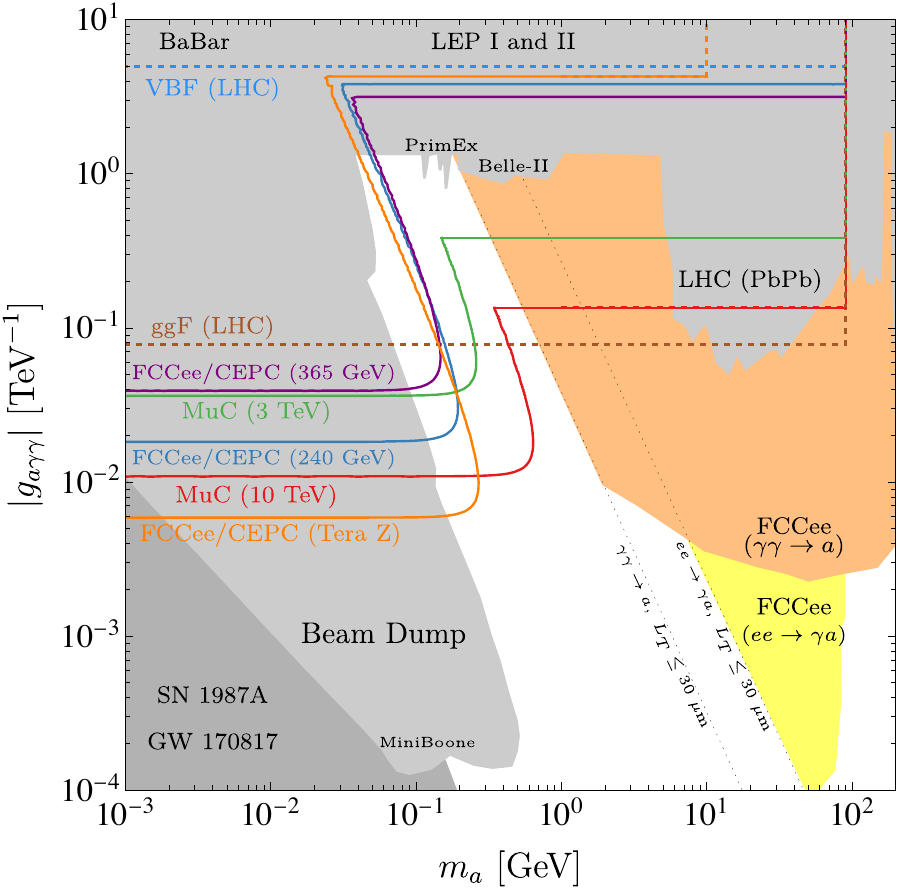}
    \includegraphics[width=0.49\textwidth]{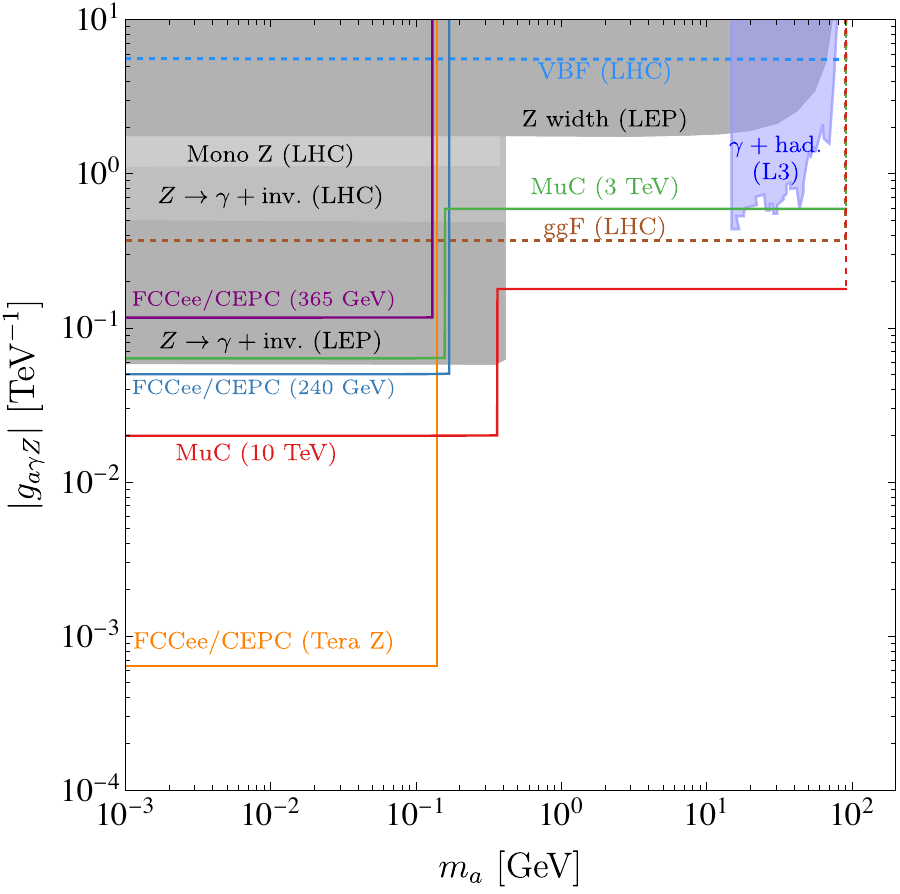}
    \includegraphics[width=0.49\textwidth]{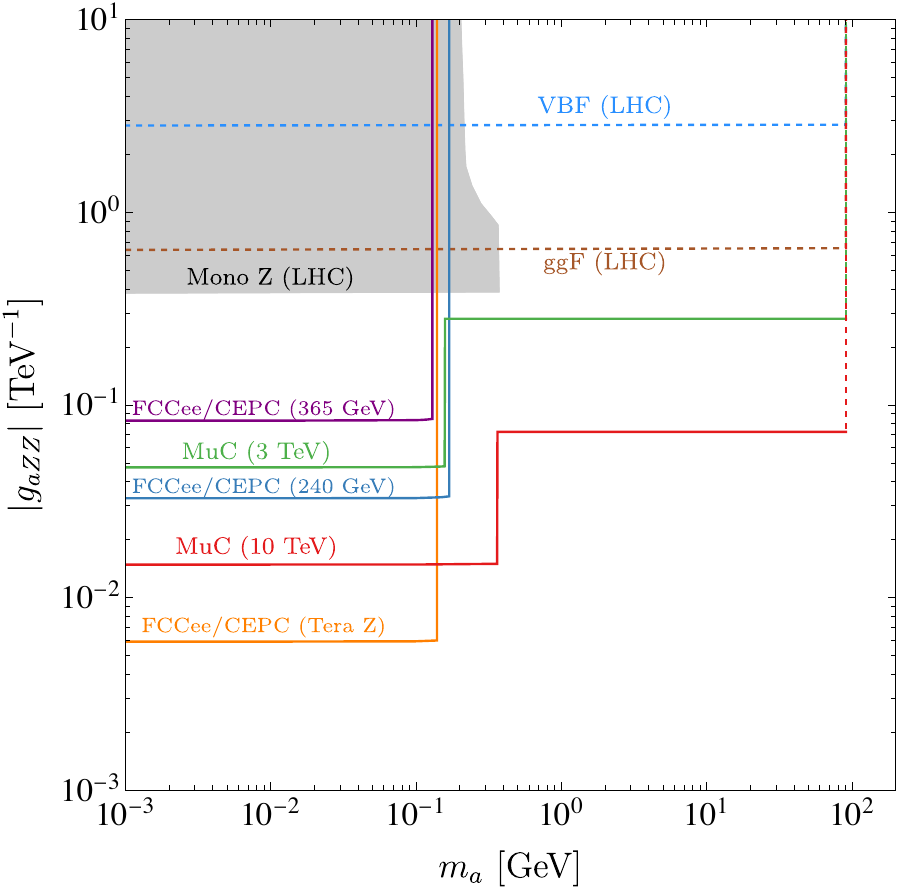}
    \includegraphics[width=0.49\textwidth]{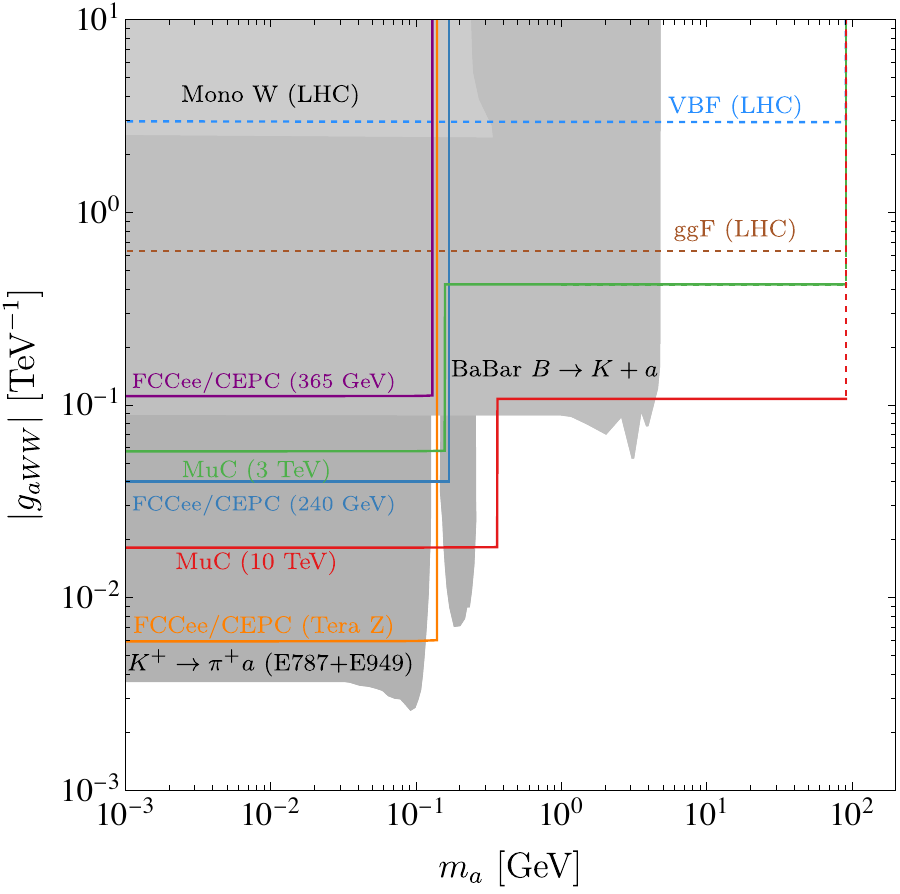}
    \caption{The 95\% CL constraints on the ALP couplings $g_{aVV}$ from the mono-$V$ and VBS scattering in this work in comparison with the existing bounds.}
    \label{fig:gaVV}
\end{figure}
 
With the translation to the $g_{aVV}$ couplings, we present the upper bounds as functions of the ALP mass $m_a$ and compare them with the existing constraints in Figure~\ref{fig:gaVV}. In this analysis, we restrict the non-resonant VBS constraints to $m_a<M_Z$, since at higher ALP masses, on-shell production processes, such as $Z\gamma\to a\to Z\gamma$, are expected to dominate and provide stronger constraints.
In the $\gamma\gamma\to a\to\gamma\gamma$ channel, this take-over occurs at a lower ALP mass whenever the ALP becomes short-lived~\cite{RebelloTeles:2023uig}.

The existing constraints on the $g_{a\gamma\gamma}$ coupling are taken from overviews in Refs.~\cite{Mimasu:2014nea, Jaeckel:2015jla, Brivio:2017ije, Bauer:2017ris, Bauer:2018uxu, dEnterria:2021ljz, Antel:2023hkf}. More specifically, the existing constraints in the MeV-GeV range is from the beam-dump experiments \cite{Riordan:1987aw,Bjorken:1988as, Blumlein:2013cua,Dobrich:2015jyk}.
The low $g_{a\gamma\gamma}$ at a low $m_a$ is constrained by the supernovae explosion energy \cite{Caputo:2021rux,Caputo:2022mah,Fiorillo:2025yzf} and gravitational wave observation \cite{Diamond:2023cto}. While the constraints in the higher $g_{a\gamma\gamma}$ coupling values or the larger $m_a$ values are from LEP~\cite{L3:1994shn,OPAL:2002vhf}, Babar~\cite{BaBar:2010eww}, LHC Pb-Pb~\cite{CMS:2018erd,ATLAS:2020hii}, PrimEx~\cite{PrimEx:2010fvg,Aloni:2019ruo}, Belle II~\cite{Dolan:2017osp,Belle-II:2020jti},
FCC-ee 3 photons~\cite{Bauer:2018uxu}, and on-shell photon-photon collision~\cite{RebelloTeles:2023uig}, respectively. 
In comparison, the strength of our analysis fills the gap between the long-lived particle searches at Beam Dump experiments~\cite{Riordan:1987aw,Bjorken:1988as,Blumlein:2013cua,Dobrich:2015jyk} and the resonance searches, i.e. $\gamma\gamma\to a\to\gamma\gamma$ and $e^+e^-\to \gamma a\to\gamma\gamma\gamma$, at FCC-ee~\cite{RebelloTeles:2023uig}.

The existing constraints on ALP-weak-boson couplings $g_{a\gamma Z}$, $g_{aZZ}$, and $g_{aWW}$ are derived from several sources: the process $e^+e^-\to\gamma +$hadrons at L3 (only for $g_{a\gamma Z}$) \cite{L3:1992kcg}, mono-$Z/W$ production at the LHC \cite{Brivio:2017ije}, and the non-observation of $Z \to \gamma + \rm{inv.}$ decays at both LEP \cite{Craig:2018kne} and the LHC \cite{ATLAS:2020uiq}. 
Additionally, the coupling of ALPs to $W$ boson contributes solely to rare meson decays at the one-loop level, resulting in stringent limits for light ALPs \cite{Izaguirre:2016dfi,BNL-E949:2009dza}. 

Assuming if the ALPs do not have to be stable particles, a more conservative bound from measuring the $Z$ decay width at LEP will replace that from $Z \to \gamma + \text{inv.}$ for $m_a \leq M_Z$ \cite{Brivio:2017ije,Craig:2018kne}.
Non-resonant searches for di-boson production at the LHC via gluon-gluon fusion \cite{Gavela:2019cmq,Carra:2021ycg,CMS:2021xor}\footnote{In these gluon-gluon fusion analyses, the $g_{aGG}$ coupling is assumed to be 1 TeV$^{-1}$.} and EW VBS \cite{Bonilla:2022pxu} also provide additional constraints on all the four $g_{aVV}$ couplings.
In comparison with these existing bounds, mono-$V$ production offers significantly extended reach for low ALP masses. As the ALP mass increases to the GeV range, EW VBS becomes dominant and provides much tighter constraints than non-resonant LHC searches \cite{Bonilla:2022pxu}.

\section{Summary}
\label{sec:summary}

High-energy colliders offer a promising avenue for the search of ALPs. In this study, we investigate the potential of future lepton colliders, including both electron-positron colliders such as the CEPC~\cite{CEPCStudyGroup:2018rmc,CEPCStudyGroup:2018ghi,An:2018dwb,CEPCAcceleratorStudyGroup:2019myu,CEPCPhysicsStudyGroup:2022uwl,CEPCStudyGroup:2023quu} and FCC-ee~\cite{FCC:2018evy,Bernardi:2022hny}, as well as multi-TeV muon colliders~\cite{MuonCollider:2022xlm,Aime:2022flm,Black:2022cth,Accettura:2023ked,Delahaye:2019omf,Bartosik:2020xwr,Schulte:2021hgo,Long:2020wfp,MuonCollider:2022nsa,MuonCollider:2022ded,MuonCollider:2022glg,InternationalMuonCollider:2025sys}, to probe ALP-EW gauge boson interactions. We focus on scenarios involving light ALPs with long-lived characteristics, where the ALPs can escape the detector before decaying. The primary constraints in these scenarios come from mono-$V$ processes.
In addition, we examine non-resonant VBS processes, $V_1 V_2 \to V_1' V_2'$, including both light-by-light scattering and EW VBS. 
The resulting constraints are largely independent on the ALP mass and serve as a complement to resonant searches at higher ALP masses~\cite{Inan:2022rcr,Buttazzo:2018qqp,Bao:2022onq,Han:2022mzp,Yue:2021iiu,RebelloTeles:2023uig}.

In the effective field theory (EFT) framework, the interactions between the ALP and EW bosons are characterized by the four couplings $g_{a\gamma\gamma}$, $g_{a\gamma Z}$, $g_{aZZ}$, and $g_{aWW}$. These can be parameterized using two Wilson coefficients, $(C_W, C_B)$, along with a universal decay constant, $f_a$. The couplings $g_{aVV}$ are then expressed in terms of $C_W$, $C_B$, and $f_a$ in Eq.~(\ref{eq:couplings}). 
Using this formulation, the existing constraints from LEP mono-photon searches~\cite{OPAL:2000puu}, non-resonant VBS at CMS Run 2~\cite{CMS:2020fqz,CMS:2020gfh,CMS:2020ypo,CMS:2021gme} and the 3 ab$^{-1}$ future HL-LHC projections~\cite{Bonilla:2022pxu}, $Z\to a\gamma$ decays~\cite{Brivio:2017ije}, and $\Upsilon\to a\gamma$ searches~\cite{Masso:1995tw,CrystalBall:1990xec} are mapped onto the $(C_W, C_B)$ parameter space, as shown in Figure~\ref{fig:alp_limit}. 
Among these, the LEP mono-photon measurement at 189 GeV provides the most stringent bounds on the Wilson coefficients and the $g_{aVV}$ couplings.  

At high-energy lepton colliders, the ALPs can be produced in association with photons and EW gauge bosons, as well as through vector boson fusion. A heavy ALP is generally short-lived and can decay into SM particles within the detector, leaving a resonance signature.
In contrast, a light ALP with decay length longer than detector size will fly all through the detector and behave as an invisible particle.
As a consequence, the vector boson associated ALP production $(V=\gamma,\,Z)$ gives a mono-$V$ signal at lepton colliders.

In the mono-photon channel, the SM background primarily arises from processes such as $\ell^+\ell^- \to \gamma + \nu\bar{\nu}$. This includes contributions from both $\gamma+Z$ production with $Z \to \nu\bar{\nu}$ and $W^*$ exchange. Additionally,  the $\ell^+\ell^- \to \gamma\ell^+\ell^-$ process through the $\gamma/Z$ exchange contributes where final-state leptons move forward without entering the detectors.
The $\gamma+Z(\nu\bar{\nu})$ background is characterized by a monochromatic photon energy and recoil mass due to its 2-to-2 kinematics, which can be well distinguished at future electron colliders. However, at multi-TeV muon colliders, the $Z$ boson becomes asymptotically massless and behaves similarly to a light ALP, making separation in the $E_\gamma$ spectrum ineffective. Nonetheless, the background itself is suppressed by a factor of $1/s$.
Applying cuts on the photon pseudorapidity ($|\eta_\gamma| < 2.5$) and photon transverse momentum ($p_{T,\gamma}$) effectively reduces backgrounds from $W^*$ and $\gamma/Z$ exchanges. By optimizing these kinematic cuts for each collider setup, we achieve the best sensitivities to the $g_{aVV}$ couplings at future lepton colliders.
The most stringent constraints on ALP couplings in the long-lived limit come from the mono-photon channel at the Tera-$Z$ phase of future electron colliders. This exceptional sensitivity is driven by three key factors: the extremely low SM background near the $Z$ pole, the resonant enhancement of the ALP signal via the on-shell $Z$ boson, and the enormous integrated luminosity of $\sim 100$ ab$^{-1}$. As a result, the Tera-$Z$ program can improve upon the existing LEP mono-photon bounds by over two orders of magnitude. Using the most recent splitting profile of the FCC-ee~\cite{janot_2024_yr3v6-dgh16}, we obtain the 95\% CL upper limits on the $g_{aVV}$ couplings
\begin{eqnarray}
    &&|g_{a\gamma\gamma}| \leq 5.87 \times 10^{-3}~{\rm TeV}^{-1},~~~|g_{a\gamma Z}| \leq 6.39 \times 10^{-4}~{\rm TeV}^{-1},\nonumber \\
    &&|g_{aZZ}| \leq 5.89 \times 10^{-3}~{\rm TeV}^{-1},~~~|g_{aWW}| \leq 5.91 \times 10^{-3}~{\rm TeV}^{-1}.\nonumber
\end{eqnarray}
The constraints from the 240/365 GeV CEPC/FCC-ee and the 3/10 TeV muon colliders are summarized in Table~\ref{tab:gconstraints_mono_photon}.

We extend the search for invisible ALPs to the mono-$Z$ channel, considering both leptonic and hadronic $Z$ decays. The hadronic decay channel offers better sensitivity due to its larger branching fraction. Compared to the mono-photon channel, mono-$Z$ searches face more complex SM backgrounds, with additional contributions from di-$W$ and di-$Z$ production processes.  
As in the mono-photon analysis, backgrounds from $W^*$ and $\gamma/Z$ exchanges can be effectively reduced with appropriate transverse momentum and energy cuts. Additionally, an invariant mass window cut on collimated or well-separated lepton or jet pairs significantly suppresses the di-$W$ background. We optimize these cuts individually for each collider and channel, achieving the sensitivities shown in Figure~\ref{fig:monoZ}.  
By combining results from both leptonic and hadronic modes of mono-$Z$ production, we obtain the overall sensitivities summarized in Table~\ref{tab:gconstraints_mono_z}.  
At future $e^+e^-$ colliders at 240/365 GeV, the mono-$Z$ constraints are about one order of magnitude weaker than those from the mono-photon channel. However, at multi-TeV muon colliders, the mono-$Z$ channel provides stronger bounds on ALP-weak-boson couplings ($g_{a\gamma Z}$, $g_{aZZ}$, and $g_{aWW}$). 

For slightly heavier ALPs, \emph{e.g.}, $m_a\sim\mathcal{O}(\GeV)$, the ALPs become short-lived and may decay within the detector. In this scenario, mono-$V$ searches are no longer applicable. However, ALPs can still participate in VBS processes, as illustrated in Figure~\ref{feyn:VBS}, including light-by-light scattering and EW VBS. This work focuses on non-resonant VBS production, where the ALP is off-shell in the scattering process, resulting in cross sections and distributions that are nearly independent of the ALP mass. 
The main background is from the $\ell^+\ell^- \to \gamma\gamma$, for which we need to tag the forward lepton as a trigger for the signal. 
With optimized cuts, we analyze the $e^+e^- \to e^+e^- \gamma\gamma$ and obtain the constraints on the $g_{a\gamma\gamma}$ coupling at the future electron colliders,  summarized in Table~\ref{tab:lbl_gavv}.
At the multi-TeV muon collider, the $g_{a\gamma Z}$ and $g_{aZZ}$ dependence is negligible in $\mu^+\mu^- \to \mu^+\mu^- \gamma\gamma$, and the obtained constraints on $g_{aVV}$ are listed in Table~\ref{tab:lbl_gavv2}.
At multi-TeV muon colliders, the collision energy is high enough for $W/Z$-induced VBS processes to yield sizable cross sections.  
We investigate non-resonant EW VBS scatterings at multi-TeV muon colliders, which offer a unique opportunity to probe the ALP-weak-boson couplings $g_{a\gamma Z}$, $g_{aZZ}$, and $g_{aWW}$, with results summarized in Table~\ref{tab:gconstraints_VBS}. We also incorporate light-by-light scattering constraints and compare them in the right column of Table~\ref{tab:gconstraints_VBS}.  
Our findings show that EW VBS processes at the muon colliders play a minimal role in constraining $g_{a\gamma\gamma}$ but provide essential sensitivity to $g_{a\gamma Z}$, $g_{aZZ}$, and $g_{aWW}$. 

Finally, we combine all constraints from mono-photon, mono-$Z$, and non-resonant VBS processes in the ALP long-lived limit at the Tera-$Z$/240/365 GeV electron colliders and the future multi-TeV muon colliders, as discussed in Sec.~\ref{sec:comb2comp}, with the upper constrains on $g_{aVV}$ summarized in Table~\ref{tab:gconstraints_combine}. A comparison among individual channels is shown in Figure~\ref{fig:contours_combine}: for long-lived light ALPs, the mono-photon channel provides the strongest constraints on $(C_W, C_B)$, while the mono-$Z$ channel offers complementary sensitivity in the $g_{aZZ}$ direction at multi-TeV muon colliders.  
Using the nominal integrated luminosities proposed for future electron colliders \cite{CEPCStudyGroup:2018ghi,CEPCStudyGroup:2018rmc,An:2018dwb,CEPCAcceleratorStudyGroup:2019myu,CEPCPhysicsStudyGroup:2022uwl,CEPCStudyGroup:2023quu} and muon colliders \cite{MuonCollider:2022xlm,Aime:2022flm,Black:2022cth,Accettura:2023ked,Delahaye:2019omf,Bartosik:2020xwr,Schulte:2021hgo,Long:2020wfp,MuonCollider:2022nsa,MuonCollider:2022ded,MuonCollider:2022glg,InternationalMuonCollider:2025sys}, we compare our best sensitivities with existing bounds in Figure~\ref{fig:CompareContours}, considering ALP masses of 1 MeV and 1 GeV, which represent the long-lived and lifetime-independent scenarios, respectively. Future lepton colliders are expected to improve upon the current best LEP bounds \cite{OPAL:2000puu} by one or two orders of magnitude. In particular, the Tera-$Z$ run of the future electron collider embraces a great capacity to probe the light ALP couplings due to its high luminosity and low background. 
Converting these results to ALP-boson couplings via Eq.~(\ref{eq:couplings}), we compare our constraints with existing ones in Figure~\ref{fig:gaVV}. 
Besides the constraints on $g_{a\gamma\gamma}$ from the mono-photon and light-by-light scattering, the mono-$Z$ as well as the non-resonant EW VBS provide additional constraining power on the ALP-weak-boson couplings $g_{aVV}$ at the low and high ALP mass, which can be extended up to $m_a\sim M_Z$ before the on-shell resonant searches take over. 

In summary, we have explored the ALP searches at the future lepton colliders through the mono-$V$ channels and non-resonant vector-boson scattering in detail. Our finding underscore the great discovery potentials of future lepton colliders, which extends the probe to the ALP couplings significantly, with respect to the existing bounds from LEP~\cite{OPAL:2000puu} as well as the current and future LHC~\cite{Bonilla:2022pxu}. 
The conclusions drawn for multi-TeV muon colliders are also applicable to TeV-scale linear $e^+e^-$ colliders, such as the 3 TeV CLIC~\cite{Aicheler:2012bya,Linssen:2012hp,Lebrun:2012hj,CLIC:2016zwp,CLICdp:2018cto,Brunner:2022usy} and 1 TeV ILC~\cite{ILC:2013jhg,ILCInternationalDevelopmentTeam:2022izu}.

\acknowledgments
The authors thank I. Brivio, C. Degrande, G. Durieux, T. Han, O. Mattelaer, and X. Wang for helpful discussions. 
This research is partially supported by the IISN-FNRS convention 4.4517.08, ``Theory of fundamental interactions.''
The work of K. Xie is supported by the U.S. National Science Foundation under Grants No.~PHY-2310291 and PHY-2310497.
Y. Ma acknowledges the support as a Postdoctoral Fellow of the Fond de la Recherche Scientifique de Belgique (F.R.S.-FNRS), Belgium.
Computational resources have been provided by the supercomputing facilities of the Universit\'e catholique de Louvain (CISM/UCL) and the Consortium des \'Equipements de Calcul Intensif en 
F\'ed\'eration Wallonie Bruxelles (C\'ECI) funded by the Fond de la Recherche Scientifique de Belgique (F.R.S.-FNRS) under convention 2.5020.11 and by the Walloon Region.
The work of Y. Wu is supported by the National Natural Science Foundation of China (NSFC) under grant No.~12305112.
The work of S. Bao and H. Zhang is supported by the National Natural Science Foundation of China (NSFC) under grant No.~12447105.
This work used the high-performance computing clusters at Michigan State University.
Y. Ma also acknowledges support from the COMETA COST Action CA22130.
\appendix

\bibliographystyle{JHEP}
\bibliography{ref.bib}

\end{document}